\documentclass[a4paper, fleqn, usenatbib, useAMS]{mnras}

\usepackage{float}
\usepackage{graphicx}
\usepackage{times}
\usepackage{amstext}
\usepackage{amsmath}
\usepackage{amssymb}	% Extra maths symbols
\usepackage{natbib}
\usepackage{url}
\usepackage{setspace}
\usepackage{tabularx}
\usepackage{mathtools}
\usepackage{mathrsfs}
\usepackage{amssymb}
\usepackage{rotating}
\usepackage{xcolor}
\usepackage{graphics}
\usepackage{multirow}

\newcommand{\kms}{\,km\,s$^{-1}$} % kilometres per second
\newcommand{\ntwoh}{N$_{2}$H$^{+}$}

\newcommand{\tco}{$^{13}$CO}

\newcommand{\amm}{NH$_3$}
\newcommand{\tone}{$J=1\rightarrow0$}

\newcommand{\vel}{km\,s$^{-1}$\,pc$^{-1}$}
\newcommand{\scouse}{{\sc scouse}}
\newcommand{\solar}{M$_{\odot}$}

\usepackage[T1]{fontenc}
\usepackage{ae,aecompl}

\usepackage{mathptmx}
\usepackage{txfonts}

\title[Gas kinematics of the CMZ]{Molecular gas kinematics within the central 250\,pc of the Milky Way}

\author[J. D. Henshaw et al.]{J. D. Henshaw$^{1}$\thanks{Contact e-mail: j.d.henshaw@ljmu.ac.uk}, S. N. Longmore$^{1}$\thanks{Contact e-mail: s.n.longmore@ljmu.ac.uk}, J.\ M.\ D.\ Kruijssen$^{2,3}$, B.\ Davies$^{1}$,
\newauthor J.\ Bally$^{4}$,  A.\ Barnes$^{1}$, C.\ Battersby$^{5}$, M.\ Burton$^{6}$, M.\ R.\ Cunningham$^{6}$, J.\ E.\ Dale$^{7}$, 
\newauthor A.\ Ginsburg$^{8}$, K.\ Immer$^{8}$, P.\ A.\ Jones$^{6}$, S.\ Kendrew$^{9}$, E.\ A.\ C.\ Mills$^{10}$, S.\ Molinari$^{11}$, 
\newauthor T.\ J.\ T.\ Moore$^{1}$, J.\ Ott$^{10}$, T.\ Pillai$^{12}$, J.\ Rathborne$^{13}$, P.\ Schilke$^{14}$, A.\ Schmiedeke$^{14}$, 
\newauthor L.\ Testi$^{8}$, D.\ Walker$^{1}$,  A.\ Walsh$^{15}$, and Q.\ Zhang$^{5}$
\\
\\
$^{1}$ Astrophysics Research Institute, Liverpool John Moores University, Liverpool, L3 5RF, UK\\
$^2$ Max-Planck Institut f\"{u}r Astrophysik, Karl-Schwarzschild-Stra\ss e 1, 85748 Garching, Germany\\
$^3$ Astronomisches Rechen-Institut, Zentrum f\"{u}r Astronomie der Universit\"{a}t Heidelberg, M\"{o}nchhofstra\ss e 12-14, 69120 Heidelberg, Germany\\
$^4$ Department of Astrophysical and Planetary Sciences, University of Colorado, UCB 389, Boulder, CO 80309, USA\\
$^5$ Harvard-Smithsonian Center for Astrophysics, 60 Garden St., Cambridge, MA 02138, USA\\
$^6$ School of Physics, University of New South Wales, Sydney, NSW 2052, Australia\\
$^7$ Excellence Cluster `Universe', Boltzmannstra\ss e \ 2, D-85748 Garching, Germany\\
$^8$ European Southern Observatory, Karl-Schwarzschild-Stra\ss e \ 2, D-85748 Garching bei M{\"u}nchen, Germany\\
$^9$ University of Oxford, Department of Physics, Denys Wilkinson Building, Keble Road, Oxford OX1 3RH, UK\\
$^{10}$ National Radio Astronomy Observatory, 1003 Lopezville Rd, Socorro, NM 87801, USA\\
$^{11}$ INAF - Istituto di Astrofisica e Planetologia Spaziali, Via Fosso del Cavaliere 100, I-00133 Roma, Italy\\
$^{12}$ Max Planck Institut f{\"u}r Radioastronomie, Auf dem H{\"u}gel 69, D-53121 Bonn, Germany\\
$^{13}$ CSIRO Astronomy and Space Science, P.O. Box 76, Epping NSW, 1710, Australia\\
$^{14}$ I. Physikalisches Institut, Universit{\"a}t zu K{\"o}ln, Z{\"u}lpicher Stra\ss e \ 77, 50937, K{\"o}ln, Germany\\
$^{15}$ International Centre for Radio Astronomy Research, Curtin University, GPO Box U1987, Perth WA 6845, Australia
}
\date{Accepted 2016 January 11. Received 2015 January 8; in original form 2015 October 6}

\pubyear{2016}

\begin{document}
\label{firstpage}
\pagerange{\pageref{firstpage}--\pageref{lastpage}}
\maketitle

\begin{abstract}
Using spectral-line observations of HNCO, \ntwoh, and HNC, we investigate the kinematics of dense gas in the central $\sim250$ pc of the Galaxy. We present \scouse \ (\textbf{S}emi-automated multi-\textbf{CO}mponent \textbf{U}niversal \textbf{S}pectral-line fitting \textbf{E}ngine), a line-fitting algorithm designed to analyse large volumes of spectral-line data efficiently and systematically. Unlike techniques which do not account for complex line profiles, \scouse \ accurately describes the $\{l,\,b,\,v_{\rm LSR}\}$ distribution of Central Molecular Zone (CMZ) gas, which is asymmetric about Sgr~A* in both position and velocity. Velocity dispersions range from $2.6\,{\rm km\,s^{-1}}<\sigma<53.1\,{\rm km\,s^{-1}}$. A median dispersion of 9.8\,\kms, translates to a Mach number, $\mathcal{M}_{\rm 3D}\ge28$. The gas is distributed throughout several ``streams'', with projected lengths $\sim100-250$\,pc. We link the streams to individual clouds and sub-regions, including Sgr~C, the 20 and 50\,\kms \ clouds, the dust ridge, and Sgr~B2. Shell-like emission features can be explained by the projection of independent molecular clouds in Sgr~C and the newly identified conical profile of Sgr~B2 in $\{l,\,b,\,v_{\rm LSR}\}$ space. These features have previously invoked supernova-driven shells and cloud-cloud collisions as explanations. We instead caution against structure identification in velocity-integrated emission maps. Three geometries describing the 3-D structure of the CMZ are investigated: i) two spiral arms; ii) a closed elliptical orbit; iii) an open stream. While two spiral arms and an open stream qualitatively reproduce the gas distribution, the most recent parameterisation of the closed elliptical orbit does not. Finally, we discuss how proper motion measurements of masers can distinguish between these geometries, and suggest that this effort should be focused on the 20\,\kms \  and 50\,\kms \ clouds and Sgr~C. 
\end{abstract}

\begin{keywords}
stars: formation -- ISM: clouds -- ISM: kinematics and dynamics -- ISM: structure -- Galaxy: centre -- galaxies: ISM.
\end{keywords}

%%%%%%%%%%%%%%%%%%%%%%%%%%%%%%%%%%%%%%%%%%%%%%%%%%

%%%%%%%%%%%%%%%%% BODY OF PAPER %%%%%%%%%%%%%%%%%%

% The MNRAS class isn't designed to include a table of contents, but for this document one is useful.
% I therefore have to do some kludging to make it work without masses of blank space.

\section{Introduction}

The inner $\sim500$\,pc of our Galaxy, the ``Central Molecular Zone'' (hereafter, CMZ) contains $\lesssim10$ per cent of the Milky Way's molecular gas but $\sim80$ per cent of the dense ($\gtrsim10^{3}$\,cm$^{-3}$) gas: a reservoir of $2-6\times10^{7}$\,\solar \ of molecular material (see e.g. \citealp{morris_1996, molinari_2014} and references therein). The physical properties of the interstellar medium in the CMZ differ substantially from those in the disc. Gas column and volume densities can be $\sim2$ orders of magnitude greater \citep{longmore_2013, kruijssen_2014b, rathborne_2014b}, velocity dispersions measured for a given physical size are larger (e.g. \citealp{bally_1988,shetty_2012}), and although there exists a significant fraction of cold dust (e.g. \citealp{molinari_2011}), gas temperatures can range from 50-400\,K (\citealp{ao_2013, mills_2013, ott_2014, ginsburg_2015}). Several of the physical, kinematic, and chemical properties of the CMZ are, by comparison to the Galactic disc, more similar to those observed in nearby and high-redshift galaxies. Understanding the star formation process within the CMZ may therefore help us to understand star formation across cosmological timescales (particularly at the epoch of peak star formation density, $z\sim2-3$; as suggested by \citealp{kruijssen_2013}). 

The distribution of material within the CMZ has been subject to intense scrutiny for several decades (e.g. \citealp{bania_1977, liszt_1977, bally_1987, bally_1988, binney_1991, sofue_1995, oka_1998, tsuboi_1999, molinari_2011, jones_2012, jones_2013, ott_2014, kruijssen_2015}). More recently, this research has been motivated, at least in part, by theoretical developments which suggest that obtaining a detailed description of the gas kinematics of the inner 250\,pc will aid our understanding of some of the key open questions in Galactic Centre research (for example, highly turbulent motions within CMZ gas may explain the low levels of star formation activity relative to the immense dense gas reservoir; \citealp{kruijssen_2014b, krumholz_2015}). 

One of the most striking features of the CMZ, noted from the earliest studies, is the complex gas distribution in position-position-velocity (PPV) space. Most dense gas within $\sim150$\,pc from the Galactic centre (corresponding to an angular distance of $\sim1\degr$ at an assumed distance of 8.3\,kpc; \citealp{reid_2014}) is distributed throughout several coherent gas streams spanning $\sim200$\,\kms \ in velocity (with significant substructure also evident). Unfortunately, our view through the Galactic plane restricts our ability to determine unique distances to all of these components. Inferring the 3-D structure of the gas therefore requires assumptions about the relative distance of emission features combined with some degree of kinematic modelling. 

Obtaining a 3-D picture of the CMZ is typically motivated by either the desire for a holistic understanding of the gas distribution and dynamics (e.g. \citealp{binney_1991, sofue_1995, tsuboi_1999, molinari_2011, kruijssen_2015}), or to aid the description of individual clouds or sub-regions (e.g. \citealp{liszt_1995, sato_2000, kendrew_2013, longmore_2013b, ott_2014, rathborne_2014}). However, a detailed description of the molecular gas distribution and kinematics, encompassing a \emph{variety} of molecular line tracers, is still lacking. As mentioned above, the gas in the CMZ spans a wide range of physical properties. Any single molecular line transition therefore only traces a fraction of the gas. While opting for molecular species and line transitions that are highly-abundant and widespread (for example, the lower $J$ transitions of CO) may seem like a solution to this problem, emission from such molecules can very quickly become optically thick, making them poor probes of the underlying kinematics in high column density regions. 

To date, the kinematics of the CMZ have been described using a combination of position-velocity diagrams (typically Galactic longitude-velocity diagrams over specific Galactic latitude ranges; e.g. \citealp{sofue_1995, oka_1998, tsuboi_1999}), channel maps (highlighting the distribution of molecular line emission integrated incrementally in velocity; e.g. \citealp{bally_1988}), and moment analysis (integrated intensity, and intensity-weighted velocities as a function of position; e.g. \citealp{sofue_1995, ott_2014}). Such techniques are advantageous as they are simple to implement and well understood. However, as many of the authors themselves acknowledge, the output from these methods can be subjective and very easily misinterpreted.

In this paper we try to overcome these challenges. We make use of data from the Mopra CMZ survey \citep{jones_2012}, which has simultaneously mapped the inner (projected) 250\,pc of the Galaxy in many molecular line transitions. We have developed a new line-analysis tool (\scouse\footnote{\scouse \ is publicly available at \url{https://github.com/jdhenshaw/SCOUSE}}; \textbf{S}emi-automated multi-\textbf{CO}mponent \textbf{U}niversal \textbf{S}pectral-line fitting \textbf{E}ngine) which systematically fits the complex emission from these lines with multiple Gaussian components. We then use the output of this line-fitting to: (i) quantify the robustness of moment analysis with respect to the fit results (\S~\ref{Section:results_linefitting}); (ii) provide a detailed description of the molecular gas kinematics (\S~\ref{Section:results_global} and \S~\ref{Section:results_local}); (iii) compare the fit results against the current competing interpretations for the 3-D structure of the CMZ (\S~\ref{Section:results_models}); (iv) show how future proper motion measurements will be able to observationally distinguish between different interpretations for the 3-D structure of the CMZ, and specify where to concentrate this research effort (\S~\ref{Section:results_obstests}).

Section~\ref{Section:data} describes the data used throughout this paper. Our methodology is described in Section~\ref{Section:method}, with Section~\ref{Section:method_scouse} providing a detailed description of \scouse \ and Section~\ref{Section:method_cmz} discussing how we have tuned the algorithm for use within the CMZ. Section~\ref{Section:results} discusses our kinematic results, and we draw our final conclusions in Section~\ref{Section:conclusions}.

\section{Data}\label{Section:data}

This paper makes use of data from the Mopra (a 22\,m radio telescope of the Australia Telescope National Facility) CMZ survey.\footnote{These data are publicly available at \url{http://atoa.atnf.csiro.au/CMZ}} Specifically, we utilise molecular lines first published by \citet{jones_2012}. The Mopra Monolithic Microwave Integrated Circuit was tuned to a central frequency of 89.3\,GHz, to cover the range 85.3 to 93.3\,GHz, resulting in the detection of 20 molecular lines. The observations cover the region $-0\fdg72<l<1\fdg80$, $-0\fdg30<b<0\fdg22$. Typical root-mean-square (rms) noise level noise level prior to post-processing (see below) is of the order $40-80$\,mK per $\sim1$\,MHz Hanning-smoothed channel. Please see \citet{jones_2012} for a full description of the observations and data reduction.

We select the HNCO 4(0,4)-3(0,3) transition (rest~freq.~$\approx~87.925$\,GHz) as our primary tracer of the kinematics. The \tone \ transitions of \ntwoh \ (rest freq. of the main ${\rm J,\,F_{1},\,F}~=~1,\,2,\,3 \rightarrow 0,\,1,\,2$ hyperfine component $\approx~93.174$\,GHz) and HNC (rest freq. of ${\rm J,\,F}~=~\,1,\,2\rightarrow0,\,1$ hyperfine component $\approx90.664$\,GHz) are also utilised. We address our line selection explicitly in \S~\ref{Section:method_cmz}. 

Additional post-processing was performed using the {\sc gildas} packages {\sc class} and {\sc mapping}. The data cubes were convolved with a Gaussian kernel and smoothed in velocity to further increase the signal-to-noise. The final data cubes have an effective spatial resolution of $\sim60$\,arcsec (corresponding to $\sim2.4$\,pc), a pixel size of 30\,arcsec\,$\times$\,30\,arcsec ($1.2\,{\rm pc}\times1.2\,{\rm pc}$), and a spectral resolution of 2\,\kms \ (which is sufficient, given the broad $\sim10$\,\kms \ velocity dispersions commonly observed throughout the CMZ; see \S\,\ref{Section:results_linewidth}). Our analysis focuses on the inner $\sim250$\,pc of the CMZ and the main gas streams. We therefore restrict the range of the observations to $-0\fdg65<l<1\fdg10$ (corresponding to a physical extent of $\sim250$\,pc) and $-0\fdg25<b<0\fdg20$ ($\sim65$\,pc). Since the absolute flux calibration is irrelevant to the present kinematic analysis, intensities remain in $T^{*}_{A}$ scale, without correction for the specific beam efficiency at the frequency of each transition. 
\section{Methodology}\label{Section:method}

The primary aim of this study is to obtain a detailed description of the molecular gas kinematics within the CMZ. To date, studies investigating the kinematics of gas in the CMZ have used techniques such as position-velocity diagrams, channel maps, and intensity-weighted velocities as a function of position (or some combination of these). Such techniques are well-established and simple to implement. However, their output can be misinterpreted. This becomes particularly pertinent in an environment as kinematically complex as the CMZ, where velocity gradients, multiple line-of-sight velocity components, and broad line-widths are commonplace. The presence of multiple velocity components, for example, can lead to artificially broadened intensity-weighted velocity dispersions (as discussed by e.g. \citealp{ott_2014}) or lead to confusion when either identifying gas structures or inferring the connectivity between multiple gas structures. 

In contrast, there has been no attempt to use line-fitting to analyse the molecular gas kinematics over the entire inner 250\,pc of the Galaxy. Line-fitting is beneficial as it enables one to account for the complex spectral profiles which are observed throughout the CMZ. The issue however, is that fitting large quantities of spectral line data ``by hand'' can be both subjective and time consuming. The region of the CMZ selected for this study contains in excess of 10000 spectra \emph{per cube} (following the post-processing steps outlined in \S~\ref{Section:data}). Additionally, obtaining reliable spectral fits necessitates the provision of reasonable estimates to model parameters to help minimization algorithms to converge towards a solution. In the CMZ, the kinematics vary \emph{substantially} as a function of position. Therefore, although line-fitting may appear an attractive alternative to techniques such as moment analysis, the challenge is to minimise time-consuming user-interaction while maintaining the ability to describe complex line-profiles in an efficient and systematic manner.

Our solution to this problem is the development of a new line-analysis tool, written in {\sc idl}, named \scouse. Building on the algorithm developed by \citet{henshaw_2014}, the primary function of \scouse \ is to fit large quantities of spectral line data in a robust, systematic, and efficient way. This is achieved by: i) systematically excluding regions of low significance from the analysis, reducing the total number of spectra to be fit and focusing the attention on the most significant emission; ii) minimizing time-consuming user-interaction by breaking a map up into small regions and requiring the user to fit \textit{only} the spatially-averaged spectra extracted from these regions; iii) using the best-fitting solutions to these spatially-averaged spectra as \emph{non-restrictive guides} when fitting the individual spectra, minimizing bias and ensuring that each spectrum is processed systematically. Provided therefore, that the best-fitting solution to a spatially-averaged spectrum is able to adequately describe the individual spectra belonging to the same region, \scouse \ can provide an efficient and systematic approach to fitting large quantities of spectral-line data. 

\begin{figure*}
\begin{center}
\includegraphics[trim = 0mm 0mm 0mm 0mm, clip, width = 0.98\textwidth]{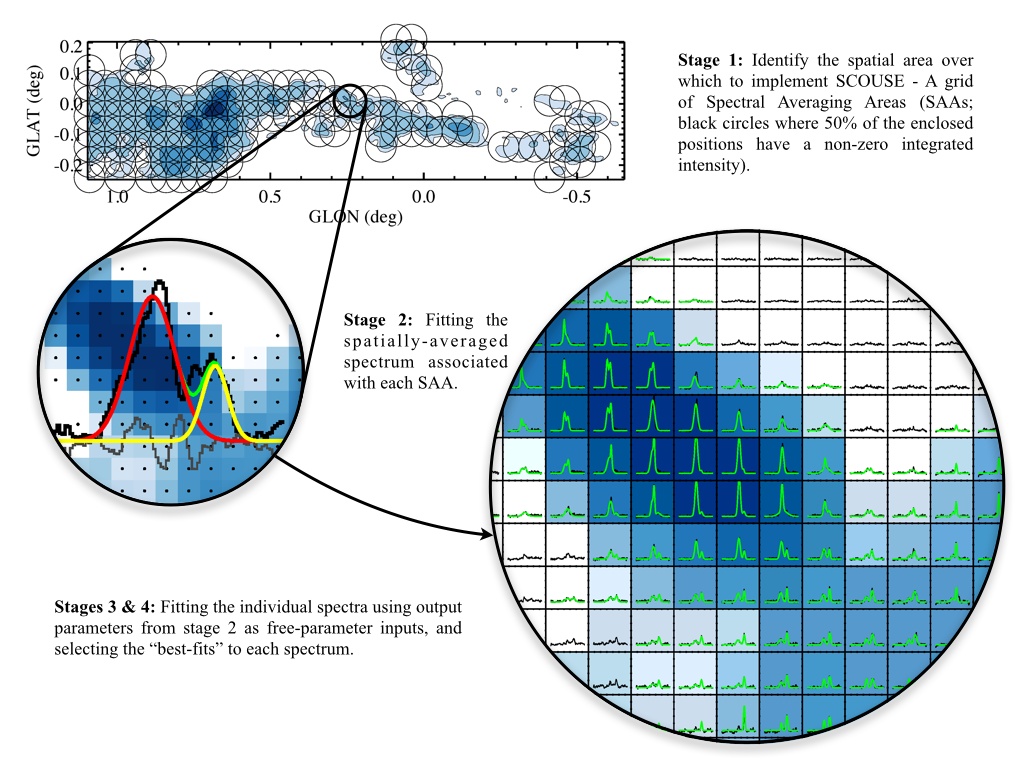}
\end{center}
\caption{A schematic figure highlighting the main stages of \scouse. All data shown correspond to the HNCO 4(0,4)-3(0,3) transition from the Mopra CMZ survey \citep{jones_2012}. The top left image displays a contour map of the integrated emission of HNCO throughout the CMZ and refers to stage~1 (see \S~\ref{Section:method_coverage} for details). The filled contours represent 1, 5, 25, 45, and 65\,per cent of peak (integrated) emission ($\sim122$\,K\,\kms). Black circles are the SAAs that make up the coverage (regions whereby 50 per cent of the enclosed spectra have non-zero integrated emission). The centre-left image is a close up image of the SAA associated with G0.253+0.016, an IRDC with known kinematic complexity (e.g. \citealp{kauffmann_2013, bally_2014, johnston_2014, rathborne_2014, rathborne_2015, mills_2015}). The spatially-averaged spectrum taken from this SAA is shown in black. The velocity range displayed is $-110.0\,{\rm km\,s^{-1}}<v_{\rm LSR}<170.0\,{\rm km\,s^{-1}}$. The line profile is dominated by at least two distinct components (the residual spectrum in grey suggests that additional velocity components may be present). The minimization algorithm {\sc mpfit} \citep{markwardt_2009} finds the best fitting solution to Equation~\ref{Equation:gauss_func}, following the insertion of (user-provided) initial estimates for the parameters that describe each Gaussian component (stage~2; see \S~\ref{Section:method_SAAfitting} for more details). The two Gaussian profiles (seen in red and yellow), and total profile (seen in green), are the result of this process. The bottom-right image represents stages~3 and 4. It depicts the same SAA however, now the resultant fit to each of the pixels marked by black dots in the stage~2 image is displayed. The observed line profiles are the result of applying the algorithm described in Figure~\ref{Figure:flowchart} (stage~3; see \S~\ref{Section:method_indivfits}). Alternative solutions to each composite spectrum are provided from neighbouring SAAs. In this way, the final results are not restricted to two velocity components. Instead, these are the best solutions selected from a sample of overlapping SAAs (stage~4; see \S~\ref{Section:method_bestfits}). }
\label{Figure:cartoon}
\end{figure*}

The following section is dedicated to discussing the main stages of the fitting procedure. Figure~\ref{Figure:cartoon} is a cartoon representation of the routines presented in \S~\ref{Section:method_coverage}-\S~\ref{Section:method_bestfits}, and each will be described in more detail in their respective sections. 

\subsection{SCOUSE: Semi-automated multi-COmponent Universal Spectral-line fitting Engine}\label{Section:method_scouse}

% Figure
% Figure

\subsubsection{STAGE 1: Identifying the spatial area over which to implement \scouse}\label{Section:method_coverage}

The first stage identifies which regions of a 3-D spectral line data cube are to be analysed. This optimises the fitting routine, ensuring that only regions of significant (user-defined) emission are included in the fitting process.\footnote{This also aids in removing spectra that exhibit baseline artefacts such as the ripples noted by \citet{jones_2012}. These artefacts have the potential to mimic low-level emission and influence our line-fitting. Although this step does not remove all spectra with poor baselines, our results are not affected significantly (regions with low-level emission appear to be influenced more).} The area over which \scouse \ is implemented is referred to as the ``coverage'', and it is defined as follows:

\begin{enumerate}

\item The user is required to provide several input values: 1) ranges in position and velocity over which to fit the data; 2) an estimate of the typical spectral rms, $\sigma^{\rm est}_{\rm rms}$; 3) a radius, $R_{\rm SAA}$, that defines the size of the ``spectral averaging areas'' (SAAs; see definition below) used in the next step of the fitting procedure. \label{S1:inputs} \\

\item An estimate of the spectral rms is required in order to calculate an integrated intensity (zeroth order moment) map. All voxels\footnote{A pixel in position-position-velocity space.} where the intensity is less than 3\,$\sigma^{\rm est}_{\rm rms}$ are set to 0\,K before calculating the integrated intensity. Locations where no voxels exhibit emission greater than $3\sigma^{\rm est}_{\rm rms}$ are found, flagged, and the integrated emission is assigned a value of 0\,K\,\kms. \\

\item An integrated emission map (integrated over the velocity range provided by the user in step\,\ref{S1:inputs} above) is produced from the non-flagged map locations. \label{cov:ii} \\

\item \scouse \ identifies the spatial area covered by the significant emission in the resulting map. \\

\item This area is segmented into a grid of smaller regions, which we refer to as SAAs, whose spacing is equivalent to $R_{\rm SAA}$. This spacing ensures a smooth transition between adjacent SAAs. \label{cov:nyquist}\\

\item SAAs where $\geq50$ per cent of the enclosed pixels have non-significant integrated emission are rejected from the fitting procedure. \label{cov:SAA}\\

\item The remaining SAAs define the coverage, the area over which the fitting routine is to be implemented.

\end{enumerate}

The top-left of Figure~\ref{Figure:cartoon} indicates the coverage produced from the HNCO molecular line emission used throughout this paper (for more information on our line selection, refer to \S~\ref{Section:method_cmz}). The image in the background is the integrated emission map created in step~\ref{cov:ii} above. The black circles indicate the locations of the SAAs that make up the coverage (see step~\ref{cov:SAA}).
\subsubsection{STAGE 2: Fitting the spatially-averaged spectrum extracted from each SAA}\label{Section:method_SAAfitting}
 
\begin{enumerate}

\item A spatially-averaged spectrum is created from all pixels contained within an SAA. \label{fit:avspec}\\

\item Using this spectrum as a guide, the user is asked to provide a range of line-free channels over which the spectral rms will be calculated, $\sigma_{\rm rms}$ (defined as the standard deviation of the data over the user-defined velocity range). This velocity range is stored and used during stage\,3  (see \S~\ref{Section:method_indivfits}). \label{S2:rms} \\

\item The user is prompted to specify how many velocity components are evident in the spatially-averaged spectrum. For each component, an estimate of the intensity, centroid velocity, and dispersion must be provided. \label{fit:ncomp} \\

\item The minimization algorithm, {\sc mpfit} \citep{markwardt_2009},\footnote{{\sc mpfit} is available for download at: \url{http://purl.com/net/mpfit}} uses these input parameters to find the best-fitting solution to the following function:
\begin{equation}
I(v)=\sum_{i=0}^{n} I_{\rm peak, i}\exp{\big[-(v-v_{\rm 0, i})^{2})/(2\sigma_{\rm i}^{2})\big]},
\label{Equation:gauss_func}
\end{equation}
whereby $I_{\rm peak, i}$, $v_{\rm 0, i}$, $\sigma_{\rm i}$ represent the peak intensity, centroid velocity, and velocity dispersion of the $i^{\rm th}$ spectral component, respectively (and $n$ refers to the total number of components identified). \label{fit:markwardt}\\

\item The best-fitting solution is displayed, along with the residual spectrum and $\chi^{2}_{\rm red}$. If deemed satisfactory, the best-fitting solution and residual spectrum are written to file. Alternatively, steps~\ref{fit:ncomp}-\ref{fit:markwardt} can be repeated. \label{fit:bestfit}

\end{enumerate} 
 
Steps~\ref{fit:avspec}-\ref{fit:bestfit} are then repeated until each SAA has a representative solution. 

An example SAA spectrum is shown in the centre-left of Figure~\ref{Figure:cartoon}, accompanied by its best-fitting solution (and the residual spectrum). Here, the line-profile of the selected SAA is dominated by (at least) two spectral components. The best-fitting solution to Equation~\ref{Equation:gauss_func} is shown in green (with the two individual components shown in red and yellow, respectively). 
\subsubsection{STAGE 3: Fitting individual spectra}\label{Section:method_indivfits}

% Figure
\begin{figure}
\begin{center}
\includegraphics[trim = 20mm 0mm 0mm 0mm, clip, width = 0.55\textwidth]{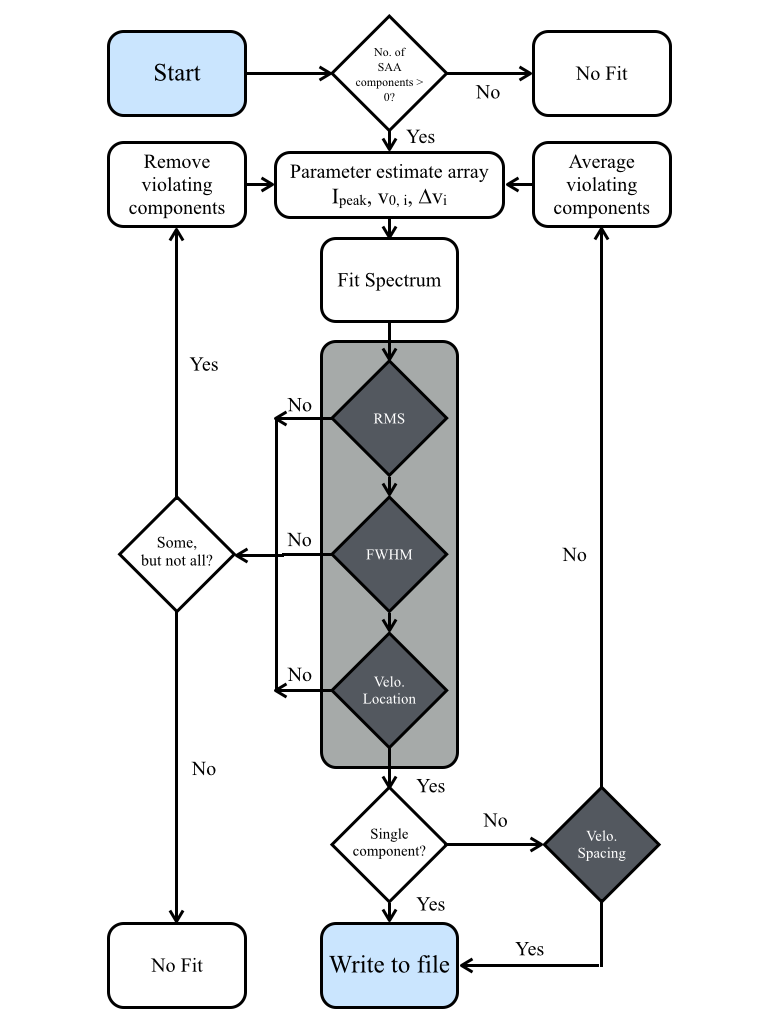}
\end{center}
\caption[]{A flowchart illustrating the algorithm governing stage~3 of the \scouse \ fitting routine (\S~\ref{Section:method_indivfits}). Decisions are indicated by diamonds, with tolerance-dependent decisions shaded in grey.}
\label{Figure:flowchart}
\end{figure}
% Figure

To fit individual spectra contained within a given SAA, \scouse \ implements the following steps (these are further depicted in the flow-chart in Figure~\ref{Figure:flowchart}):

\begin{enumerate}

\item An array of estimates (and associated uncertainties) to the parameters of Equation~\ref{Equation:gauss_func} is created. If this is the first iteration, the best-fitting parameters of the spatially-averaged spectrum extracted from the parent SAA populate the array. \label{fit:array}\\

\item These are passed to {\sc mpfit}, which uses these input parameters to find the best-fitting solution to Equation~\ref{Equation:gauss_func} (these are indicated with a superscript `bf' in the following discussion). \label{fit:mpfit} \\

\item For a ``good'' fit to an individual spectrum, the best-fitting parameters extracted in step~\ref{fit:mpfit} must satisfy several conditions. Each condition is dependent on user-defined tolerance level. The dark box in the centre of Figure~\ref{Figure:flowchart} signifies the criteria that \emph{all} identified components must meet in order for the algorithm to progress towards a satisfactory solution. These conditions are as follows: \label{fit:conditions} \\

\begin{description}

\item[\textbf{Condition 1:} ``RMS'':] The peak intensity of each identified Gaussian component, $I^{\rm bf}_{\rm peak, i}$, must satisfy:
\begin{equation}
I^{\rm bf}_{\rm peak, i}>T_{1}\sigma_{\rm rms},
\end{equation}
where $T_{1}$ is a unitless tolerance level, defined as a multiple of the local rms noise, $\sigma_{\rm rms}$.\\ 

\item[\textbf{Condition 2:} ``FWHM'':] The full-width at half-maximum line-width of each identified component, $\Delta v^{\rm bf}_{\rm i}$, must satisfy the following condition: 
\begin{equation}
T_{2}v_{\rm res}<\Delta v^{\rm bf}_{\rm i}<T_{3}\Delta v^{\rm SAA}_{\rm i}, 
\end{equation}
where $T_{2}$ and $T_{3}$ are unitless tolerance levels, corresponding to multiples of the spectral resolution of the data, $v_{\rm res}$, and the FWHM of feature with the most similar $I_{\rm peak}$, $v_{\rm 0}$, and $\sigma$ in the parent SAA solution, $\Delta v^{\rm SAA}_{\rm i}$, respectively.\\

\item[\textbf{Condition 3:} ``Velocity Location'':] The centroid velocity of each identified component, $v^{\rm bf}_{0,i}$, must satisfy the condition:
\begin{equation}
[v^{{\rm SAA}}_{\rm 0,i}-(T_{4}\sigma^{\rm SAA}_{\rm i})]<v^{{\rm bf}}_{\rm 0,i}<[v^{{\rm SAA}}_{\rm 0,i}+(T_{4}\sigma^{\rm SAA}_{\rm i})],
\end{equation}
where $T_{4}$ is a unitless tolerance level, corresponding to a multiple of the velocity dispersion of the closest matching spectral feature in the parent SAA solution, $\sigma^{\rm SAA}_{\rm i}$, and $v^{{\rm SAA}}_{0,i}$ is the centroid velocity of that same feature.\\

\end{description}

If some, but not all, spectral components satisfy conditions~1--3, those which do satisfy the conditions form a new array of parameter estimates for use with {\sc mpfit}, whilst those which do not, are discarded. Steps~\ref{fit:array}--\ref{fit:conditions} are repeated until either \emph{none} or \emph{all} spectral components in the current best-fitting solution satisfy conditions\,1--3. In this way, the array of parameter estimates is updated with each iteration. 

At this stage, if the spectrum is best described by a single Gaussian component, all requirements are deemed to be satisfied and the resultant fit parameters are written to file. \\

\item If multiple components are present then a final condition is implemented: \label{fit:velsep} 

\begin{description}

\item[\textbf{Condition 4:} ``Velocity Spacing'':] The separation in centroid velocity between two adjacent Gaussian components must satisfy:
\begin{equation}
|v^{{\rm bf}}_{\rm 0,i}-v^{{\rm bf}}_{\rm 0,min}|>T_{5}\Delta v^{\rm bf}_{\rm min}
\end{equation}
where $v^{{\rm bf}}_{\rm 0,i}$ and $v^{{\rm bf}}_{\rm 0,min}$ refer to the centroid velocities of the $i^{\rm th}$ component and that which has the smallest velocity separation from the $i^{\rm th}$ component, respectively, and $T_{5}$ is a unitless tolerance level. \scouse \ identifies which component (either the $i^{\rm th}$ or that which is closest in velocity to the $i^{\rm th}$ component) has the smallest FWHM line-width, and this value is used to define $\Delta v^{\rm bf}_{\rm min}$. If this condition is not satisfied the two components are deemed \emph{indistinguishable}.

\end{description}

In the event that condition~4 is not met, \scouse \ takes the average of the two (or more) violating components. These form a new array of parameter estimates. Steps~\ref{fit:array}-\ref{fit:velsep} [or Steps~\ref{fit:array}-\ref{fit:conditions}, depending on whether the best-fitting solution is reduced to a single Gaussian] are repeated until \emph{all} spectral components satisfy the requisite conditions. \\

\item Once all conditions are satisfied, the resultant fit parameters are written to file. The user has the option to revisit spectra with no best-fitting solution, as a result of the above algorithm, during the next stage of the process. \\

\item In addition to the best-fitting parameters to Equation~\ref{Equation:gauss_func}, the following information is also written to file:

\begin{enumerate}

\item The spectral rms, $\sigma_{\rm rms}$.\\
\item The residual value, $\sigma_{\rm resid}$ (defined as the standard deviation of the residual spectrum). \\
\item $\chi^{2}$, the number of degrees of freedom, and $\chi^{2}_{\rm red}$. All of which are provided by {\sc mpfit}. \\
\item The Akaike Information Criterion \citep{akaike_1974}, defined as:
\begin{equation}
	AIC = \chi^{2}+2k+\frac{2k(2k+1)}{(N-k-1)},
\end{equation}
where $k$ refers to the number of free-parameters needed to describe the best-fitting solution and $N$ is the number of measurements. This provides a statistical method by which \scouse \ can select a preferred model to represent the data when faced with several choices (i.e. best-fitting solutions with different numbers of velocity components). \\

\end{enumerate}

\end{enumerate}

This process is repeated for each spectrum contained within a parent SAA, before moving on to the next SAA.

\subsubsection{STAGE 4: Selecting the best fits}\label{Section:method_bestfits}

As discussed in \S~\ref{Section:method_coverage}, the coverage grid is arranged such that a region of overlap exists between adjacent SAAs. Consequently, there are (often) multiple fits to any given spectrum. This is a novel feature of \scouse \ which becomes particularly important when considering adjacent SAAs for which the spatially-averaged spectra are best described by a different number of velocity components (i.e. those SAAs where a ``transition region'' exists between the two). In this way, it is possible to identify the best-fitting solution to any given spectrum when faced with several choices. To find the best-fitting solution to any given spectrum \scouse \ follows the following procedure:

\begin{enumerate}
\item Where the spatially-averaged spectra of adjacent SAAs are best-described by the same number of velocity components, {\sc mpfit} will often converge to the same solution. In such instances, \scouse \ retains only one solution and removes any duplicates.\\
\item Best-fitting solutions whose composite Gaussian components have either $I^{\rm bf}_{\rm i}$ or $\sigma^{\rm bf}_{\rm i}$ less than 5\,$\times$ the parameter uncertainty are discarded. This is a quality control measure to ensure only statistically significant solutions are considered.\\
\item Out of the remaining available solutions, that which has the smallest value of the $AIC$ (see \S~\ref{Section:method_indivfits}) is deemed to provide the best description of the spectral line profile.
\end{enumerate}

This process is repeated for all positions. As an optional final step, all resultant fits are then checked (by eye). The user has the option to amend the result for any given spectrum. This can be done in three main ways: 1) the user can seek an alternative solution to a spectrum where multiple solutions are available (a result of our SAA sampling); 2) manually refit the spectrum; 3) remove completely the current solution from the data file. 

The final image in the schematic (see bottom-right of Figure~\ref{Figure:cartoon}) depicts the best-fitting solutions to the spectra in the selected SAA. \scouse \ then produces a file containing the following parameters for each spectral component: X-position, Y-position, line intensity (with uncertainty), centroid velocity (with uncertainty), FWHM (with uncertainty), $\sigma_{\rm rms}$, $\sigma_{\rm resid}$, $\chi^{2}$, the number of degrees of freedom, $\chi^{2}_{\rm red}$, $AIC$.

\begin{figure}
\begin{center}
\includegraphics[trim = 18mm 15mm 0mm 40mm, clip, width=0.48\textwidth]{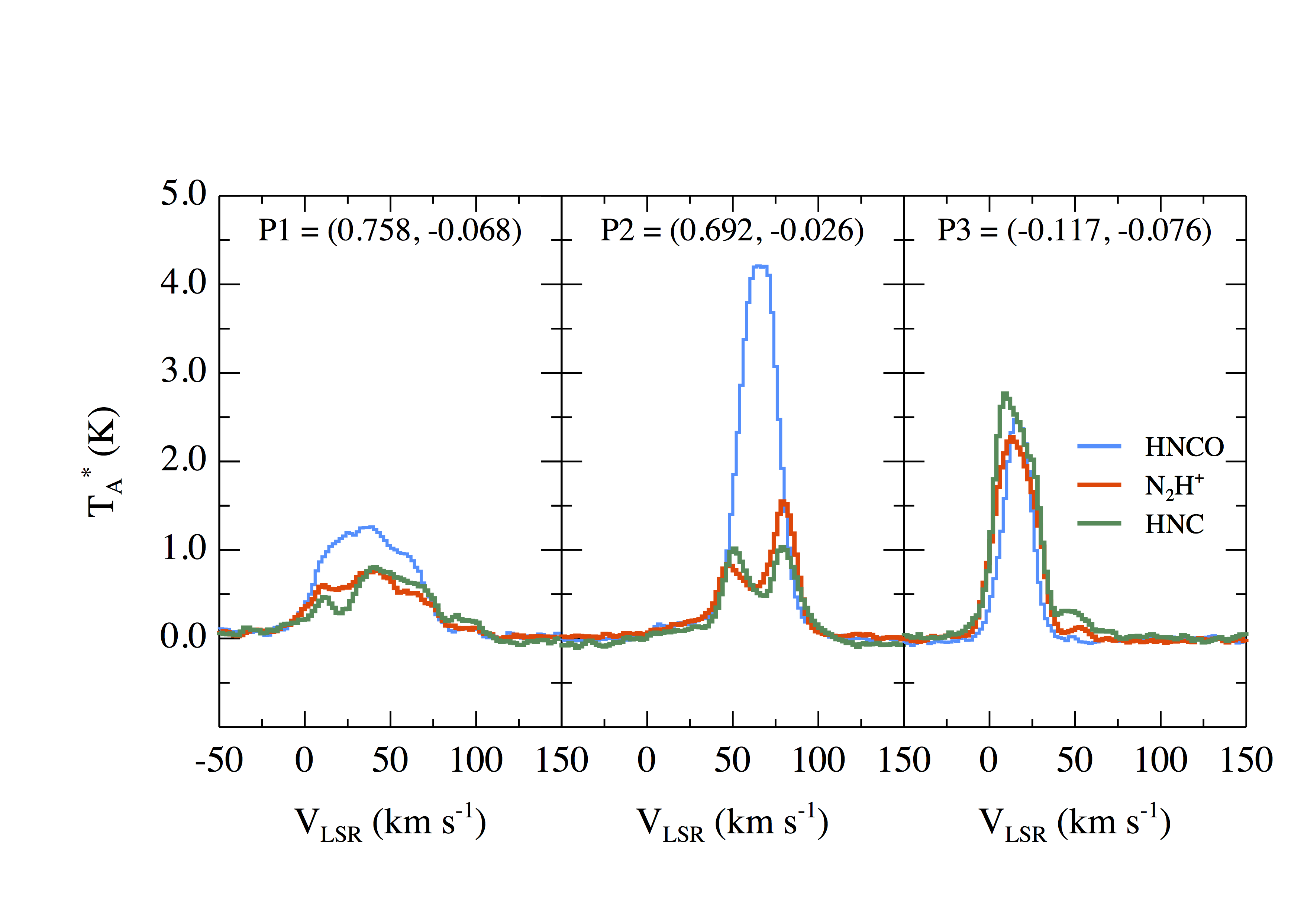}
\end{center}
\caption{Selected spectra from the Mopra CMZ survey (the positions of these spectra are highlighted with `+' symbols in Figure\,\ref{Figure:ncomp}). The spectra of three molecular lines are shown within each panel, HNCO (blue), \ntwoh \ (red), and HNC (light green). P1 and P2 are close to the Sgr B2 region (with P2 representing the location of peak HNCO emission), and P3 represents the peak in HNCO emission at negative longitudes (at $\sim$\,the location of the 20\,\kms \ cloud). Note the double-peaked profiles (possibly indicating self-absorption) in all but the HNCO line at P2. HNCO is selected as our primary tracer of the kinematics within the CMZ. }
\label{Figure:spectra}
\end{figure}

\subsection{SCOUSE: Application to the Central Molecular Zone}\label{Section:method_cmz}

\scouse \ is designed to work with spectral-line data taken from a variety of different regions. Consequently, a number of considerations must first be made in order to optimise the algorithm for use in a particular environment. 

\begin{figure*}
\begin{center}
\includegraphics[trim = 20mm 40mm 0mm 80mm, clip, width = 0.98\textwidth]{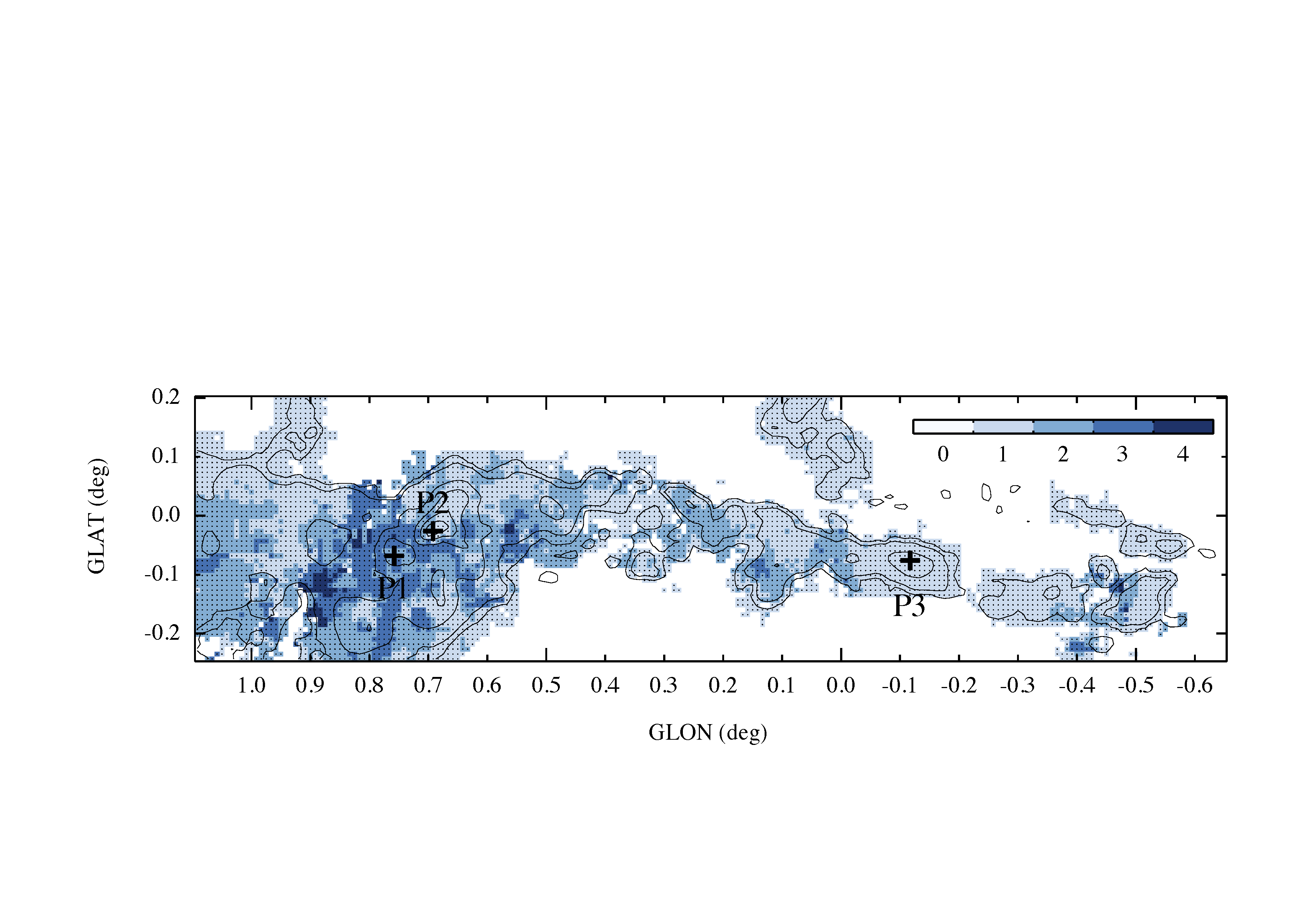}
\end{center}
\caption{Highlighting the spatial distribution of HNCO within the CMZ. Each coloured pixel refers to a location with a corresponding best-fitting solution. Each pixel is colour-coded according to the number of velocity components identified at that location (note the scale bar in the top-right hand corner). Contours show the integrated intensity of HNCO. Contour levels are, 1, 5, 25, 45, 65, and 85 per cent of peak (integrated) emission ($\sim$\,122\,K\,\kms). Positions indicated with `+' symbols, P1, P2, and P3 correspond to the locations of selected spectra in Figure\,\ref{Figure:spectra}.}
\label{Figure:ncomp}
\end{figure*}

An important consideration is the line selection. As discussed in \S~\ref{Section:data}, the Mopra CMZ survey provides a total of 20 molecular lines at 3\,mm. \citet{jones_2012} note that molecular species exhibiting strong emission over most of the imaged area (e.g. HCN, HCO$^{+}$, HNC), are typically optically thick in the highest column density regions (e.g. the Sgr B2 cloud complex). Conversely, lines that do not suffer prominently from optical depth effects, typically cover a smaller area. The aim therefore, is to identify a tracer that covers a significant fraction of the CMZ that also suffers minimally from self-absorption. Selecting an optically thin line ensures a more accurate description of the line-of-sight kinematics. 

Figure~\ref{Figure:spectra} shows spectra taken from three locations within the CMZ. P1 and P2 (left and centre panels, respectively) are located towards the Sgr B2 molecular cloud complex (with P2 representing the location of peak HNCO emission over the region and P1 a secondary peak), whereas P3 (right panel) represents the peak in HNCO emission at negative longitudes (at the approximate location of the 20\,\kms \ cloud). At each position, the spectra from the HNCO (blue), \ntwoh \ (red), and HNC (green) lines at 3\,mm are shown. 

Of the 20 molecular lines provided by the Mopra CMZ survey we select HNCO as our \emph{primary} tracer of the gas kinematics. Our line selection is based on the following factors:

\begin{enumerate}

\item Towards the Galactic centre, HNCO is extended (e.g. \citealp{dahmen_1997, jones_2012}).\footnote{HNCO is also extended towards the nuclei of other galaxies e.g. IC~342 \citep{meier_2005}.} This allows us to study the kinematics of a significant fraction of CMZ gas. Furthermore, HNCO is observed to be abundant across a wide variety of conditions, from shocked gas, to more quiescent dense molecular clouds, to low- and high-mass star-forming regions \citep{jackson_1984, blake_1987, van-dishoeck_1995, zinchenko_2000, rodriguez_2010, li_2013}. \\

\item In the spectrum extracted at the position P2 (central panel, Figure~\ref{Figure:spectra}), both the \ntwoh \ and HNC line profiles appear to suffer from self-absorption, whereas the HNCO line profile is singly-peaked. The P2 and P3 locations correspond to local peaks in the molecular hydrogen column density maps derived from Hi-GAL (\emph{Herschel} infrared Galactic plane survey; \citealp{molinari_2010}) data (\citealp{battersby_2011}, Battersby et al.~in prep.). \emph{The observed lack of self-absorption does not necessarily mean that the HNCO line is optically thin}. However, this is an encouraging sign that HNCO provides an accurate description of the line-centroid velocity.

Using principal component analysis, \citet{jones_2012} also infer that the differences between emission from the 3\,mm molecular lines such as HCN, HCO$^{+}$, HNC, and \ntwoh \  to that of HNCO at the locations of the Sgr B2 and Sgr A cloud complexes, are at least partly due to the latter being less \emph{likely} to be optically thick. \\

\item \citet{jones_2012} identify strong absorption features in the highly abundant molecules (HCN, HCO$^{+}$, and HNC) at velocities of $-52$, $-28$, and $-3$\,\kms, due to line-of-sight Galactic features (see also \citealp{greaves_1994,wirstrom_2010}). Even though HNCO is also abundant throughout the mapped region, line-of-sight confusion and absorption are less prominent for the selected transition which has a critical density, $n_{\rm crit}\sim10^{6}\,{\rm cm^{-3}}$ (both the \ntwoh \ and HNC lines have $n_{\rm crit}\sim3\times10^{5}\,{\rm cm^{-3}}$; \citealp{rathborne_2014}). \\

\end{enumerate}

\begin{table}
	\caption{\scouse: Input parameters and global statistics \vspace{0.6cm}} 
	\centering  
	\tabcolsep=0.15cm \normalsize{
	\begin{tabular}{ l  c }
	\hline
	Input parameter & Value \\ 
	\hline  
	
	$R_{\rm SAA}$ & 0\fdg05/7.25\,pc\\ [0.5ex]
	$T_{1}$ & 3.0 \\ [1.0ex] 
	$T_{2}$ & 1.0 \\ [1.0ex]
	$T_{3}$ & 4.0 \\ [1.0ex]
	$T_{4}$ & 1.0 \\ [1.0ex]
	$T_{5}$ & 0.5 \\ [2.0ex]
	
 	Output statistic &  \\ [2.0ex]
	
	$N_{\rm tot}$ & 11339 \\ [1.0ex] 
	$N_{\rm tot,\,SAA }$ & 6546 \\ [1.0ex]
	$N_{\rm SAA }$ & 132 \\ [1.0ex]
	$N_{\rm fit}$ & 5224 \\ [1.0ex]
	$N_{\rm comp}$ & 8206 \\ [1.0ex]
	$N_{\rm comp}$/$N_{\rm fit}$ & 1.6 \\ [1.0ex]
	
	\hline
	\end{tabular}
	\vspace{0.5cm}
	
\begin{minipage}{0,5\textwidth}\footnotesize{
	\centering  
	\tabcolsep=0.1cm
	\begin{tabular}{ l  l }
	$N_{\rm tot}$ & Total number of positions in the mapped area.\\   
	$N_{\rm tot,\,SAA}$ & Total number of positions included in the coverage (\S\,\ref{Section:method_coverage}).\\
	$N_{\rm SAA}$ & Total number of spectral averaging areas (\S\,\ref{Section:method_coverage}).\\
	$N_{\rm fit}$ & Total number of positions fit. \\ 
	$N_{\rm comp}$ & Total number of components fit. \\
	$N_{\rm comp}$/$N_{\rm fit}$ & Mean number of Gaussian components per position. \\  
	\end{tabular}
}
\end{minipage}
}
\label{Table:global_stats}
\end{table}

Another consideration relates to the input parameters supplied to \scouse \ during the fitting procedure. These are summarised in Table~\ref{Table:global_stats} (each input parameter is described in \S~\ref{Section:method}). Each has been selected to ensure reliable fit results to the HNCO data. In summary, we use $R_{\rm SAA}=0\fdg05$ such that the line-profile of each spatially averaged spectrum (see \S~\ref{Section:method_SAAfitting}) is representative of its composite spectra (deviations from this are evident at $R_{\rm SAA}\sim0\fdg10$). All velocity components identified must be above 3\,$\sigma_{\rm rms}$ ($I_{\rm tol}$, where $\sigma_{\rm rms}$ refers to the local rms noise level) and have a FWHM line-width of at least $v_{\rm res}$. The line-width and centroid velocity of each component must be smaller than $4\Delta v_{\rm i}^{\rm SAA}$ and be separated by no more than $\sigma_{\rm i}^{\rm SAA}$ from the most closely matching spectral feature in the parent SAA solution, respectively. These two tolerances avoid artificially narrow or broad components. Finally, if multiple components are evident then two adjacent components must be separated in velocity by at least 0.5\,$\times$ the line-width of the narrowest of the two. This does not necessarily remove instances where a narrow component is projected on top of a much broader component (although it can), as it is the line-width of the narrow component which determines the separation limit. All input parameters have largely been determined empirically and will ultimately vary with different data sets.

Table~\ref{Table:global_stats} also includes the global statistics output following the fitting of the HNCO line. This includes information on the number of SAAs whose spatially averaged spectra were fit during stage~2 ($N_{\rm SAA}$), the total number of locations fit ($N_{\rm fit}$), the total number of Gaussian components ($N_{\rm comp}$) and hence the degree of multiplicity of spectral components ($N_{\rm comp}$/$N_{\rm fit}$). These values are obtained following a manual inspection of the best-fitting solutions (see \S~\ref{Section:method_bestfits}). It is important to reaffirm that the user has the option to seek alternative solutions if necessary. For the HNCO data presented in this work, we estimate that $\ll1$ per cent of the best-fitting solutions to high-SN singly-peaked spectra (i.e. those which are fit with ease) required alternate solutions. In general however, we chose to revisit $\sim5-10$ per cent of all spectra. Analysing the residuals of our final best-fitting solutions, we find a median value of $\sigma_{\rm resid}/\sigma_{\rm rms}\sim1.1$, with 75 per cent of all positions having $\sigma_{\rm resid}/\sigma_{\rm rms}<1.4$. Spectral rms values, for reference, vary between $0.01\,{\rm K}<\sigma_{\rm rms}<0.10\,{\rm K}$, with $\sim75$ per cent of spectra having $\sigma_{\rm rms}\leq0.02\,{\rm K}$. 

Figure~\ref{Figure:ncomp} indicates the locations which have best-fitting solutions. Each pixel in Figure~\ref{Figure:ncomp} is colour-coded according the number of spectral components identified at that location. Multiple component fits are required to describe the spectral line profiles over a significant ($\sim45$ per cent) portion of the map. This provides some indication as to the complexity of the velocity structure (which will be described more fully in \S~\ref{Section:results_global} and \S~\ref{Section:results_local}). \\

\section{Results and discussion}\label{Section:results}

Having described our methodology in Section~\ref{Section:method}, the following sections present the results obtained from this analysis. In Section~\ref{Section:results_linefitting} we compare the kinematic properties extracted using line-fitting with those obtained through moment analysis. Section~\ref{Section:results_global} looks at the kinematics of the CMZ as a whole, discussing both the line-of-sight velocity structure and the velocity dispersion. Section~\ref{Section:results_local}, by comparison, investigates the properties of individual clouds and sub-regions. Section~\ref{Section:results_models} discusses our line-fitting results in the context of three currently competing interpretations for the 3-D structure of the CMZ. Finally, in Section~\ref{Section:results_obstests} we discuss how proper motion measurements of masers could distinguish between the different interpretations discussed in Section~\ref{Section:results_models}, and suggest areas to focus this effort.

\begin{figure*}
\begin{center}
\includegraphics[trim = 0mm 20mm 0mm 35mm, clip, width = 0.9\textwidth]{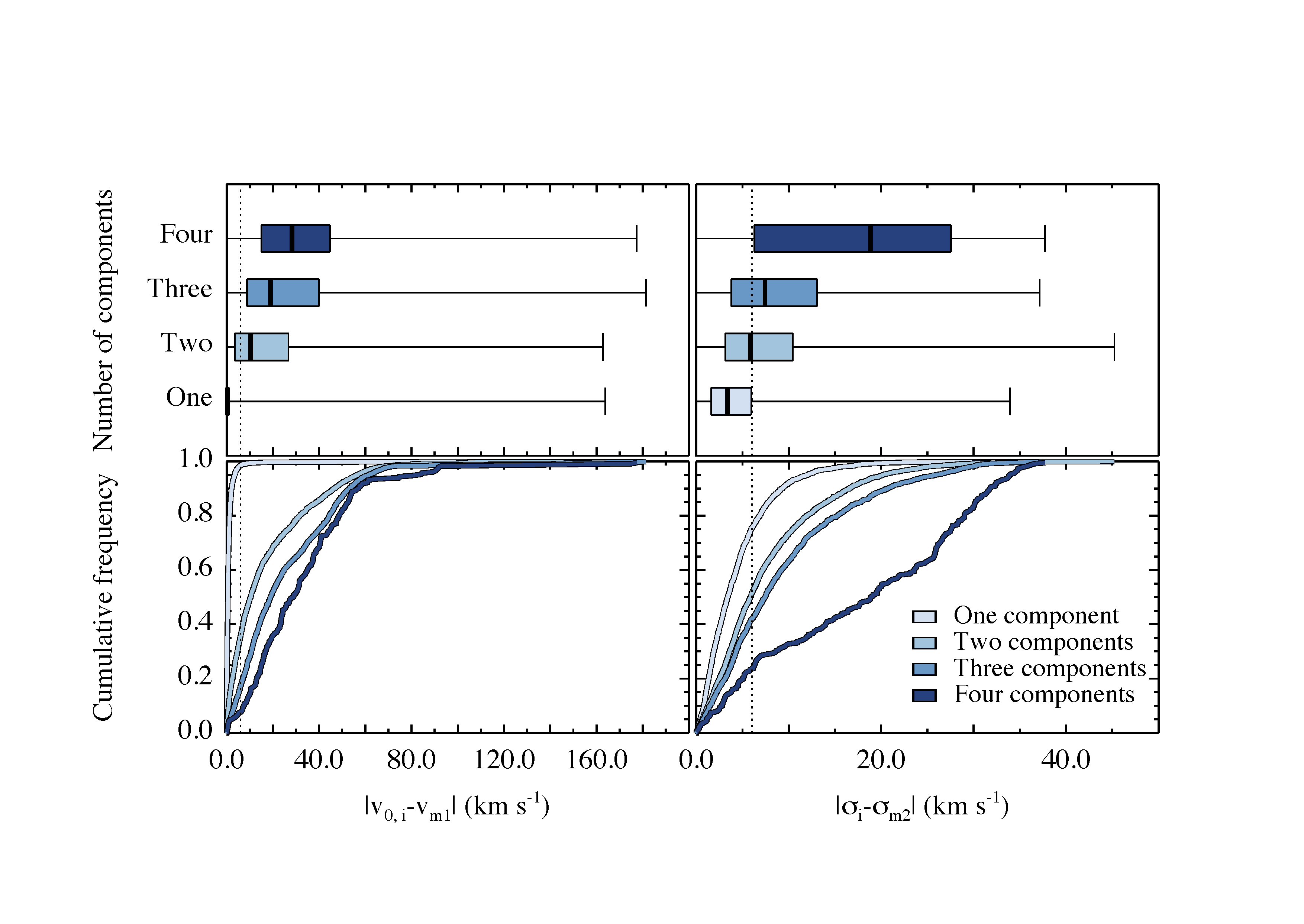}
\end{center}
\caption{Quantifying the difference between line-fitting and moment analysis. The top panels are box plots showing: (left) the difference between the first order moment ($v_{\rm m1}$; an intensity-weighted velocity) and the centroid velocity of the spectral component(s) identified at the same location ($v_{\rm 0,\, i}$); (right) the same but for the second order moment ($\sigma_{\rm m2}$; an intensity-weighted velocity dispersion) and the velocity dispersion of the extracted component(s) ($\sigma_{\rm i}$). The data are grouped according to the number of velocity components identified at each location (the colours match those presented in Figure~\ref{Figure:ncomp}). The plots highlight the range of data, the interquartile range (the box itself), and the median values (the thick lines within each box). The absolute values are presented in Table\,\ref{Table:lf_versus_mom}. The bottom panels are corresponding cumulative histograms.   }
\label{Figure:lf_versus_mom}
\end{figure*}

\subsection{Line-fitting versus moment analysis}\label{Section:results_linefitting}

Techniques such as moment analysis provide a quantitative measure of the average kinematic properties of molecular clouds. However, their accuracy diminishes in regions with complex line-of-sight velocity structure, leading to misleading results. While moment analysis has previously been used within the CMZ, our study is the first to use line-fitting to investigate the kinematics of the molecular gas throughout the entire $1\fdg7\times0\fdg4$ region encompassing the inner $\sim250$\,pc of the Milky Way. This is therefore an ideal opportunity to compare the accuracy of the two techniques. 

Moment analysis is performed using the HNCO data over the velocity range $-200\,{\rm km\,s^{-1}}~\leq~v_{\rm LSR}~\leq~200\,{\rm km\,s^{-1}}$. Only voxels above $3\langle\sigma_{\rm rms}\rangle$ are considered (where $\langle\sigma_{\rm rms}\rangle\sim0.02$\,K and refers to the mean $\sigma_{\rm rms}$ extracted by \scouse \ across all spectra). The first and second order moments are given by 
\begin{align}
v_{\rm m1}=\frac{\int{T^{*}_{A}(v)v\delta v}}{\int{T^{*}_{A}(v)\delta v}}; && \sigma_{\rm m2}=\bigg[{\frac{\int{T^{*}_{A}(v)(v-v_{\rm m1})^{2}\delta v}}{\int{T^{*}_{A}(v)\delta v}}}\bigg]^{1/2},
\end{align}
respectively. 

\begin{table*}
	\caption{Line-fitting versus moment analysis: Statistics. The columns represent the relative difference in velocity (either centroid or dispersion) between line-fitting and moment analysis for different percentiles. These are highlighted by the box plots in the upper panels of Figure\,\ref{Figure:lf_versus_mom}.  } \vspace{0.3cm}
	\centering  
	\tabcolsep=0.4cm \normalsize{
	\begin{tabular}{ c  c  c  c  c  c  c  c }
	\hline
	\multicolumn{2}{ c }{} & \multicolumn{5}{ c }{Percentile} & \\ [0.5ex]
	 & 
	 Number of & 
	 0 & 
	 25 & 
	 50 (Median) & 
	 75 & 
	 100 & 
	 Mean  
	 \\ [0.5ex]
	 	 
	 & 
	 components & 
	 (\kms) & 
	 (\kms) & 
	 (\kms) & 
	 (\kms) & 
	 (\kms) & 
	 (\kms)  
	 \\ [0.5ex]
	 
	\hline
	
	\multirow{4}*{$|v_{\rm 0,\, i}-v_{\rm m1}|$} & 1 & 0.0 & 0.2 & 0.5 & 1.1 & 163.8 & 1.1 \\ [0.5ex]
	& 2 & 0.0 & 3.5 & 10.4 & 26.8 & 162.9 & 17.7 \\ [0.5ex]
	& 3 & 0.0 & 8.8 & 19.0 & 40.1 & 181.4 & 25.7 \\ [0.5ex]
	& 4 & 0.1 & 15.0 & 28.3 & 44.7 & 177.4 & 32.4 \\ [2.0ex]
		
	\multirow{4}*{$|\sigma_{\rm 0,\, i}-\sigma_{\rm m1}|$} & 1 & 0.0 & 1.6 & 3.4 & 5.9 & 33.9 & 4.4 \\ [0.5ex]
	& 2 & 0.0 & 3.1 & 5.9 & 10.4 & 45.2 & 7.6 \\ [0.5ex]
	& 3 & 0.0 & 3.8 & 7.4 & 13.1 & 37.2 & 9.4 \\ [0.5ex]
	& 4 & 0.0 & 6.3 & 18.8 & 27.6 & 37.7 & 17.7 \\ [0.5ex]
	\hline
	\end{tabular}
	\vspace{0.5cm}	
}
\label{Table:lf_versus_mom}
\end{table*}

The top panels of Figure~\ref{Figure:lf_versus_mom} display box plots of: (left) the absolute difference between the first order moment ($v_{\rm m1}$, an intensity-weighted velocity) and the centroid velocity of all components identified at the same location; (right) the absolute difference between the second order moment ($\sigma_{\rm m2}$, an intensity-weighted velocity dispersion) and the velocity dispersion of all components identified at the same location (the absolute values of the main statistical features displayed in each panel are presented in Table~\ref{Table:lf_versus_mom}). The bottom panels are the corresponding cumulative histograms. In both panels, results are grouped based on the number of components identified at each location. 

Where only a single velocity component is present, results from the two techniques largely agree. In $\sim75$ (98) per cent of cases, the difference in velocity dispersion (centroid velocity) measurements is less than 6\,\kms \ (see Table~\ref{Table:lf_versus_mom}). Therefore where only one component is evident, moment analysis provides a good representation of the gas kinematics. Although we have endeavoured to validate the \scouse \ fits by eye, this also gives us confidence that our best-fitting solutions are quantitatively robust. 

The top-left panel of Figure~\ref{Figure:lf_versus_mom} reveals that the difference in the centroid velocity can deviate by as much as $\sim164$\,\kms \ (for ``single component'' spectra). Such instances are rare. These typically correspond to multiply peaked spectra where one of the features has been rejected by \scouse \ during line-fitting (due to large parameter uncertainties or a poor fit), but still contributes to the average spectral properties determined by moment analysis, which can lead to a large discrepancy. 

We can use the close relationship between the two techniques for singly-peaked spectra as a way of quantifying the deviation where multiple components are present. Very simplistically, we define a limit of 6\,\kms \ (see above) to signify good agreement between the two techniques (vertical dotted line in Figure~\ref{Figure:lf_versus_mom}). Where two components are identified, only $\sim36$ per cent and $\sim51$ per cent of results agree for the centroid velocity and dispersion, respectively. This falls to $\sim17$ per cent and $\sim42$ per cent for three components, and to $\sim7$ per cent and $\sim24$ per cent where four components are identified. The deviation between the two techniques increases with an increasing number of spectral features (seen also in the monotonically increasing median deviation in the top-panels of Figure~\ref{Figure:lf_versus_mom} and in Table~\ref{Table:lf_versus_mom}). Given that multiple component fits are required to reproduce the line profiles of $\sim45$ per cent of the (HNCO) data, we conclude that moment analysis provides an inadequate description of the kinematics of the CMZ.

\subsection{The kinematics of the CMZ I: Global gas properties}\label{Section:results_global}

\subsubsection{The line-of-sight velocity structure of the CMZ}\label{Section:results_velocity}

\begin{figure*}
\begin{center}
\includegraphics[trim = 15mm 5mm 10mm 20mm, clip, width = 0.98\textwidth]{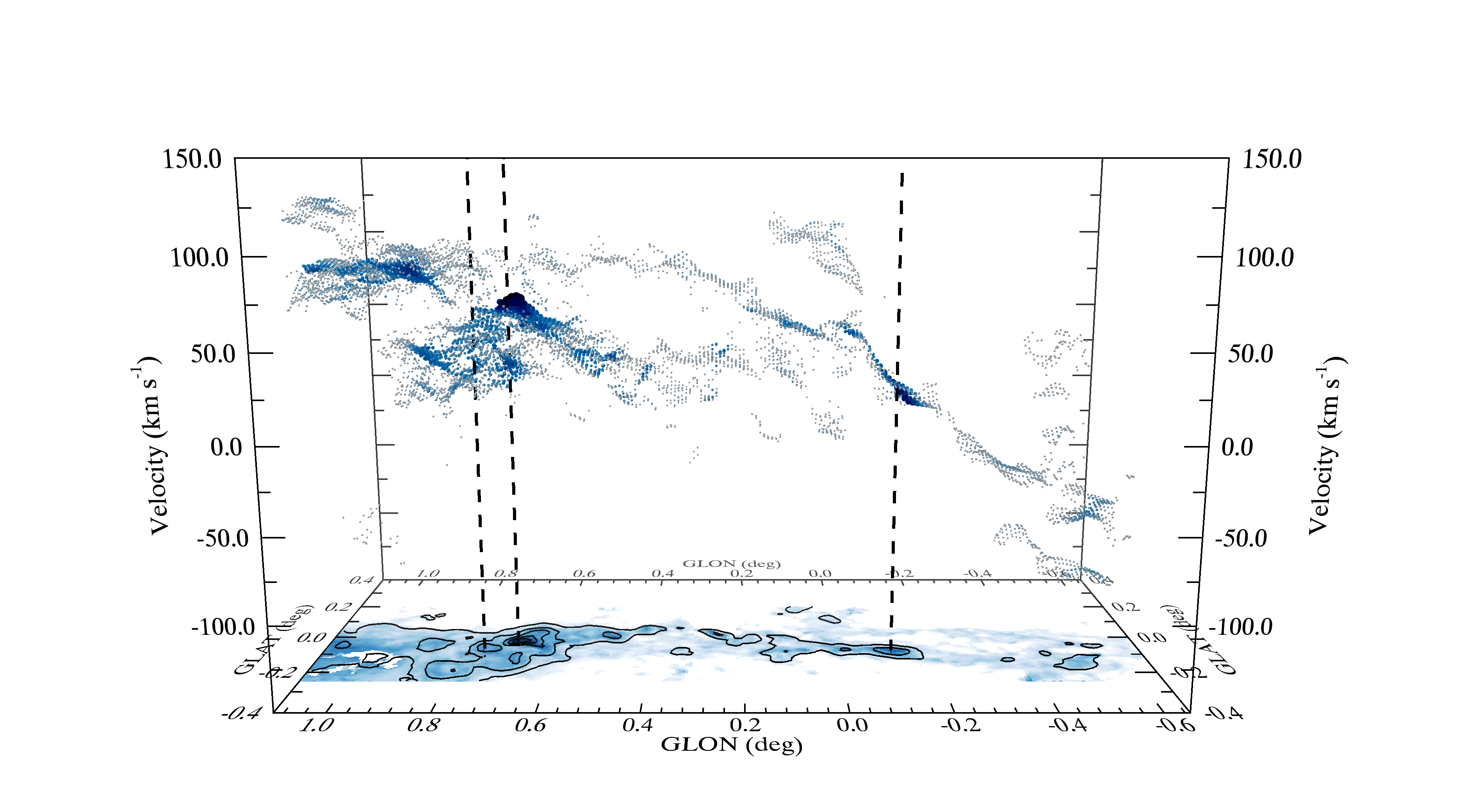}
\end{center}
\caption{A position-position-velocity (PPV) diagram highlighting the large-scale kinematic structure of the CMZ. Each pixel denotes the location and centroid velocity of a Gaussian component identified in HNCO emission and extracted using \scouse. The size (from small to large) and colour (from light to dark) of each data point is proportional to the peak intensity of the corresponding spectral component. At the base of the plot, in logarithmic colour-scale, is the \emph{Herschel}-derived molecular hydrogen column density map (Battersby et al., in prep.). Column densities range from 10$^{22}$--10$^{24}$\,cm$^{-2}$. Overlaid are contours highlighting the distribution of HNCO integrated emission. Contour levels are, 5, 25, 45, 65, and 85 per cent of peak (integrated) emission (cf. Figure\,\ref{Figure:ncomp}). The three vertical dotted lines refer to the locations P1, P2, and P3, also highlighted in Figure\,\ref{Figure:ncomp} (the corresponding spectra of which are shown in Figure\,\ref{Figure:spectra}). A movie that shows this figure in 3-D is available in the online supplementary material. }
\label{Figure:PPV_all}
\end{figure*}

Figure~\ref{Figure:PPV_all} is a 3-D position-position-velocity (PPV) diagram that highlights the distribution of HNCO emission throughout the CMZ. Each pixel within Figure~\ref{Figure:PPV_all} refers to the $\{l, b, v_{\rm LSR}\}$ coordinates of an identified spectral component. The size (from small to large) and colour (from light to dark) of each pixel is proportional to the peak intensity of the detected HNCO spectral component at that location. 

As noted by \citet{bally_1988}, the distribution of molecular gas is asymmetric about Sgr~A*. We can also demonstrate this using the output from \scouse. Within the presented region, and only considering emission from HNCO, we measure the distribution of material relative to the position $\{l,\,b\,,v_{\rm LSR}\}~=~\{~-~0\fdg056,\,~-~0\fdg046\,,-14.0\,~{\rm km\,s^{-1}}\}$, i.e. the position \citep{petrov_2011} and velocity of Sgr A* (accounting for the sun's radial velocity towards the Galactic centre; \citealp{schonrich_2012}). We find that $\sim87$ per cent of all identified velocity components are situated at positive (offset with respect to Sgr~A*) longitudes (this rises to $\sim92$ per cent when each component is weighted according to its integrated intensity), $\sim52$ per cent of which lie at negative (offset) latitudes. Additionally, the emission is also asymmetric in velocity, with $\sim99$ per cent of the data situated at positive (offset) longitudes is also at positive (relative) velocities (rising to $\sim99.8$ per cent when each component is weighted according to its integrated intensity). 

A large-scale ($\sim200$\,pc) velocity gradient is observed, with the velocity increasing in the direction of increasing Galactic longitudes. The magnitude of this gradient can be estimated (assuming it is linear) using (e.g. \citealp{goodman_1993}):
\begin{equation}
v_{\rm LSR} = v_{\rm LSR, 0}+\nabla v_{l}\Delta l+\nabla v_{b}\Delta b.
\label{Equation:linear}
\end{equation}
Here, $v_{\rm LSR, 0}$ is the systemic velocity of the mapped region with respect to the local standard of rest, $\Delta l$ and $\Delta b$ are the offset Galactic longitude and latitude (expressed in radians), and $\nabla v_{l}$ and $\nabla v_{b}$ refer to the magnitudes of the velocity gradients in the $l$ and $b$ directions, respectively (in \kms\,rad$^{-1}$). The magnitude of the velocity gradient ($\mathscr{G}$), and its direction ($\Theta_{\mathscr{G}}$), are then estimated using:
\begin{align}
\mathscr{G} \equiv |\nabla v_{l,b}| = \frac{(\nabla v_{l}^{2}+\nabla v_{b}^{2})^{1/2}}{D},
\label{Equation:grad_calc}
\end{align}
and
\begin{equation}
\Theta_{\mathscr{G}} \equiv {\rm tan}^{-1}\bigg(\frac{\nabla v_{l}}{\nabla v_{b}}\bigg),
\label{Equation:dir_calc}
\end{equation}
whereby $D$ is the distance to the object in pc (assuming all gas is equidistant from Earth). 

The least-squares minimization routine {\sc mpfit} has been used in order to find a best-fitting solution to Equation~\ref{Equation:linear} over \emph{all} velocity components extracted using \scouse \ in the mapped region. The magnitude of this velocity gradient is estimated to be $\mathscr{G}~\sim~0.6$\,\vel \ in a direction $\Theta_{\mathscr{G}}\sim58$\degr \ east of Galactic north (with Galactic east in the direction of positive longitudes). This magnitude is similar to that found by \citet{jones_2012}, who studied the peak velocity of \ntwoh \ emission from the Mopra CMZ survey, finding $\mathscr{G}\sim0.7$\,\vel. 

Although this provides a global overview of the gas distribution, we caution that deriving the velocity gradient in this way incorporates all of the identified HNCO velocity components. This method is therefore insensitive to localised variations in the velocity gradient. In \S~\ref{Section:results_local} however, we expand on this investigation, exploit the density of information provided by \scouse, and study the velocity structure of individual sub-regions in more detail.

\subsubsection{Velocity dispersion}\label{Section:results_linewidth}

\begin{figure}
\begin{center}
\includegraphics[trim = 22mm 15mm 0mm 20mm, clip, width = 0.48\textwidth]{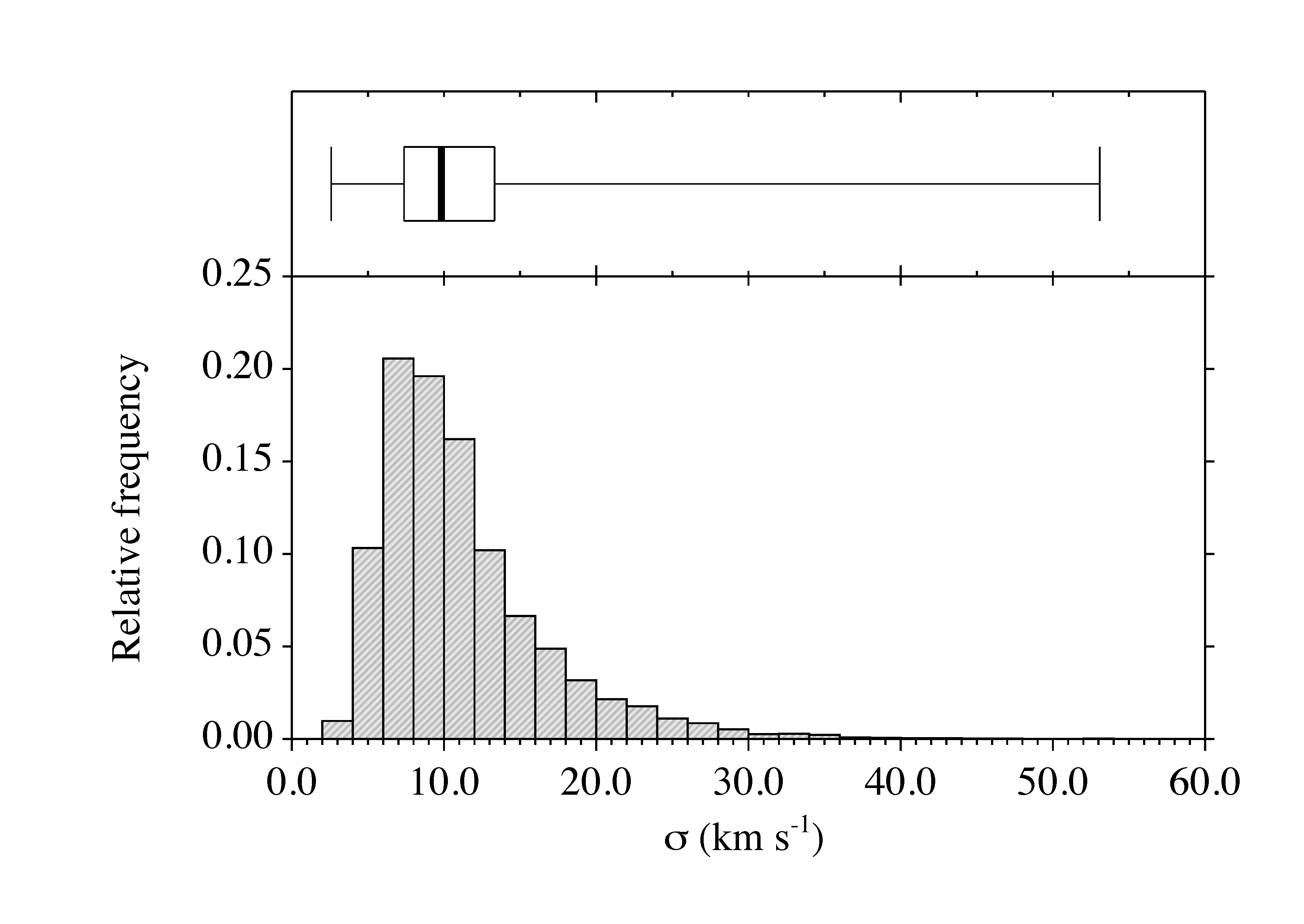}
\end{center}
\caption{(Top) A box plot of the velocity dispersion for all spectral components extracted from the HNCO spectral line data using \scouse. This highlights the range in velocity dispersion, the interquartile range (the box itself) and the median velocity dispersion ($\sim$\,9.8\,\kms, vertical thick black line). See \S~\ref{Section:results_linewidth} for more details. (Bottom) A histogram of the velocity dispersion. 
 }
\label{Figure:disp_histo}
\end{figure}

\begin{figure*}
\begin{center}
 \includegraphics[trim = 15mm 5mm 10mm 20mm, clip, width = 0.98\textwidth]{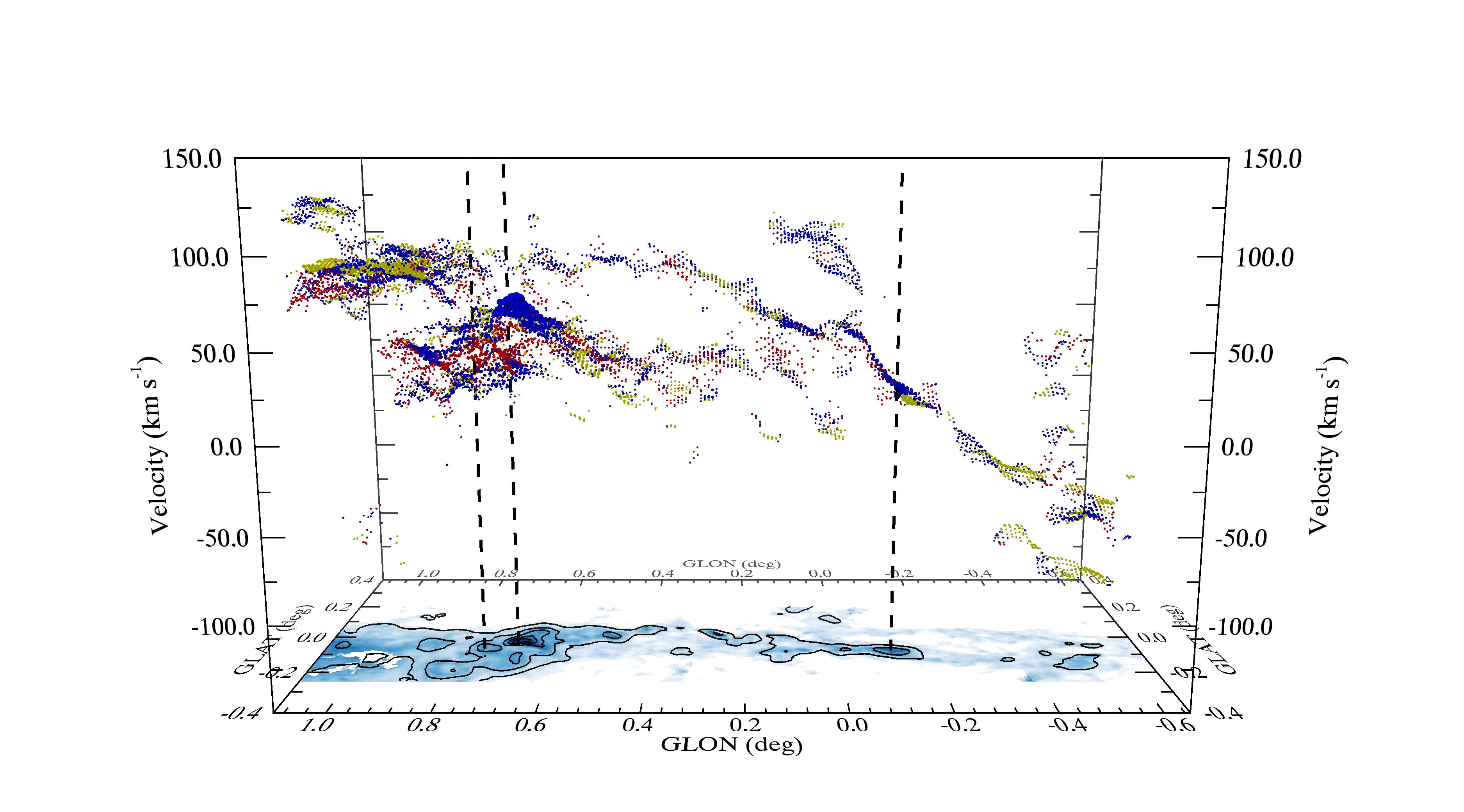}
\end{center}
\caption{The same as Figure~\ref{Figure:PPV_all}. However, each pixel is now colour-coded according to specific ranges in velocity dispersion (see \S~\ref{Section:results_linewidth} for more information on the selection of these ranges). \textbf{Yellow}: $2.6\,{\rm km\,s^{-1}}<\sigma<7.4\,{\rm km\,s^{-1}}$;  \textbf{Blue}: $7.4\,{\rm km\,s^{-1}}<\sigma<13.3\,{\rm km\,s^{-1}}$; \textbf{Red}: $13.3\,{\rm km\,s^{-1}}<\sigma<53.1\,{\rm km\,s^{-1}}$. Note the contrast in velocity dispersion between the Sgr B2 (where $\sim45$ per cent of the high-dispersion gas lies) and Sgr C (which predominantly exhibits low- or intermediate dispersion gas) molecular cloud complexes situated at $\{l,\,b,\,v_{\rm LSR}\}\sim\{0\fdg70,-0\fdg05, 60\,{\rm km\,s^{-1}}\}$ and $\sim\{-0\fdg50,-0\fdg10, -55\,{\rm km\,s^{-1}}\}$, respectively.}
\label{Figure:PPV_all_disp}
\end{figure*}

Figure~\ref{Figure:disp_histo} shows the distribution of velocity dispersions extracted from the HNCO spectral line data using \scouse. The top panel is a box plot of the measured velocity dispersion, $\sigma$, at each location in the map. Highlighted are some of the main statistical features. The size of the box depicts the interquartile range ($IQR~\equiv~Q_{3}~-~Q_{1}$; whereby $Q_{1}=7.4$\,\kms \ and $Q_{3}=13.3$\,\kms \ are first and third quartiles, respectively). The thick vertical line highlights the median velocity dispersion $=9.8$\,\kms \ (the mean value of the velocity dispersion is not shown, but has a value $11.1$\,\kms). The lower and upper limit to the velocity dispersion are $2.6$\,\kms \ and $53.1$\,\kms, respectively. By comparison, the lower panel of Figure\,\ref{Figure:disp_histo} is a histogram of all velocity dispersions. Figure~\ref{Figure:PPV_all_disp} is a PPV diagram similar to that presented in Figure~\ref{Figure:PPV_all}. However, the colour of each data point refers to three ranges in velocity dispersion: $2.6-7.4$\,~\kms \ (yellow; $0-25$ per cent); $7.4-13.3$\,~\kms \ (blue; $25-75$ per cent); $7.4-13.3$\,~\kms \ (red; $75-100$ per cent).

Figure~\ref{Figure:disp_histo} indicates that there is a broad range in the observed velocity dispersion. We find no trend between the velocity dispersion and peak temperature across the distribution of identified velocity components. At the lower end of this distribution, velocity dispersions of $2.6$\,\kms \ may imply the detection of contaminant line-of-sight molecular clouds by \scouse. Spectral components with $\sigma<5$\,\kms \ are generally spread out throughout the CMZ, making up less than 5 per cent of the data. However, there is a notable concentration of these components towards $\{l,\,b,\,v_{\rm LSR}\}\approx\{1\fdg0,\,0\fdg0,\,80.0\,{\rm km\,s^{-1}}\}$ (see Figure~\ref{Figure:PPV_all_disp}). At the other end of the scale, the broadest identified component is situated at $\{l,\,b\}=\{0\fdg80,\,0\fdg04\}$ towards the Sgr B2 cloud complex. Approximately $45$ per cent of the broad velocity dispersion gas ($13.3\,{\rm km\,s^{-1}}~<~\sigma~<~53.1\,{\rm km\,s^{-1}}$) is located towards the Sgr B2 region. 

\citet{shetty_2012} investigate the dispersion-size relationship for the CMZ. Their diagnostic line tracers include \ntwoh, HCN, H$^{13}$CN, and HCO$^{+}$. We can compare our measured velocity dispersion with that predicted by the dispersion-size relationships of \citet{shetty_2012} for an equivalent physical extent. For a cloud diameter of 2.4\,pc ($\sim$\,the smoothed spatial resolution of these data, see \S~\ref{Section:data}), the \citet{shetty_2012} relationships predict velocity dispersions of $\sim2.0-3.3$\,\kms. This is $\sim3.5\times$ smaller than the median velocity dispersion extracted via \scouse \ (and $\sim3\times$ smaller than the median intensity-weighted velocity dispersion extracted using moment analysis). 

A possible explanation for this difference is that \citet{shetty_2012} use dendrograms \citep{rosolowski_2008} to identify structure within the CMZ. By design, dendrograms identify nested structure, with each level in the hierarchy corresponding to a unique isosurface. The dendrogram approach excludes pixels below a given threshold from each isosurface, whereas the gaussian fitting approach includes these pixels as part of the ``line wing''. Truncating the line wing, as is done with dendrograms (and other isosurface-based structure finding approaches), results in an artificially decreased line-width. The quantitative effects of this truncation are discussed in \citet{rosolowski_2005} and \citet{rosolowski_2008}. Without information on the dendrogram hierarchy, we are unable to make a more quantitative assessment of the discrepancy. However, qualitatively, the observed trend is as we would expect.

The total observed velocity dispersion includes contributions from both thermal and non-thermal motions. The contribution that non-thermal motions make to the observed velocity dispersion can be estimated (assuming both have Gaussian distributions) using:
\begin{equation}
\sigma_{\rm NT}\,=\,\sqrt{(\sigma)^2-(\sigma_{\rm T})^2},
\end{equation}
where $\sigma_{\rm NT}$, $\sigma_{\rm T}$, and $\sigma$, represent the non-thermal, the thermal, and the observed dispersion (extracted using \scouse), respectively. This can then be related to the Mach number in the following way (assuming $\mathcal{M}_{x}\approx\mathcal{M}_{y}\approx\mathcal{M}_{\rm z}$):
\begin{equation}
\mathcal{M}_{\rm 3D}\,\approx\sqrt{3}\,\frac{\sigma_{\rm NT}}{c_{\rm s}}\,=\,\sqrt{3}\bigg[\bigg(\frac{\sigma}{c_{\rm s}}\bigg)^2-\bigg(\frac{\overline{\mu}_{\rm p}}{\mu_{\rm obs}}\bigg)\bigg]^{1/2},
\label{Equation:sigmant}
\end{equation}
where $c_{\rm s}=(k_{\rm B}T_{\rm kin}/{\overline\mu}_{\rm p} m_{\rm H})^{0.5}$, is the isothermal sound speed for a gas with kinetic temperature, $T_{\rm kin}$, and mean molecular mass, $\overline{\mu}_{\rm p}~=~2.33$\,amu ($k_{\rm B}$ and $m_{\rm H}$ are the Boltzmann constant and the mass of atomic hydrogen, respectively), and $\mu_{\rm obs}$ is the molecular mass of the observed molecule (43\,amu in the case of HNCO). 

For a fiducial temperature range of $60-100$\,K (suitable for the dense gas probed in this study; \citealp{ao_2013, mills_2013, ott_2014, ginsburg_2015}), the corresponding range in isothermal sound speed is $c_{\rm s}=0.46-0.60$\,\kms. Assuming that the median value of $\sigma=9.8$\,\kms \ is representative of gas motions within the CMZ on size scales of $\sim2.4$\,pc, the Mach number ranges between $\mathcal{M}_{\rm 3D}\sim28-37$ (for the maximum and minimum of the fiducial gas temperature range, respectively). For the minimum detected velocity dispersion of 2.6\,\kms, $\mathcal{M}_{\rm 3D}\sim8$ (for $T_{\rm kin}=100$\,K). 

This confirms that, at the parsec scales probed by these observations, gas motions within the CMZ are inherently supersonic. However, this should be taken as an upper bound on the level of turbulent motion, since our assumptions do not take into account the contribution from coherent motions or substructure within the Mopra beam. Such substructure could result in a large apparent line-width, even though the line-widths of individual clumps may be narrow, by comparison. Higher-angular and spectral resolution data are needed to asses this further.

\subsection{The kinematics of the CMZ II: Individual clouds and sub-regions}\label{Section:results_local}

\begin{figure*}
\begin{center}
\includegraphics[trim = 20mm 20mm 15mm 25mm, clip, width = 0.96\textwidth]{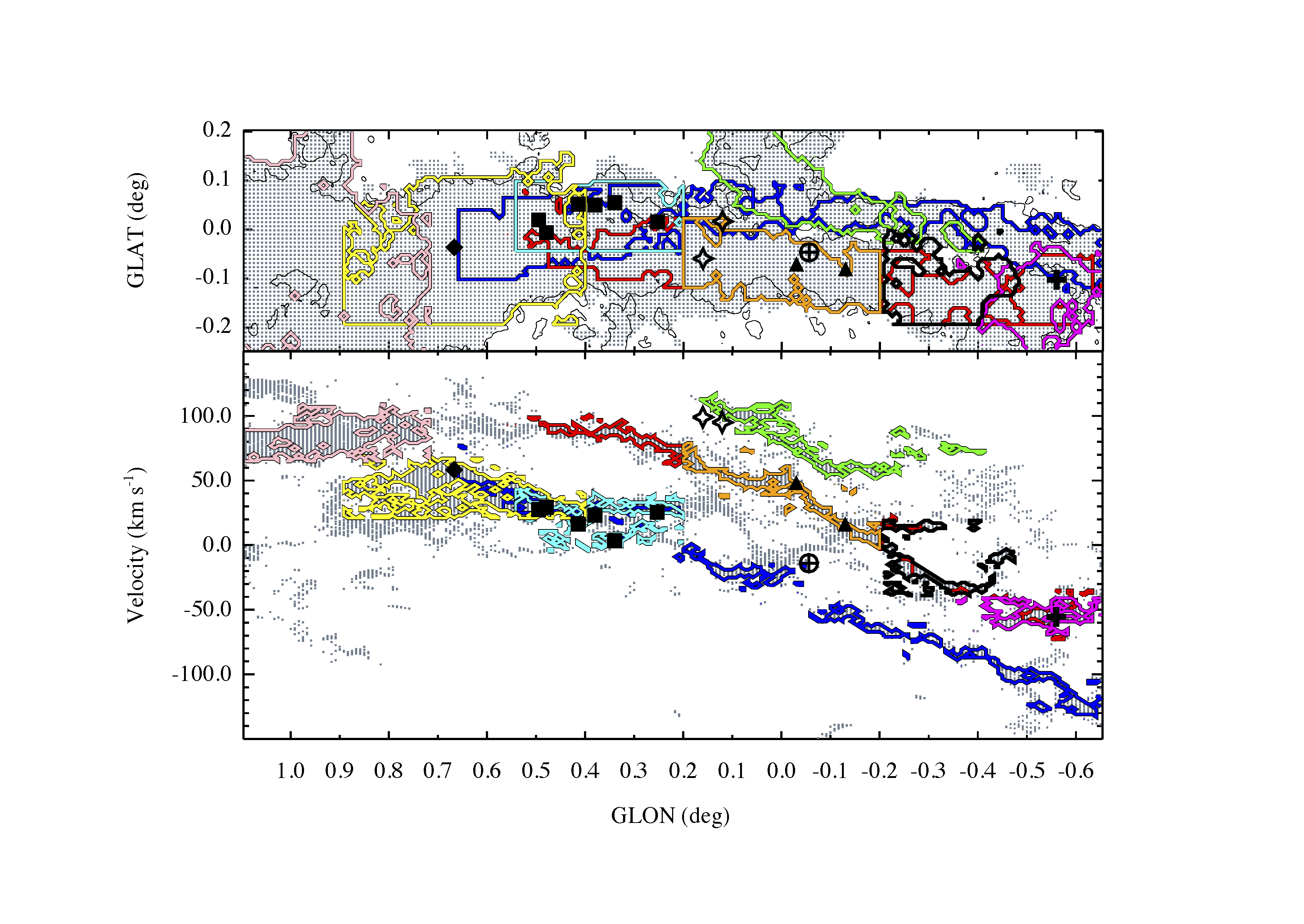}
\end{center}
\caption{Combining molecular gas kinematics at 3\,mm. The top panel depicts the spatial distribution ($\{l, b\}$) of dense gas within the CMZ, including HNCO, \ntwoh, and HNC. Each pixel denotes a location with a corresponding best-fitting solution. The single contour refers to a molecular hydrogen column density of 10$^{22}$\,cm$^{-2}$ (Battersby et al., in prep.). The bottom panel is a position-velocity ($\{l, v_{\rm LSR}\}$) diagram with each pixel referring to the Galactic longitude and centroid velocity of a spectral component. The black circle with a plus denotes the location of Sgr A*, and the open stars represent the Arches and Quintuplet clusters (the Arches cluster has the smallest projected distance from Sgr A*). Additionally, the locations of prominent molecular clouds are overlaid. In order of increasing Galactic longitude (from right to left): The Sgr C complex (black plus); the 20\,\kms \ and 50\,\kms \ clouds (black upward triangles); G0.256+0.016 (black square); Clouds B-F (black squares); The Sgr B2 complex (black diamond). The position and velocity limits of the molecular clouds can be found in Table~\ref{Table:positions}. The coloured contours highlight some of the most prominent PPV-structures. The \textbf{blue contour} shows ``Arm~{\sc i}'' (\citealp{sofue_1995}; also known as the ``negative velocity filament'' in \citealp{bally_1988}). The \textbf{red contour} shows ``Arm~{\sc ii}'' (\citealp{sofue_1995}). Both of these structures, as well as their possible connection to the PPV-structure highlighted by the pink contour, are discussed in \S\,\ref{Section:results_continuous}. The \textbf{green contour} highlights ``Arm\,{\sc iii}''  (also known as the ``Polar arc'' in \citealp{bally_1988}) and it is also discussed in \S~\ref{Section:results_continuous}. The \textbf{magenta} and \textbf{black contours} refer to molecular gas associated with the Sgr C complex (see \S~\ref{Section:results_sgrc} and Figures~\ref{Figure:PPV_sgrc} and \ref{Figure:PPV_shell}, respectively). The \textbf{orange contour} shows the 20\,\kms \ and 50\,\kms \ clouds (see \S~\ref{Section:results_2050}). The \textbf{cyan contour} highlights the ``dust ridge'' (\citealp{lis_1994,lis_2001}; see \S~\ref{Section:results_dustridge} and Figure~\ref{Figure:PPV_dustridge}). The \textbf{yellow contour} highlights the Sgr B2 cloud complex (see \S~\ref{Section:results_sgrb2} and Figure~\ref{Figure:PPV_sgrb2}). Note that there is overlap in the data associated with several of these PPV-structures since they are not all independent. This is particularly evident throughout Arm~{\sc ii}, which also incorporates data associated with the 20\,\kms \ and 50\,\kms \ clouds and the Sgr C molecular cloud complex.  } 
\label{Figure:pp_pv}
\end{figure*}

\begin{table*}
	\caption{The kinematic properties derived towards prominent molecular clouds identified within the CMZ. The physical extent of each cloud is estimated from the \emph{Herschel} molecular hydrogen column density map of Battersby et al. (in prep.) and should serve only as an approximation. All spectral components identified by \scouse \ within these physical limits, and within the velocity limits of each source's respective PPV diagram (see Figures~\ref{Figure:PPV_sgrc}, \ref{Figure:PPV_2050}, \ref{Figure:PPV_dustridge}, and \ref{Figure:PPV_sgrb2}) are included in the table. The uncertainties are shown in parentheses. Also included are the Intensity-weighted mean centroid velocity ($\langle v_{\rm 0}\rangle$) and intensity-weighted mean velocity dispersion ($\langle\sigma\rangle$).} \vspace{0.3cm}
	\centering  
	\tabcolsep=0.15cm \normalsize{
	\begin{tabular}{ l  c  c  c  c  c  c  c  c  c  c }
	\hline
	 & \multicolumn{4}{ c }{Physical limits}& \multicolumn{6}{ c }{Measured values} \\ [0.5ex]
	Source & $l_{\rm min}$  & $l_{\rm max}$  & $b_{\rm min}$ & $b_{\rm max}$ & $v_{\rm 0, min}$ & $\langle$$v_{\rm 0}$$\rangle$ & $v_{\rm 0, max}$ & $\sigma_{\rm min}$ & $\langle$$\sigma$$\rangle$ & $\sigma_{\rm max}$ \\ [0.5ex]
	&  (deg) &  (deg) & (deg) &  (deg) &  (\kms) &  (\kms) &  (\kms) & (\kms) &  (\kms) & (\kms)  \\ [0.5ex]
	\hline  
          Sgr C &  -0.57 &  -0.46 &  -0.19 &  -0.09 & -69.19 (0.92) & -55.50 (0.20) & -41.77 (2.08) &   2.97 (0.43) &   7.41 (0.01) &  22.91 (1.43) \\ [0.5ex]
      20 \,\kms &  -0.15 &  -0.07 &  -0.11 &  -0.06 &   5.29 (0.05) &  16.25 (0.05) &  31.56 (0.13) &   5.53 (0.04) &   8.27 (0.01) &  13.19 (0.55) \\ [0.5ex]
       50\,\kms &  -0.04 &   0.00 &  -0.08 &  -0.05 & -17.52 (0.58) &  48.44 (0.76) &  54.63 (0.19) &   4.19 (0.58) &   9.96 (0.07) &  11.74 (0.19) \\ [0.5ex]
   G0.253+0.016 &   0.22 &   0.28 &   0.00 &   0.06 &   1.16 (0.26) &  25.46 (0.16) &  45.46 (1.08) &   4.41 (0.12) &  11.00 (0.03) &  32.82 (1.62) \\ [0.5ex]
        Cloud B &   0.32 &   0.36 &   0.04 &   0.08 &  -3.40 (0.41) &   3.39 (0.18) &  32.53 (0.97) &   4.08 (1.13) &  14.66 (0.11) &  40.25 (1.46) \\ [0.5ex]
        Cloud C &   0.36 &   0.39 &   0.03 &   0.07 &  -0.01 (1.30) &  23.29 (0.42) &  40.06 (0.37) &   3.38 (0.41) &  11.68 (0.08) &  33.10 (1.74) \\ [0.5ex]
        Cloud D &   0.39 &   0.44 &   0.01 &   0.08 &  -4.81 (0.39) &  16.19 (0.14) &  32.81 (0.61) &   4.17 (0.67) &   9.68 (0.04) &  23.39 (1.29) \\ [0.5ex]
        Cloud E &   0.46 &   0.50 &  -0.03 &   0.01 &   0.87 (0.85) &  29.28 (0.24) &  31.34 (0.14) &   5.75 (0.09) &   7.92 (0.03) &  19.23 (0.71) \\ [0.5ex]
        Cloud F &   0.48 &   0.52 &   0.01 &   0.05 &   9.06 (0.69) &  27.56 (0.10) &  40.92 (0.27) &   4.14 (0.08) &   9.52 (0.01) &  13.56 (0.12) \\ [0.5ex]
         Sgr B2 &   0.61 &   0.72 &  -0.08 &   0.01 &  19.61 (0.44) &  58.20 (0.14) &  77.94 (0.59) &   3.93 (0.18) &  10.70 (0.01) &  35.11 (0.68) \\ [0.5ex]	
        \hline
	\end{tabular}
}
\label{Table:positions}
\end{table*}

As shown in Figure~\ref{Figure:PPV_all}, HNCO emission covers a significant fraction of the CMZ in both position and velocity. However, there are notable discontinuities. One way to establish whether or not these gaps represent true discontinuities (i.e. regions devoid of molecular gas) is to utilise the wealth of molecular line information made available in the Mopra CMZ survey \citep{jones_2012}. In the following sections we complement our HNCO data with the centroid velocity information extracted by \scouse \ for the \ntwoh \ and HNC lines at 3\,mm. The critical density of both of these molecular line transitions is smaller than that of the HNCO line ($n_{\rm crit}\sim3\times10^{5}\,{\rm cm^{-3}}$ versus $\sim10^{6}\,{\rm cm^{-3}}$; \citealp{rathborne_2014}), and their emission covers a greater portion of the CMZ. Adding this information therefore allows us to investigate the line-of-sight velocity structure in regions not traced by HNCO.

Figure~\ref{Figure:pp_pv} displays the result of merging kinematic information extracted using \scouse \ for HNCO, \ntwoh, and HNC (this process is discussed in more detail in Appendix~\ref{App:merge}). Two different projections are shown, $\{l,\,b\}$ (top panel) and $\{l,\,v_{\rm LSR}\}$ (bottom panel). Each grey pixel refers to the location of an identified spectral component. Comparing Figures~\ref{Figure:PPV_all} and \ref{Figure:pp_pv} shows that seemingly discontinuous features observed in HNCO emission (alone) may in fact become continuous when additional molecular species are included. 

In the following sections we explore how the gas kinematics change as a function of position within the CMZ. The first section describes three of the most prominent features evident in Figure~\ref{Figure:pp_pv} (those highlighted by the blue, red, and green contours, respectively). We then turn our attention to the different sub-regions and individual clouds, beginning with the Sgr C complex and ending at Sgr B2, proceeding in the direction of increasing Galactic longitude. The projected extent and kinematic information of the individual sub-regions are presented in Table~\ref{Table:positions}.

\subsubsection{Continuous PPV-structures within the CMZ}\label{Section:results_continuous}

\begin{figure*}
\begin{center}
	\includegraphics[trim = 15mm 5mm 5mm 5mm, clip, width = 0.48\textwidth]{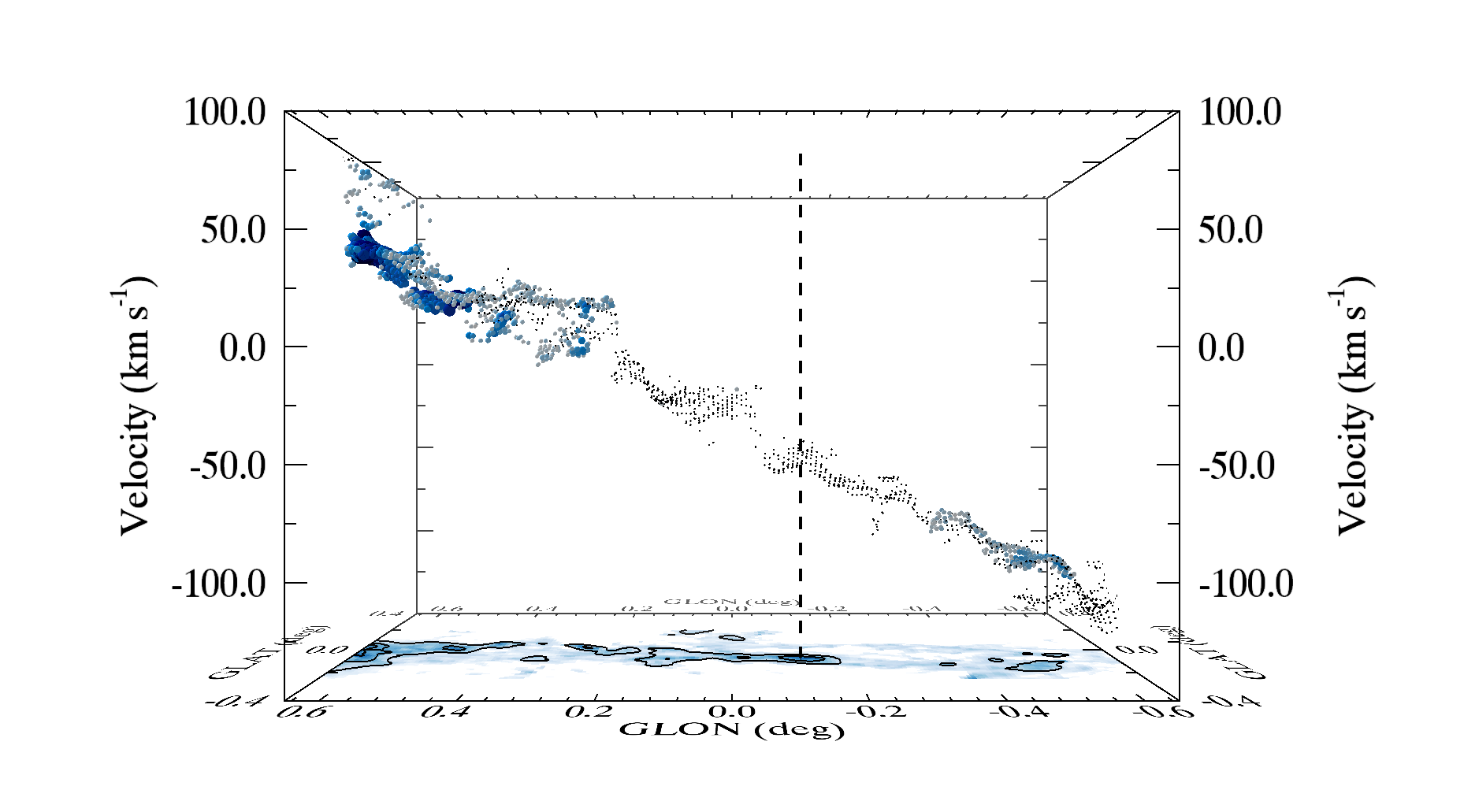}
	\hspace{3mm}
	\includegraphics[trim = 15mm 5mm 5mm 5mm, clip, width = 0.48\textwidth]{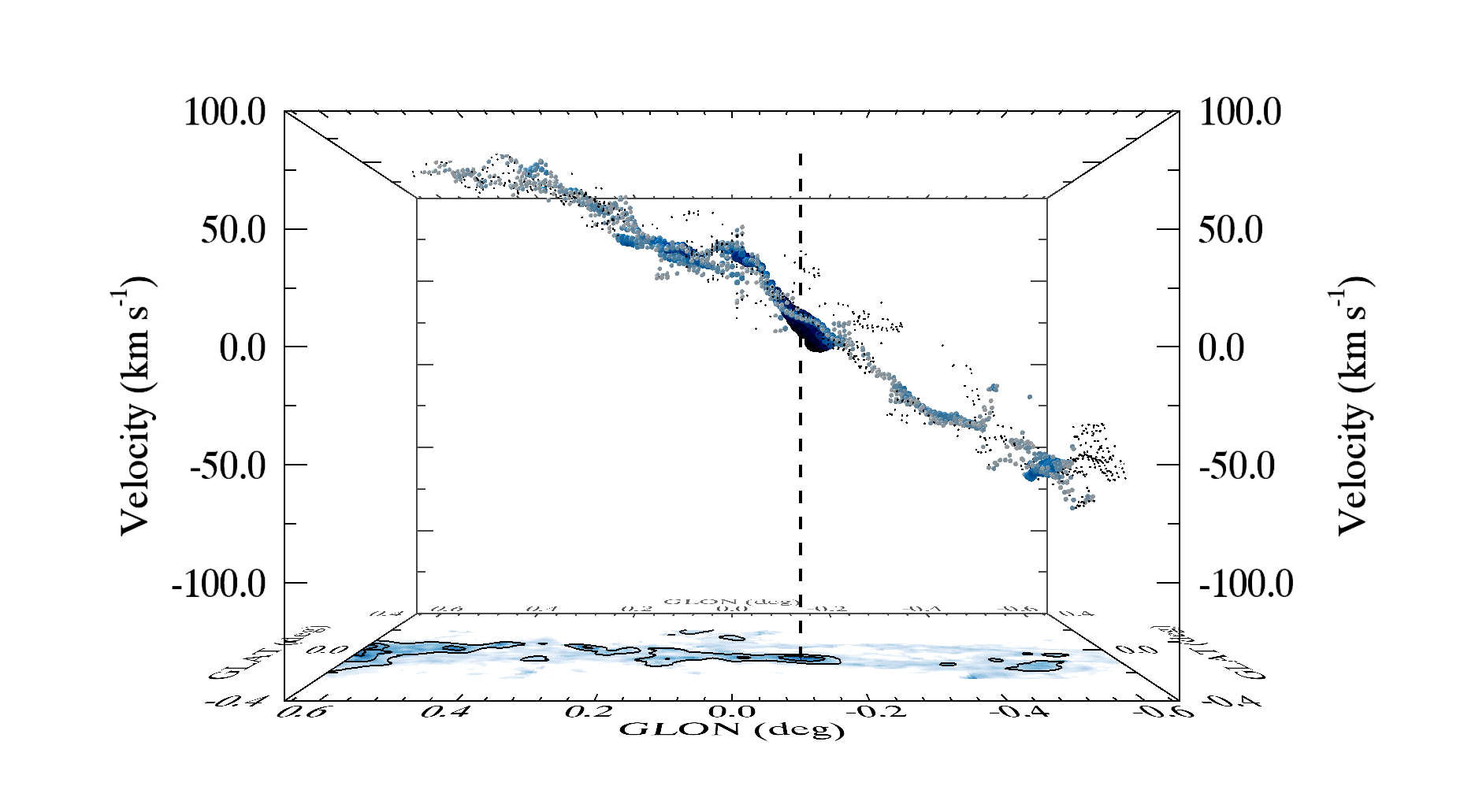}
\end{center}
\vspace{-6mm}
\begin{flushleft}
	\hspace{3.6mm}
	\includegraphics[trim = 10mm 7mm 0mm 80mm, clip, width = 0.423\textwidth]{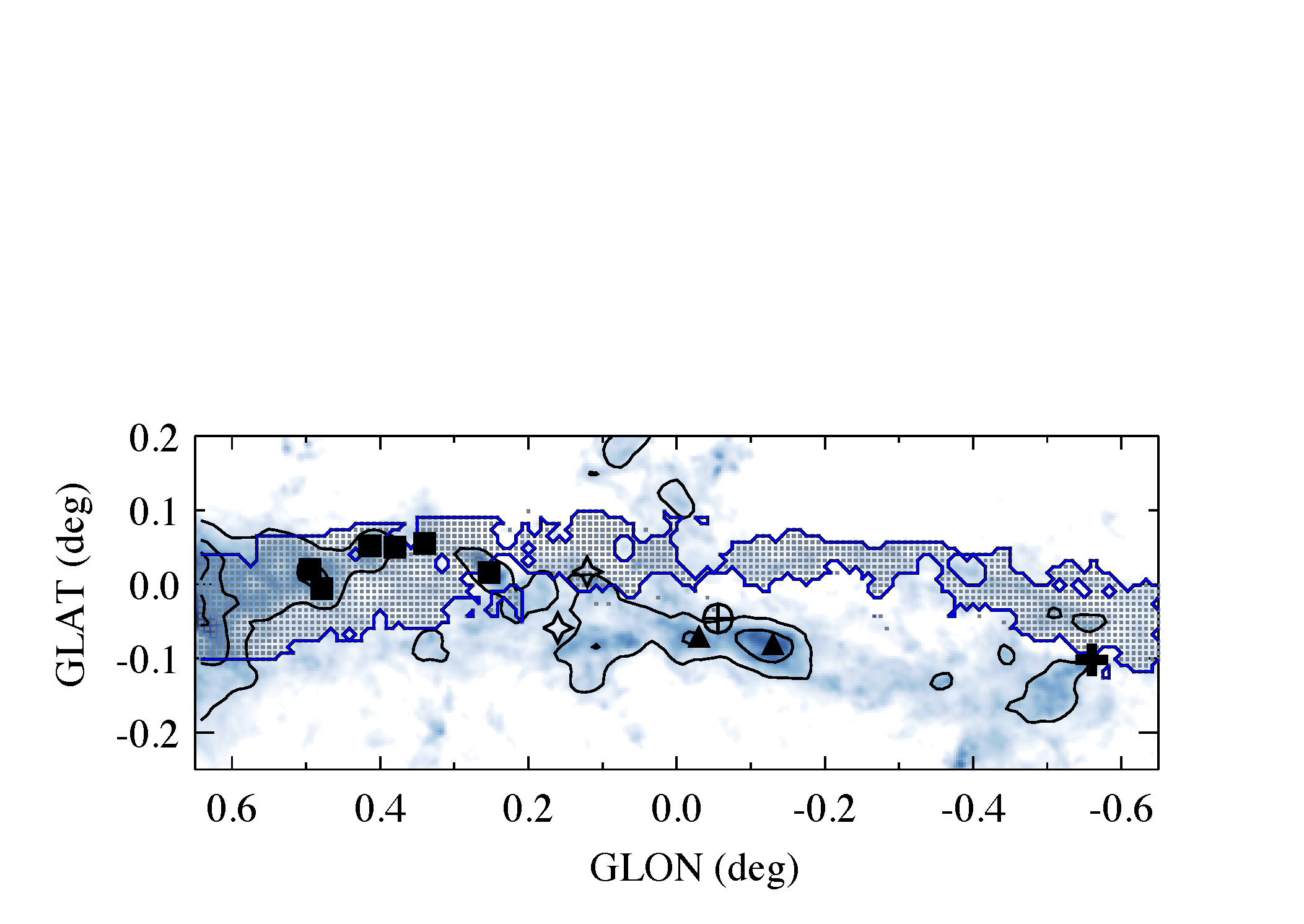}
	\hspace{13.2mm}
	\includegraphics[trim = 10mm 7mm 0mm 80mm, clip, width = 0.423\textwidth]{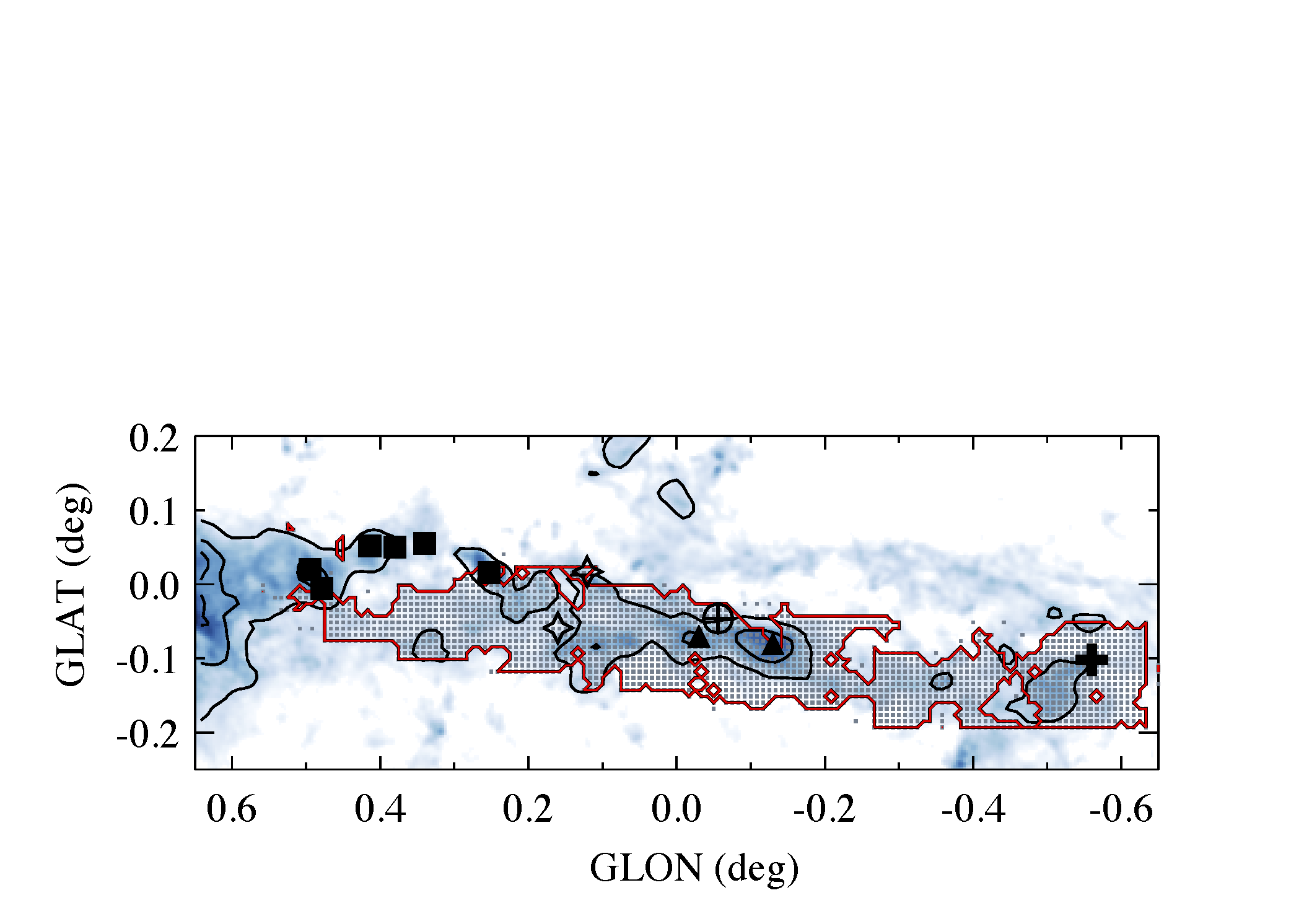}
\end{flushleft}

\caption{Top panels: PPV diagrams of the two extended features noted in the $\{l, v_{\rm LSR}\}$ projections of Figure~\ref{Figure:pp_pv} and discussed in more detail in \S~\ref{Section:results_continuous}. The coloured data points are equivalent to those presented in Figure~\ref{Figure:PPV_all}. Black data points highlight the $\{l,\,b,\,v_{\rm LSR}\}$ location of both the \ntwoh \ and HNC components. Bottom panels: Maps indicating the spatial distribution of the data points shown in the top panels. The left panels show ``Arm\,{\sc i}'' (\citealp{sofue_1995}; also known as the ``negative velocity filament'' in \citealp{bally_1988}). \citet{kruijssen_2015} argue that the data points shown in the left panels are distributed into two physically independent structures that are situated on the near- and far-sides of the Galactic centre, respectively (rather than a single spiral arm; \citealp{sofue_1995}). The right panels depict ``Arm\,{\sc ii}'' of \citealp{sofue_1995}. We discuss how the structures in this figure may contribute to the 3-D structure of the CMZ in \S\,\ref{Section:results_models}). The blue and red contours, and black symbols are equivalent to those shown in Figure~\ref{Figure:pp_pv}. The background image and contours in all panels are equivalent to those in Figure\,\ref{Figure:PPV_all}.}
\label{Figure:PPV_streams}
\end{figure*}

Perhaps the most immediately obvious features in Figure~\ref{Figure:pp_pv} are the two extended, and almost parallel (in $\{l,\,v_{\rm LSR}\}$ space), features evident between $-0\fdg65\leq~l~\leq~0\fdg5$ and $-150.0\,{\rm km\,s^{-1}}\leq~v_{\rm LSR}~\leq~100.0\,{\rm km\,s^{-1}}$. These features are isolated and displayed in Figure~\ref{Figure:PPV_streams}. The top panels of Figure~\ref{Figure:PPV_streams} represent PPV-diagrams equivalent to Figure~\ref{Figure:PPV_all}. However, in this instance, the black data points indicate the location of both the \ntwoh \ and HNC velocity components. The bottom panels depict the spatial distribution of both of these PPV-structures, where the coloured contour and symbols in each of the panels is equivalent to that in Figure~\ref{Figure:pp_pv}).

The low-velocity PPV-structure\footnote{We avoid using the term ``structure'' outright given that projection effects can cause uncertainty in the direct translation of PPV to PPP space (e.g. \citealp{beaumont_2013}). Instead we favour the term ``PPV-structure'' to describe a region of gas that is coherent in both the plane of the sky and in its kinematics. } (left panels in Figure~\ref{Figure:PPV_streams} and blue contours in Figure~\ref{Figure:pp_pv}) was identified by \citet{bally_1988} in their $^{13}$CO study of the CMZ, and dubbed ``the negative velocity filament''. Referring back to Figure\,\ref{Figure:PPV_all}, there is a significant ($\sim90$\,pc) gap between the HNCO emission associated with the dust ridge ($l>0\fdg2$) and that which is prominent at negative longitudes. The inclusion of \ntwoh \ and HNC emission features in the left panels of Figure~\ref{Figure:PPV_streams} reduces the projected extent of the emission gap. 

It is worth noting that this discontinuity coincides with lower column density material ($N[{\rm H_{2}}]\sim10^{22}$\,cm$^{-2}$). For a relatively constant abundance of HNCO with respect to H$_{2}$, this would imply that any HNCO emission is below the detection threshold. Alternatively, the lack of HNCO emission could be caused by abundance variations. HNCO is enhanced towards regions of shocked gas (see \S~\ref{Section:method_cmz}). This may therefore be indicative of relatively quiescent material (in comparison to those regions where HNCO emission is brightest, e.g. Sgr B2; see \S~\ref{Section:results_sgrb2}). This is supported by the fact that the HNCO emission at negative longitudes also exhibits a small velocity dispersion $\sigma\sim6.1$\,\kms \ (see Figure~\ref{Figure:PPV_all_disp}). \citet{kruijssen_2015} argue that the data presented in the left-hand panels of Figure\,\ref{Figure:PPV_streams} is neither coherent nor continuous in 3-D positional space. Instead, they suggest that there are two physically independent structures within this $\{l,\,b,\,v_{\rm LSR}\}$ range, and that they are situated on the near and far sides of the Galactic centre, respectively. Differing conditions throughout the Galactic centre gas stream may therefore explain the emission pattern. 

There are remarkable oscillatory patterns in the $\{l,\,b,\,v_{\rm LSR}\}$ data presented in the top-panels of Figure~\ref{Figure:PPV_streams} (also throughout the CMZ). This is particularly pronounced in the top-left panel, where we observe five full wavelengths of (projected) magnitude $\lambda_{\rm v}\sim20$\,pc (four if we include only the data at negative longitudes, taking into account the uncertainty in the coherency of this gas stream). The characteristic properties and origins of this oscillatory pattern will be explored in a future publication (Henshaw et al.,~in prep.). However, we note that the oscillation may be related to the same process that produces the characteristic spacing in \amm \ (1,1) emission features studied by \citet{kruijssen_2015}, which they attribute to the Jeans length.

Close to the dust ridge molecular clouds (cyan contour; Figure~\ref{Figure:PPV_dustridge}), the low-velocity PPV-structure appears to bifurcate (note: this not shown in Figure~\ref{Figure:PPV_streams}, please refer instead to the bottom panel of Figure~\ref{Figure:pp_pv}). This gives the appearance of a bubble-like feature centred on $\{l,\,v_{\rm LSR}\}~\sim\{0\fdg12,\,0.0\,{\rm km\,s^{-1}}\}$. A visual inspection of this feature in PPV space indicates that the high-velocity portion of this ``bubble'' is offset in Galactic latitude from the gas shown in Figure~\ref{Figure:PPV_streams} ($b\sim~-0\fdg1$). We suspect therefore that this represents the projected alignment of two spatially independent features. 

\citet{sofue_1995} also identify the second feature shown in the right panels of Figure~\ref{Figure:PPV_streams} (this feature is systematically red-shifted with respect to the first and highlighted by the red contour in Figure~\ref{Figure:pp_pv}). \citet{sofue_1995} refers to these two structures as ``Arm\,{\sc i}'' (left panels; Figure~\ref{Figure:PPV_streams}) and ``Arm\,{\sc ii}'' (right panels; Figure~\ref{Figure:PPV_streams}), in reference to the inferred three dimensional structure. Emission from HNCO is more prominent throughout the high-velocity PPV-structure, dominated by emission associated with the 20\,\kms \ and 50\,\kms \ molecular clouds (see \S~\ref{Section:results_2050}). The velocity dispersion stays relatively constant throughout this feature, within the range $7.4\,{\rm km\,s^{-1}}~<~\sigma~<~13.3\,{\rm km\,s^{-1}}$ (see Figure~\ref{Figure:PPV_all_disp}). Once again, gaps in the HNCO emission coincide with lower column density material. Only when the additional molecular species, \ntwoh \ and HNC, are added to the analysis does this feature appear continuous. 

In Figures~\ref{Figure:PPV_all} and \ref{Figure:pp_pv} it is evident that the molecular line emission in the velocity range $60.0\,{\rm km\,s^{-1}}~<v_{\rm LSR}<~100.0\,{\rm km\,s^{-1}}$, extends to Galactic longitudes of $l>0\fdg7$. This emission is highlighted in Figure~\ref{Figure:pp_pv} with a pink contour, and may extend as far as the $1\fdg3$ cloud complex (Longmore et al.,~in prep). Previous studies have drawn a connection between this emission and the extended features discussed above. \citet{tsuboi_1999} suggest that this represents a continuation of Arm\,{\sc i} (left panels; Figure~\ref{Figure:PPV_streams}), and refer to the entirety of this feature as the ``Galactic centre bow''. \citet{sawada_2004} on the other hand (see also \citealp{johnston_2014}), infer that this emission is physically connected to Arm\,{\sc ii} (right panels; Figure\,\ref{Figure:PPV_streams}). 

\begin{figure}
\begin{center}
	\includegraphics[trim = 30mm 5mm 20mm 12mm, clip, width = 0.49\textwidth]{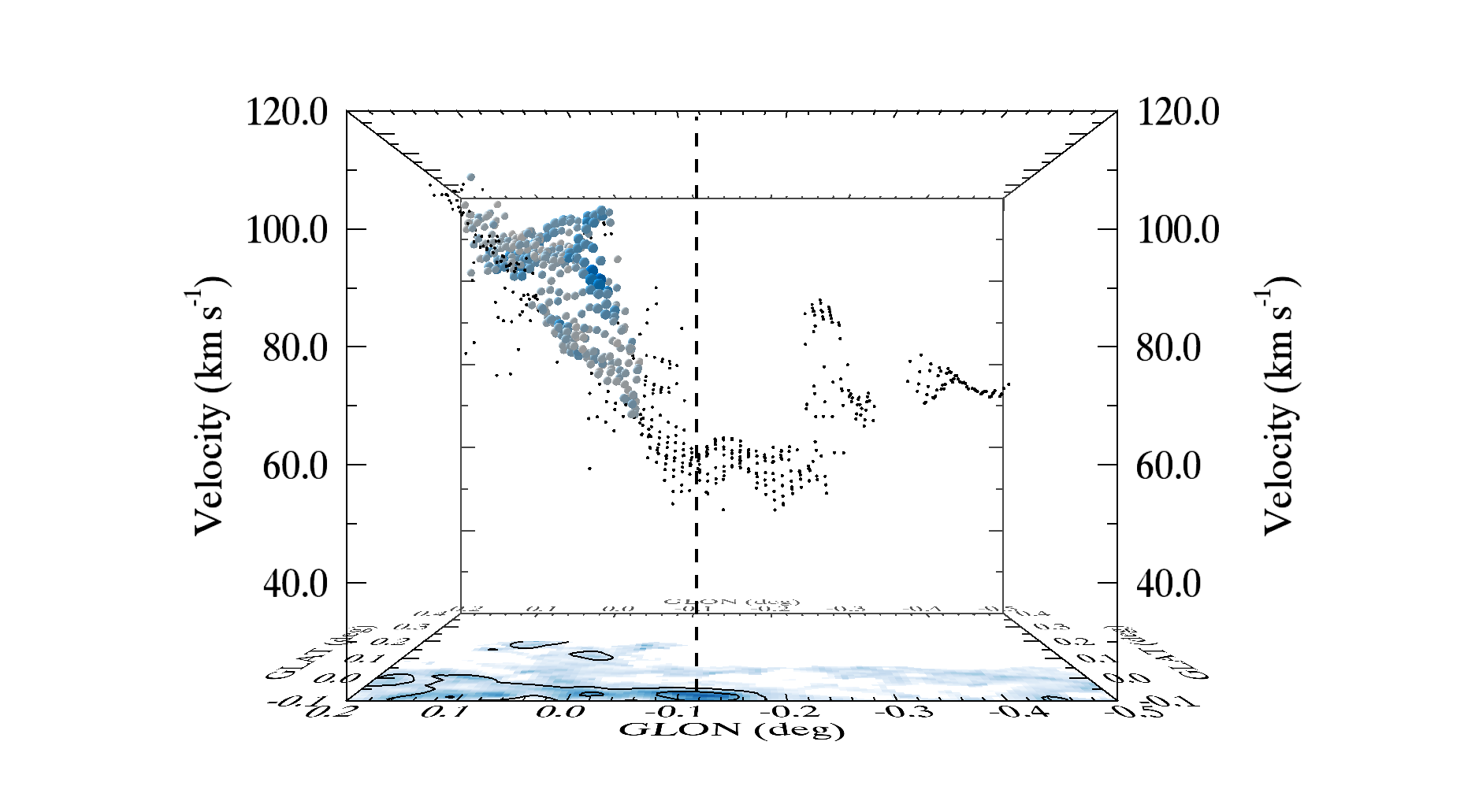}
	\includegraphics[trim = 5mm 5mm -14mm 65mm, clip, width = 0.45\textwidth]{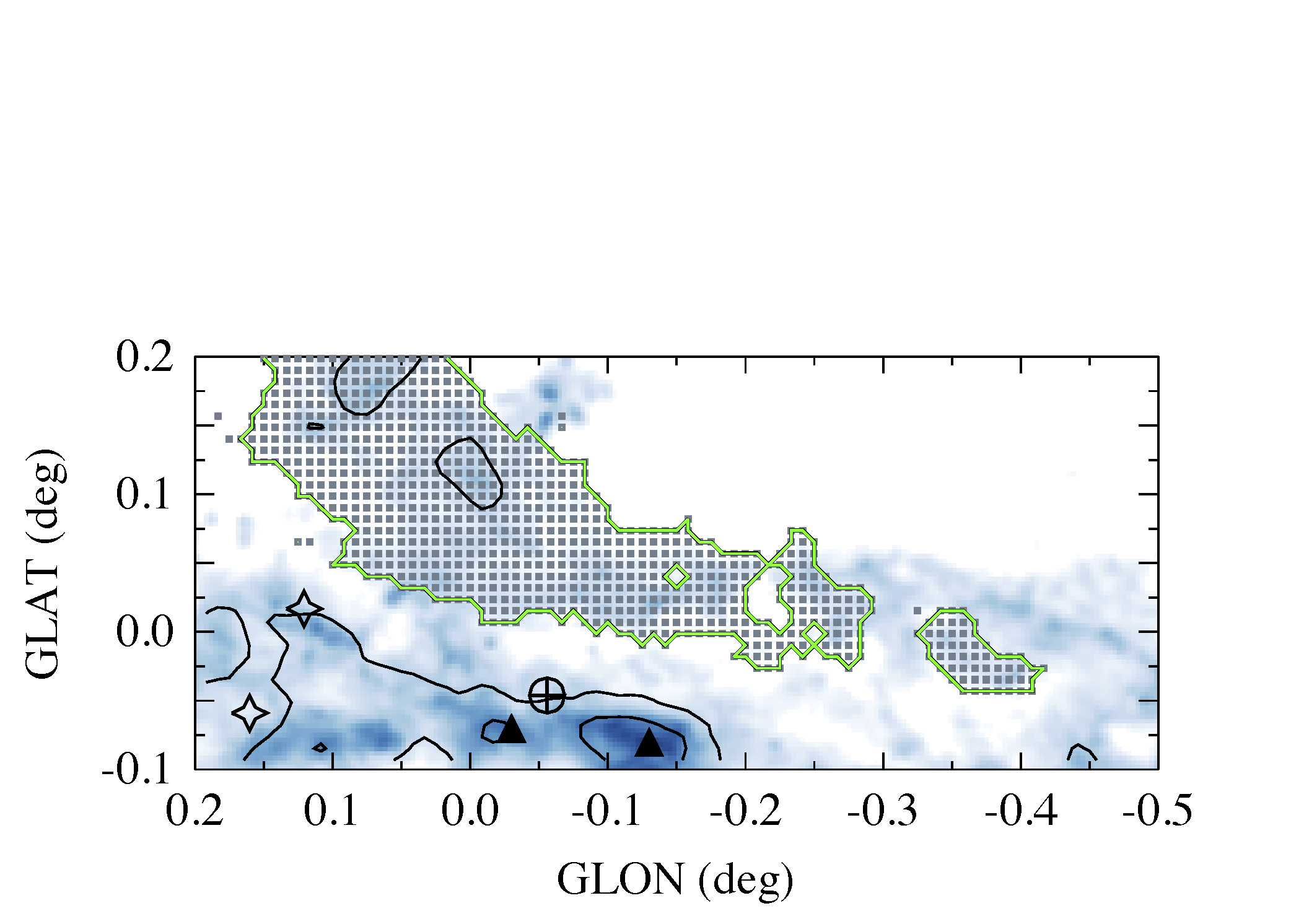}
\end{center}
\caption{Top panel: A PPV image of ``Arm\,{\sc iii}'' (\citealp{sofue_1995}; also known as the ``polar arc'' in \citealp{bally_1988}, see \S~\ref{Section:results_continuous}). The coloured data points are equivalent to those presented in Figure~\ref{Figure:PPV_streams}. Bottom panel: A map indicating the spatial distribution of the data points shown in the PPV diagram. The green contour and symbols are equivalent to those in Figure~\ref{Figure:pp_pv}. The background image and contours in both panels are equivalent to those in Figure\,\ref{Figure:PPV_all}.}
\label{Figure:PPV_polararc}
\end{figure}

Figure~\ref{Figure:PPV_polararc} shows ``Arm\,{\sc iii}'' of \citet{sofue_1995}. This was originally referred to as the ``polar arc'' in \citet{bally_1988}, who noted that the feature is inclined by $>40\degr$ with respect to the Galactic plane. Although associated with the CMZ, as is evidenced by its significant median velocity dispersion (9.6\,\kms; see Figure~\ref{Figure:PPV_all_disp}), the inclination of Arm\,{\sc iii} makes it difficult to place within the context of the prominent features discussed above. Both \citet{bally_1988} and \citet{sofue_1995} suggest that the Arm\,{\sc iii} merges with the gas at lower velocities (Arm\,{\sc ii}; right panels of Figure~\ref{Figure:PPV_streams}). However, Figure~\ref{Figure:PPV_polararc} shows that there is a turnover in velocity gradient at $\{l,\,v_{{\rm LSR}}\}\sim~\{-0\fdg2,\,55.0\,{\rm km\,s^{-1}}\}$ \ towards increasingly positive velocities (if tracing the structure from positive to negative longitudes). Although the analysis of more abundant molecular species (for example, \tco) would make for a better comparison, we find that Arm\,{\sc iii} appears to be spatially coincident with Arm\,{\sc i} (although separated in velocity) rather than Arm\,{\sc ii} (cf. Figure~\ref{Figure:PPV_streams} and the bottom panel of Figure~\ref{Figure:PPV_polararc}). 

Both \citet{molinari_2011} and \citet{kruijssen_2015}, omit the polar arc from their respective orbital studies. \citet{kruijssen_2015} discuss the possibility that this feature is physically unrelated to the PPV-structures discussed above. Instead they speculate that this material has been ejected from one of the extended PPV-structures by feedback. The spatial coincidence between the polar arc and the infrared shells blown by the Arches and Quintuplet clusters, as well as their similar line-of-sight velocities, is cited as evidence in support of this claim.

The prominence of the main features discussed above has inspired several different interpretations for how each structure contributes to the 3-D structure of the inner CMZ. In \S\,\ref{Section:results_models} we revisit this topic and discuss the above features in the context of some of the theories competing for a holistic description of the structure of the CMZ.  

\subsubsection{The Sagittarius C molecular cloud complex (and surrounding region)}\label{Section:results_sgrc}

The molecular gas distribution surrounding the Sgr C H\,{\sc ii} region (denoted by the black plus sign in Figure~\ref{Figure:pp_pv}, $\{l, b\}~=~\{-0\fdg57, -0\fdg09\}$; \citealp{kuchar_1997}) is divided into several coherent PPV-structures. Many of these overlap in position but are distinguishable in velocity. 

\begin{figure}
\begin{center}
	\includegraphics[trim = 50mm 20mm 50mm 25mm, clip, width = 0.48\textwidth]{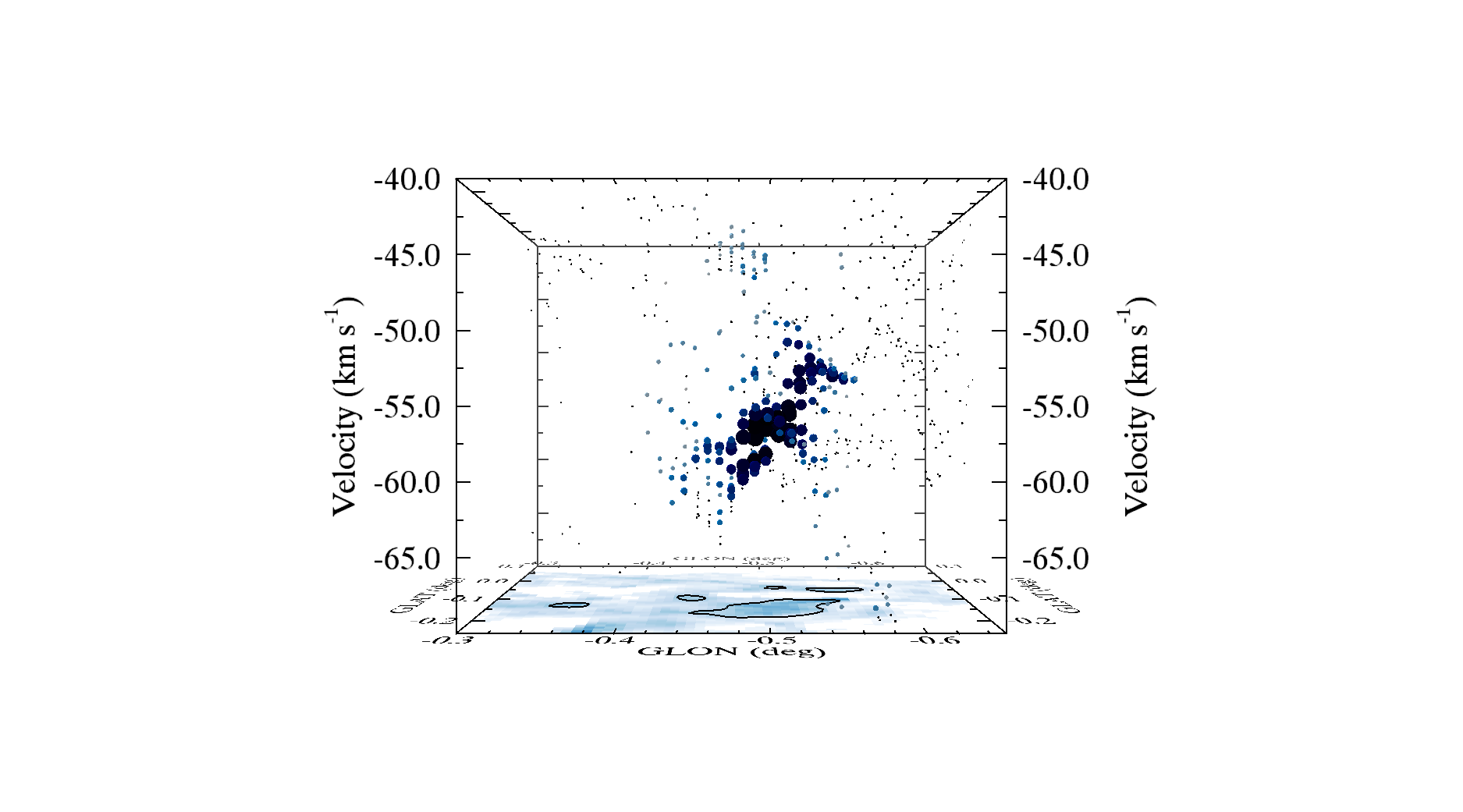}
	\includegraphics[trim = 7mm 5mm 76mm 25mm, clip, width = 0.47\textwidth]{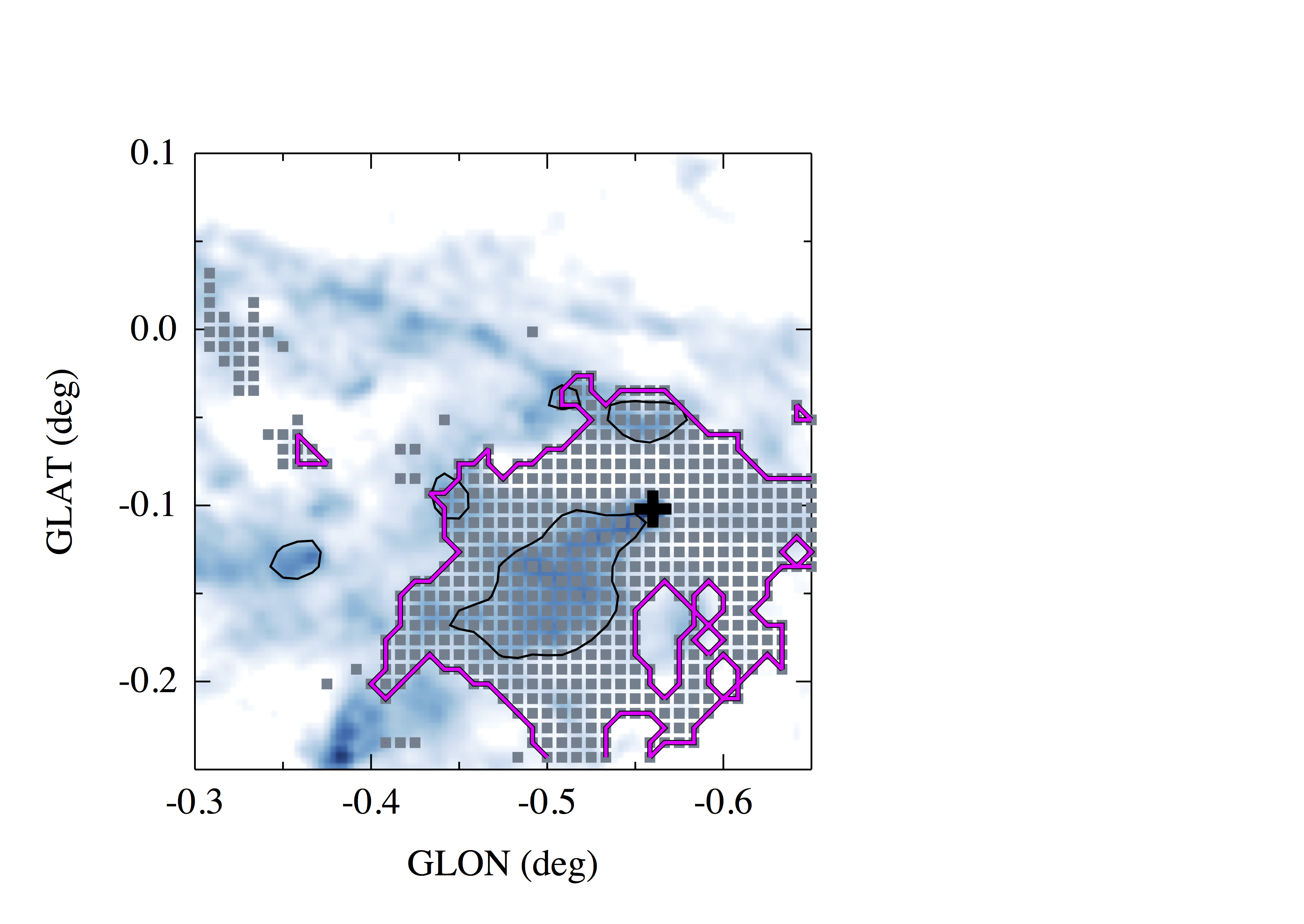}
\end{center}
\caption{Top panel: A PPV diagram of the molecular gas distribution between $-0\fdg65\,<l<-0\fdg30$ and $-0\fdg25<b<0\fdg10$ in position and $-70.0\,{\rm km\,s^{-1}}<l<40.0\,{\rm km\,s^{-1}}$ in velocity, as part of the Sgr C molecular cloud complex (see \S~\ref{Section:results_sgrc}). The data points have equivalent meaning to those presented in Figure\,\ref{Figure:PPV_streams}. Bottom panel: Highlighting the distribution of data points in the top panel. The magenta contour and black symbol are equivalent to those in Figure~\ref{Figure:pp_pv}. The background image and contours in both panels are equivalent to those in Figure\,\ref{Figure:PPV_all}.}
\label{Figure:PPV_sgrc}
\end{figure}

Figure~\ref{Figure:PPV_sgrc} is a PPV diagram highlighting the molecular gas distribution of one of the more prominent features. This focuses on the emission between $-~0\fdg65~<~l~<~-0\fdg35$ and $-0\fdg25<~b~<-~0\fdg1$ in position and $-70.0\,{\rm km\,s^{-1}}<~v_{\rm LSR}~<~-40.0\,{\rm km\,s^{-1}}$ in velocity (highlighted with the magenta contour in Figure~\ref{Figure:pp_pv}). The total mass of this $\sim16\,{\rm pc}\times16\,{\rm pc}$ cloud (Table~\ref{Table:positions}) is estimated to be $\sim10^{5}$\,\solar \ \citep{lis_1994b, kendrew_2013}. At the location of the Sgr C H\,{\sc ii} region, a velocity dispersion of $\sigma=3.1\,{\rm km\,s^{-1}}\pm0.3\,{\rm km\,s^{-1}}$ ($v_{\rm LSR}~=~-53.0\,{\rm km\,s^{-1}}~\pm~0.1\,{\rm km\,s^{-1}}$) is reported. The median velocity dispersion of all spectral components presented in Figure~\ref{Figure:PPV_sgrc} is $\sim7.8$\,\kms.

The northern portion of this cloud ($v_{\rm LSR}\sim-52$\,\kms), is thought to be associated with high-mass star formation \citep{yusef-zadeh_2009}. At the location of G359.44--0.102, in close proximity to an Extended Green Object (EGO; \citealp{cyganowski_2008}), we report a velocity dispersion of $\sigma=5.6\,{\rm km\,s^{-1}}~\pm~0.3\,{\rm km\,s^{-1}}$ ($v_{\rm LSR}~=-52.8\,{\rm km\,s^{-1}}~\pm~0.3\,{\rm km\,s^{-1}}$). This is similar to the value presented by \citet{kendrew_2013}, and the slight difference in dispersion may be due to our choice of molecular line or our spatial smoothing of the data (see \S~\ref{Section:data}).\footnote{\citet{kendrew_2013} comment that the similarity in velocity of the near 3\,kpc spiral arm and cloud at this location makes it difficult to discern the origin of the star formation signatures. Indeed, \citet{green_2009} associate the CH$_{3}$OH masers with the 3\,kpc arm rather than the Sgr C complex. However, given the broad velocity dispersion and widespread emission from shocked gas tracers, we are confident that the bulk of the material is at the Galactic centre distance.}

The southern portion of the cloud ($v_{\rm LSR}\sim\,-57$\,\kms) by comparison, has no obvious signatures of active star formation (evidenced via a lack of 8, 24, 70\,\micron \ emission; \citealp{kendrew_2013}). Given the limited angular and spectral resolution of these smoothed observations, it is not clear from Figure~\ref{Figure:PPV_sgrc} whether the HNCO emission traces a single cloud exhibiting a velocity gradient, or two independent components that are separated by $\sim5$\,\kms. Higher angular and spectral resolution observations would help to remove this ambiguity. 

\begin{figure}
\begin{center}
	\includegraphics[trim = 45mm 15mm 30mm 22mm, clip, width = 0.5\textwidth]{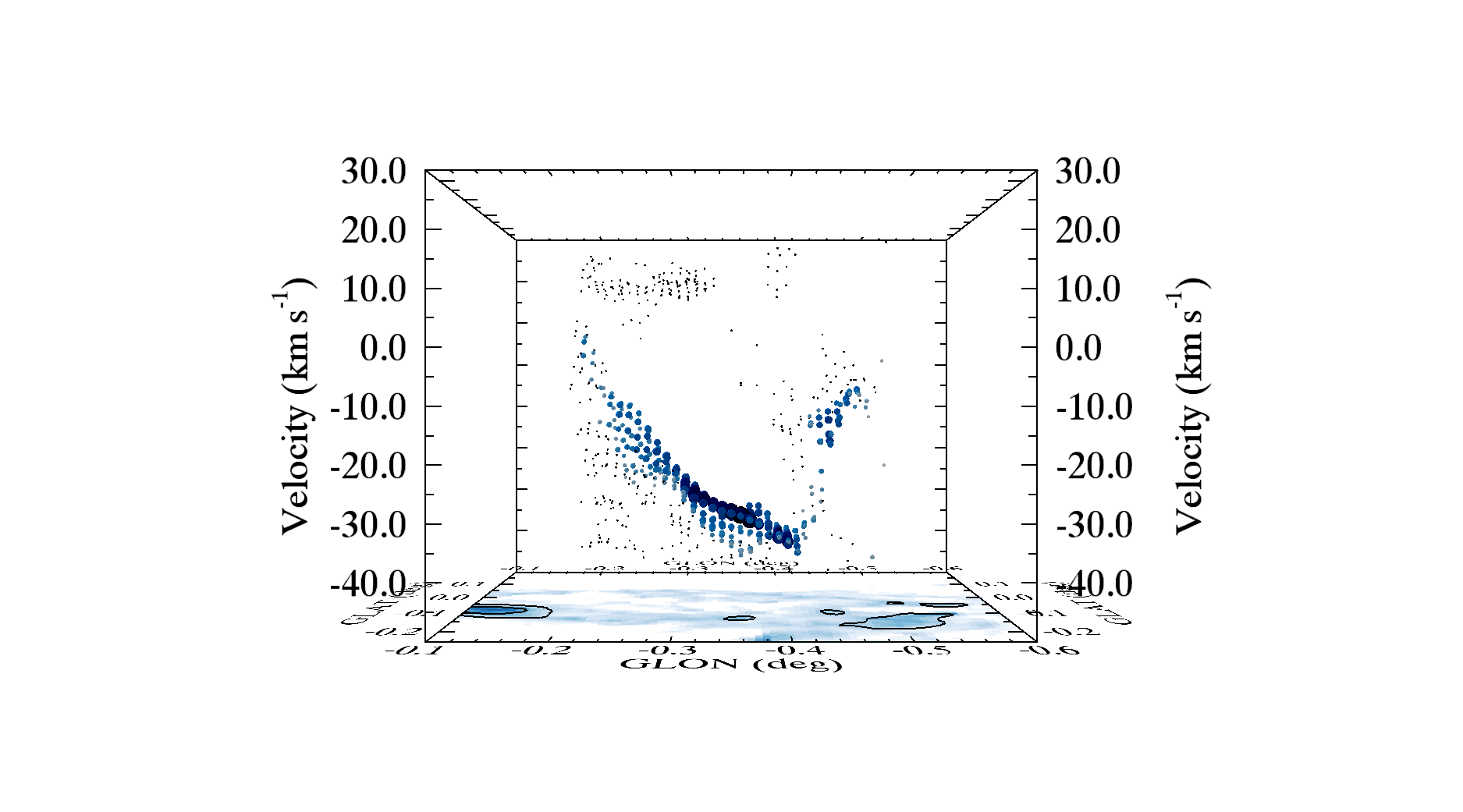}
	\includegraphics[trim = 5mm 0mm -2mm 25mm, clip, width = 0.44\textwidth]{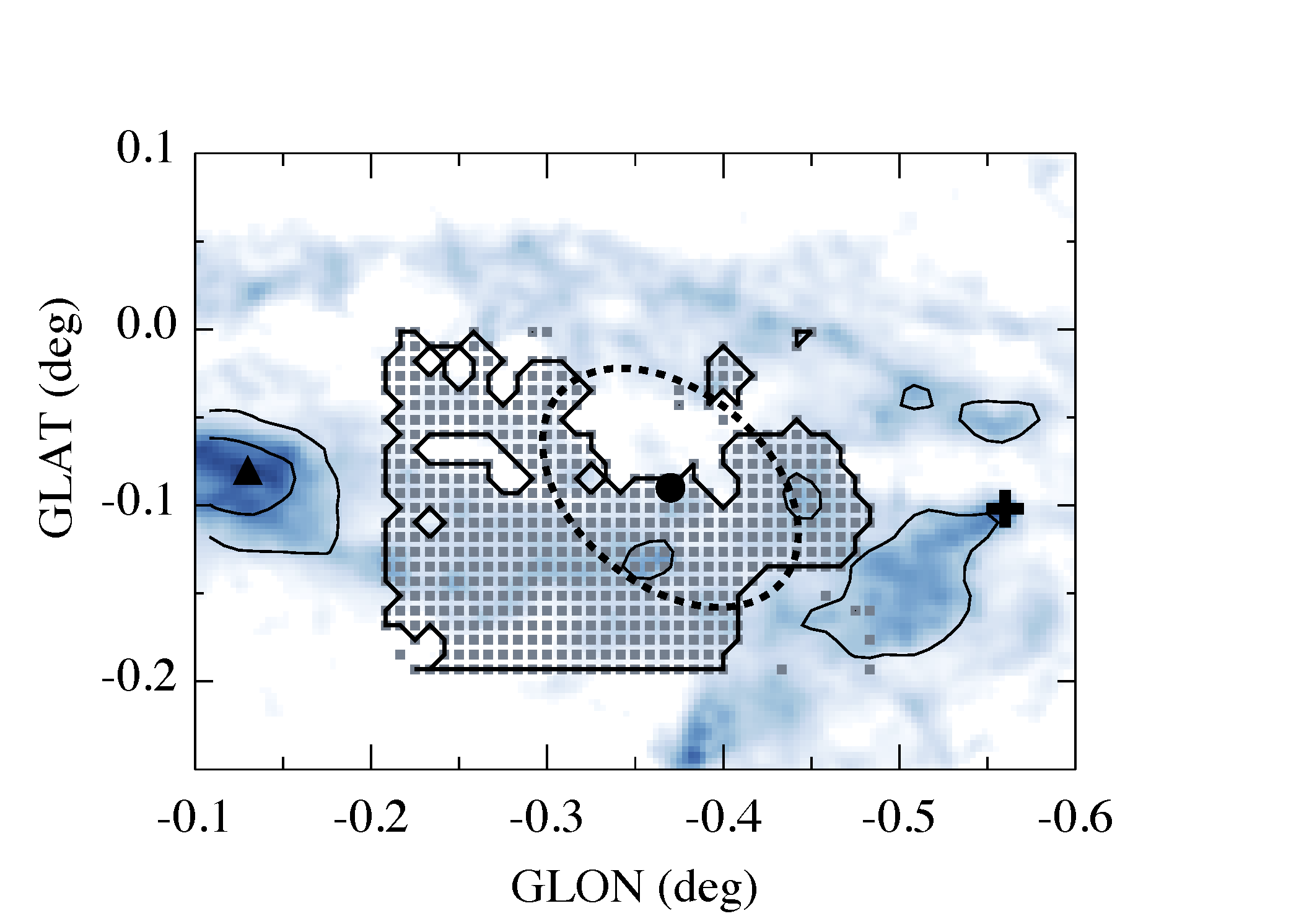}
\end{center}
\caption{Top panel: A PPV diagram of the molecular gas distribution between $-0\fdg5\,<l<-0\fdg2$ and $-0\fdg2<b<0\fdg0$ in position and $-40.0\,{\rm km\,s^{-1}}<l<20.0\,{\rm km\,s^{-1}}$ in velocity, as part of the Sgr C molecular cloud complex. The data points have equivalent meaning to those presented in Figure\,\ref{Figure:PPV_streams}. Bottom panel: Highlighting the distribution of data points in the top panel. The filled black circle and dashed black ellipse indicate the centre position and physical extent of a shell of molecular emission reported in \citet{tanaka_2014} and discussed in \S~\ref{Section:results_sgrc}. The background image and contours in both panels are equivalent to those in Figure\,\ref{Figure:PPV_all}.  }
\label{Figure:PPV_shell}
\end{figure}

Elsewhere within the Sgr C complex, \citet{tanaka_2014} report the presence of two shell-like structures seen in HCN (4-3) emission (integrated over a velocity range $-60\,{\rm km\,s^{-1}}~<~v_{\rm LSR~}<~50\,{\rm km\,s^{-1}}$). The shells are centred on $\{l, b\}=\{-0\fdg37,\,-0\fdg06\}$ and $\{l, b\}=\{-0\fdg46,\,-0\fdg21\}$, and have projected spatial separations from the centre of the Sgr C H\,{\sc ii} region of $\sim29$\,pc and 24\,pc, respectively. The top panel of Figure~\ref{Figure:PPV_shell} is a PPV diagram of the molecular gas emission between $-0\fdg5\,<l<-0\fdg2$ and $-0\fdg2<b<0\fdg0$ in position and $-40.0\,{\rm km\,s^{-1}}<v_{\rm LSR}<20.0{\rm km\,s^{-1}}$ in velocity. This serves as the approximate location of the first shell discussed in \citet{tanaka_2014}, the physical extent of which is displayed in the bottom panel as a black dashed ellipse.

\citet{tanaka_2014} speculate that the shell structures are molecular superbubbles produced by supernova explosions, concluding that more than ten supernovae (or two hypernovae) within the last $2\times10^{5}$ years would be needed to explain the kinetic energy of the expansion motion of each shell. From this, they infer a star formation rate of $(0.5-3)\times10^{-2}$\,\solar\,yr$^{-1}$. 

It is possible to test this hypothesis, assuming that the supernova explosions are related to a stellar cluster. To do this we first have to estimate the rate at which supernovae occur within a star cluster of given mass, $M_{\rm cl}$. This can then be related to the rate inferred by \citet{tanaka_2014} (for detailed information on our methodology please see Appendix~\ref{App:SNrate}). Figure~\ref{Figure:SNrate} shows the estimated supernova rate as a function of cluster age for clusters of masses $10^{4}$\,\solar \ (black) and $10^{5}$\,\solar \ (blue). Given that the Arches and Quintuplet clusters have respective ages $\tau=3.5\pm0.7$\,Myr and $4.8\pm1.1$\,Myr \citep{schneider_2014}, and they are both gas-poor, a cluster within this region would have to be younger by comparison. Assuming a conservative upper limit to the age of 5\,Myr, we estimate that a cluster of mass $M_{\rm cl}\sim5\times10^{4}$\,\solar \ would be required to produce the inferred supernova rate of \citet{tanaka_2014}.  

Following the method of Barnes et al.~(in prep.),\footnote{The method estimates the mass of the highest-mass star within the region using the bolometric luminosity-mass conversions presented by \citet{davies_2011}. The embedded population mass is then extrapolated from this value, assuming the form of the initial mass function (IMF).}  we estimate that the upper mass limit of the embedded stellar population within this region is $\sim0.8\times10^{4}$\,\solar. This rules out the notion that the supernovae are related to an Arches or Quintuplet-like star cluster leading to the production of the observed shell-like structures. Consequently, this could favour the suggestion of \citet{tanaka_2014}, that the supernovae are related to a more dispersed population of high-mass stars (there are a number of 24\,\micron \ compact sources found between Sgr~C and Sgr~A; \citealp{yusef-zadeh_2009}).

However, a cursory inspection of Figure~\ref{Figure:PPV_shell} reveals that no complete shell is evident when only considering emission from HNCO (blue data points). Additionally, when examined in the context of the CMZ as a whole, the prominent HNCO emission at this location (that which would form the lower portion of the shell in the \citealp{tanaka_2014} interpretation; see bottom panel of Figure~\ref{Figure:PPV_shell}), appears to be a small segment of the $\sim$\,150\,pc long feature discussed in \S~\ref{Section:results_continuous} and shown in the right hand panels of Figure~\ref{Figure:PPV_streams}. Integrating molecular line emission at this location over a velocity range of $>100$\,\kms \ (as in \citealp{tanaka_2014}) incorporates emission from several line-of-sight components. Shell-like features, observed in velocity-integrated emission maps of complex regions, may therefore be inferred following the superposition of several, otherwise independent, molecular clouds (particularly when integrating over significant, $\sim100$\,\kms, velocity ranges). 

\begin{figure}
\begin{center}
	\includegraphics[trim = 30mm 25mm 20mm 50mm, clip, width = 0.5\textwidth]{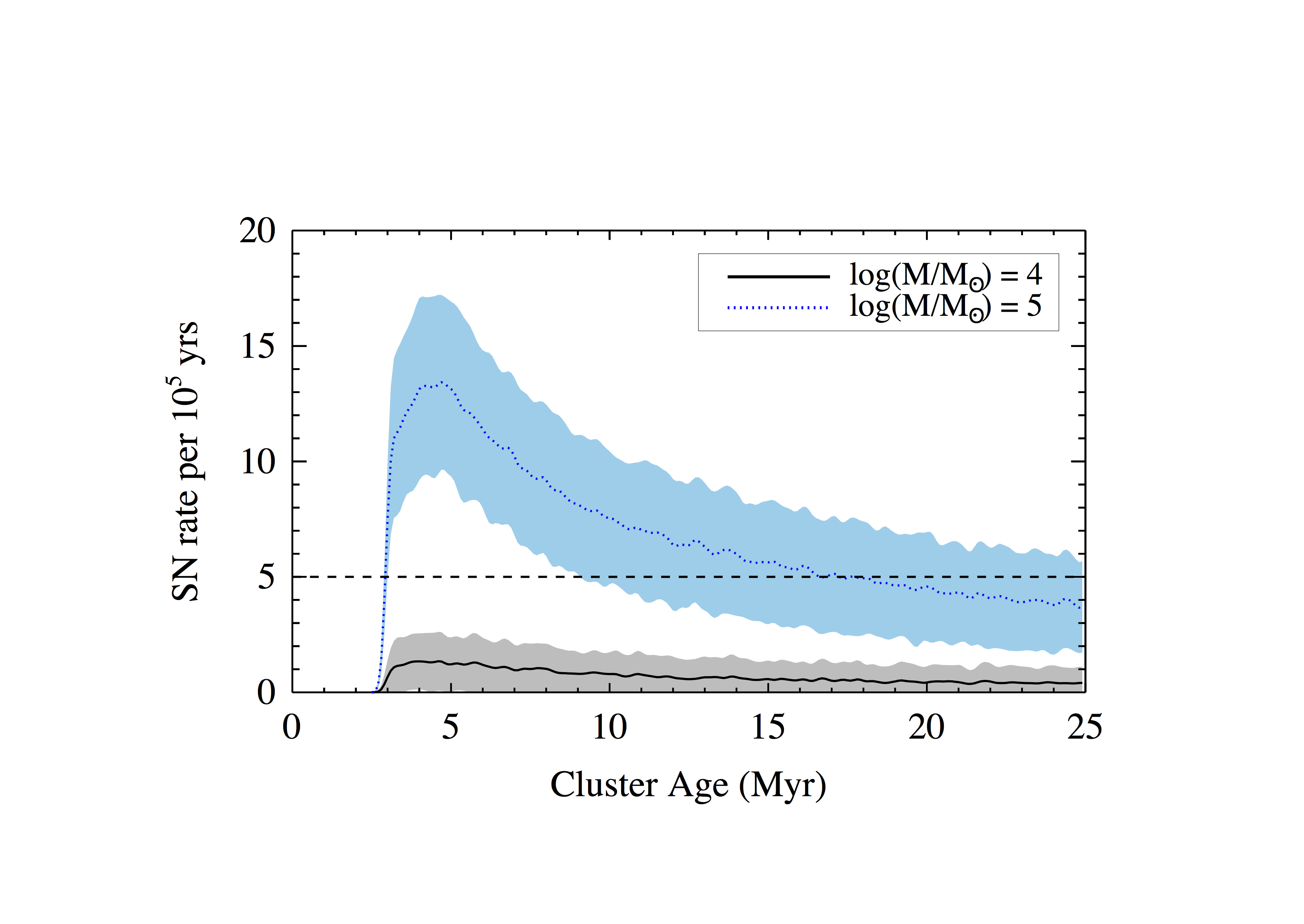}
\end{center}
\caption{Plotting the estimated supernova rate (per 10$^{5}$\,yrs) versus age for two star clusters of masses $M_{\rm cl}=10^{4}$\,\solar \ (black) and 10$^{5}$\,\solar \ (blue), respectively. The shaded region represents the standard deviation in the supernova rate over 100 runs of the simulation (for a more in-depth description of the methodology, please refer to Appendix~\ref{App:SNrate}). The horizontal dashed line indicates the supernova rate inferred by \citet{tanaka_2014} to explain the expansion motion of a shell-like feature observed in velocity-integrated emission maps towards the Sgr C molecular cloud complex ($\sim5\times10^{-5}$\,yr$^{-1}$). As is indicated, a star cluster of $M_{\rm cl}>10^{4}$\,\solar \ would be required to reproduce the supernova rate of \citet{tanaka_2014}.   }
\label{Figure:SNrate}
\end{figure}

\subsubsection{The 20\,km\,s$^{-1}$ and 50\,km\,s$^{-1}$ clouds}\label{Section:results_2050}

After the Sgr B2 region, the molecular clouds located closest in projection to Sgr A* appear brightest in HNCO emission (see Figure~\ref{Figure:PPV_all}). Situated between $-0\fdg2<l<0\fdg2$ and $-0\fdg15<b<0\fdg05$ in position and between $0.0\,{\rm km\,s^{-1}}<v_{\rm LSR}<60.0\,{\rm km\,s^{-1}}$ in velocity, the Sgr A region encompasses several molecular clouds, including the 20\,\kms \ (GCM~$~-0.13~-0.08$) and 50\,\kms \ (GCM~$~-0.02~-0.07$) clouds (denoted by the upwards facing triangles in Figure~\ref{Figure:pp_pv}) (as well as a host of other clouds; see e.g. GCM\,$0.11-0.11$ \citealp{tsuboi_2011}). 

\begin{figure}
\begin{center}
	\includegraphics[trim = 50mm 5mm 45mm 15mm, clip, width = 0.48\textwidth]{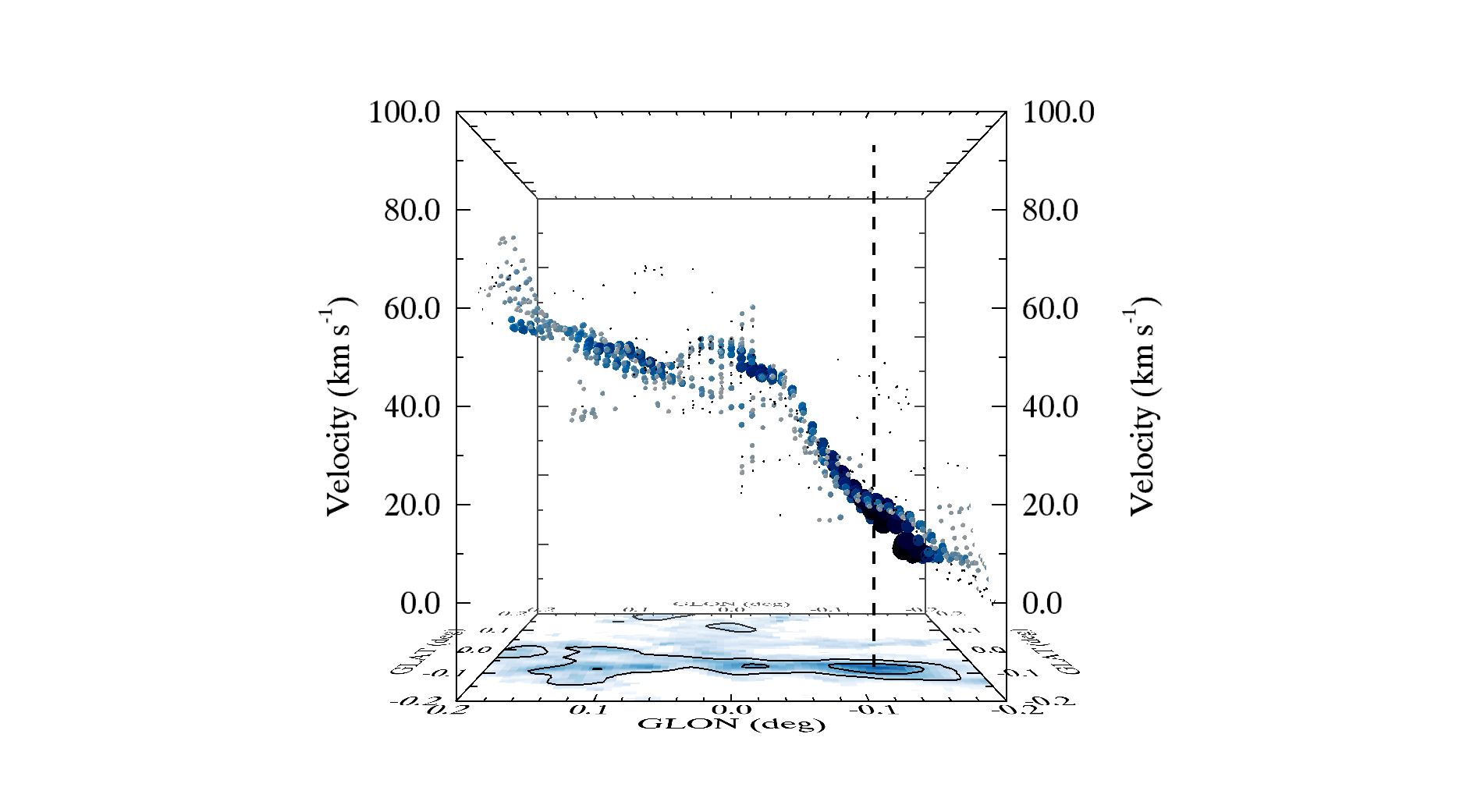}
	\includegraphics[trim = 50mm 8mm 24mm 32mm, clip, width = 0.47\textwidth]{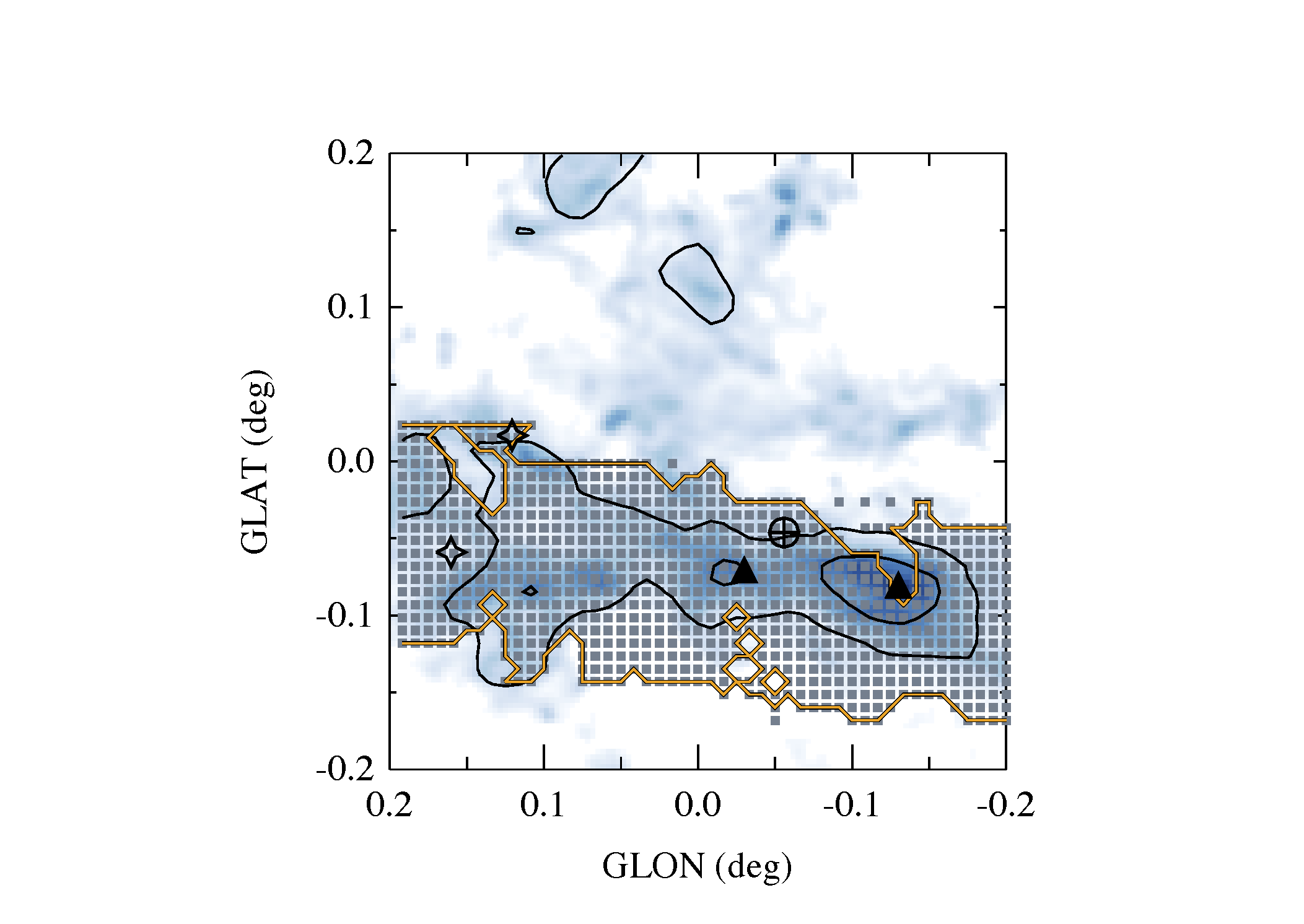}
\end{center}
\caption{Top panel: A PPV diagram of the molecular gas distribution between $-0\fdg2<l<0\fdg2$ and $-0\fdg15<b<0\fdg05$ in position and $0.0\,{\rm km\,s^{-1}}<v_{\rm LSR}<60.0\,{\rm km\,s^{-1}}$ in velocity. This region incorporates several massive molecular clouds, including the 20\,\kms \ and 50\,\kms \ clouds, and GCM\,~$0.11-0.11$. The data points have equivalent meaning to those presented in Figure~\ref{Figure:PPV_streams}. Bottom panel: Highlighting the distribution of data points in the top panel. The background image and contours in both panels are equivalent to those in Figure~\ref{Figure:PPV_all}. }
\label{Figure:PPV_2050}
\end{figure}

Figure~\ref{Figure:PPV_2050} is a PPV diagram of this region. The bottom panel highlights the spatial distribution of data points shown in the top panel. The brightest HNCO emission is associated with the 20\,\kms \ cloud. The centroid velocity and velocity dispersion measured towards the location of the 20\,\kms \ cloud at $\{l,\,b\}=\{-0\fdg133,\,-0\fdg076\}$ are $v_{0}=~9.43\pm0.04$\,\kms \ and $\sigma=~7.00\pm0.04$\,\kms, respectively. However, the intensity-weighted centroid velocity within the area $-0\fdg15<~l~<0\fdg07$ and $-0\fdg11<~b~<0\fdg06$ is $\langle~v_{\rm 0}~\rangle~=~16.25~\pm~0.05$\,\kms \ (Table~\ref{Table:positions}). 

Consistent with previous observations, the 20\,\kms \ and 50\,\kms \ clouds are connected smoothly and coherently by velocity gradient of magnitude $\sim2.3$\,\vel \ over a projected distance of $\sim17$\,pc \citep{bally_1988, sandqvist_1989, zylka_1990, oka_1998, coil_2000}. The centroid velocity and velocity dispersion measured towards the 50\,\kms \ cloud at $\{l,\,b\}=\{-0\fdg017,\,-0\fdg068\}$ are $v_{0}=~49.03\pm0.09$\,\kms \ and $\sigma=~9.79\pm0.09$\,\kms, respectively. Continuing in the direction of increasing Galactic longitude, the magnitude of the velocity gradient decreases before reaching GCM\,$0.11-0.11$ (where we measure a centroid velocity and velocity dispersion of $v_{0}~=~52.65\pm0.35$\,\kms \ and $\sigma~=~9.44\pm0.31$\,\kms, respectively).

The line-of-sight location of the 20\,\kms \ and 50\,\kms \ molecular clouds is currently debated. There are several lines of evidence which suggest that these clouds are situated close to ($\ll60$\,pc), and possibly interacting with, the circumnuclear disc orbiting Sgr~A* (e.g. \citealp{ho_1985, yusef-zadeh_1999, herrnstein_2005, sjouwerman_2008, ott_2014}). In their interpretation for the 3-D structure of the CMZ, \citet{molinari_2011} suggest that Sgr~A* must be offset from the dynamical centre of the orbit they prescribe (situated closer to the 20\,\kms \ and 50\,\kms \ clouds), in order to explain the difference in the clouds' radial velocities. In the \citet{molinari_2011} geometry, the 20\,\kms \ and 50\,\kms \ clouds are situated within $\sim20$\,pc from Sgr~A*, and are both on the near-side of the Galactic centre (this will be discussed in more detail in \S~\ref{Section:results_closedellipse}). 

The spectral resolution of the observations presented in this work is insufficient to draw any physical connection between the 20\,\kms \ and 50\,\kms \ clouds and the circumnuclear disc orbiting Sgr~A*. However, comparison between the top panel of Figure~\ref{Figure:PPV_2050} and the right-hand panel of Figure~\ref{Figure:PPV_streams} reveals that the bulk of molecular emission attributed to these two clouds is associated with the high-velocity extended gas stream discussed in \S~\ref{Section:results_continuous}. This is consistent with the work of \citet{kruijssen_2015}, whose dynamical model describing the gas kinematics of the CMZ implies a physical separation of  $\sim60$\,pc between the 20\,\kms \ and 50\,\kms \ clouds and Sgr~A* (see \S~\ref{Section:results_openstream}).

\subsubsection{The dust ridge molecular clouds}\label{Section:results_dustridge}

\begin{figure}
\begin{center}
	\includegraphics[trim = 38mm 5mm 35mm 15mm, clip, width = 0.48\textwidth]{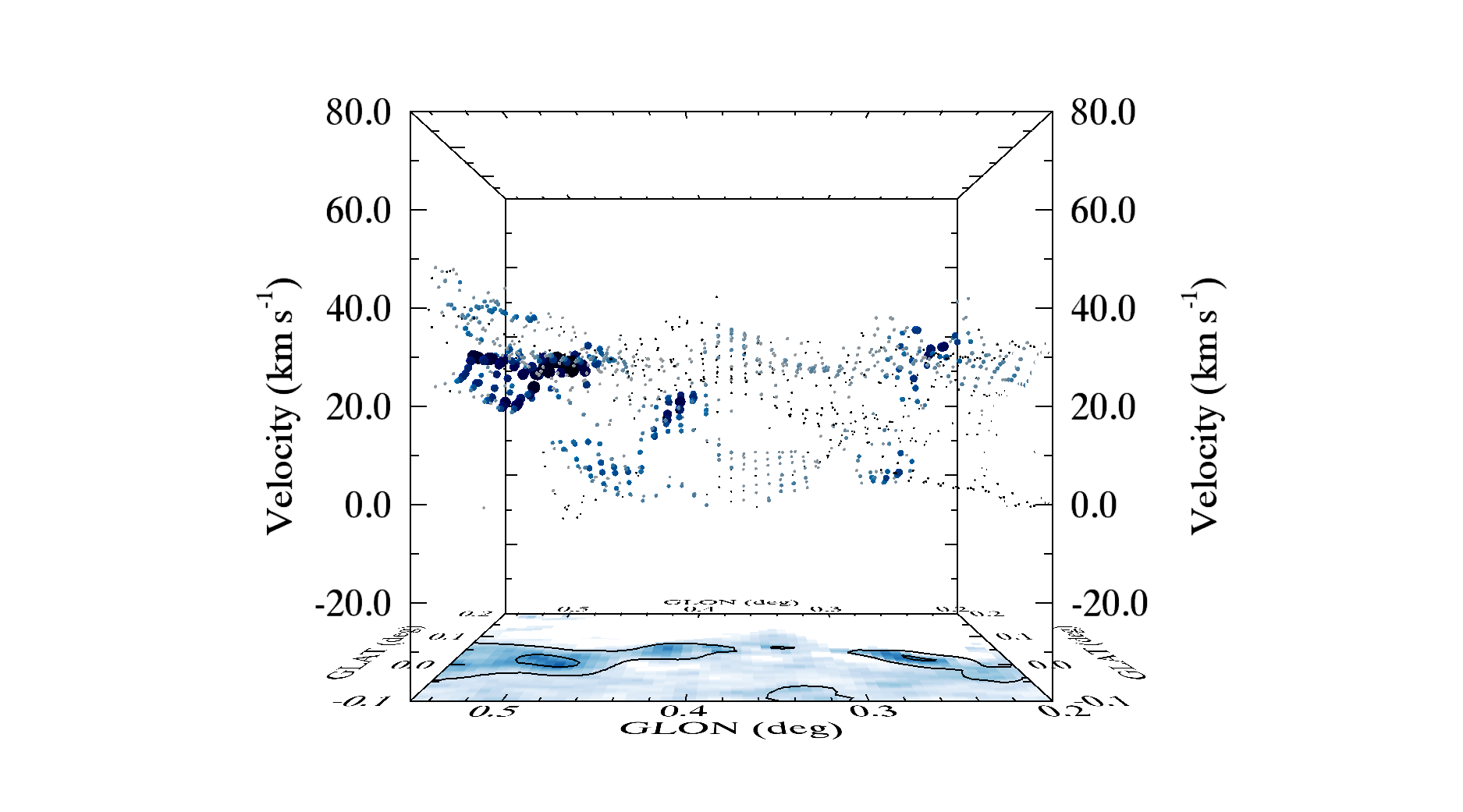}
	\includegraphics[trim = 5mm 5mm 45mm 30mm, clip, width = 0.45\textwidth]{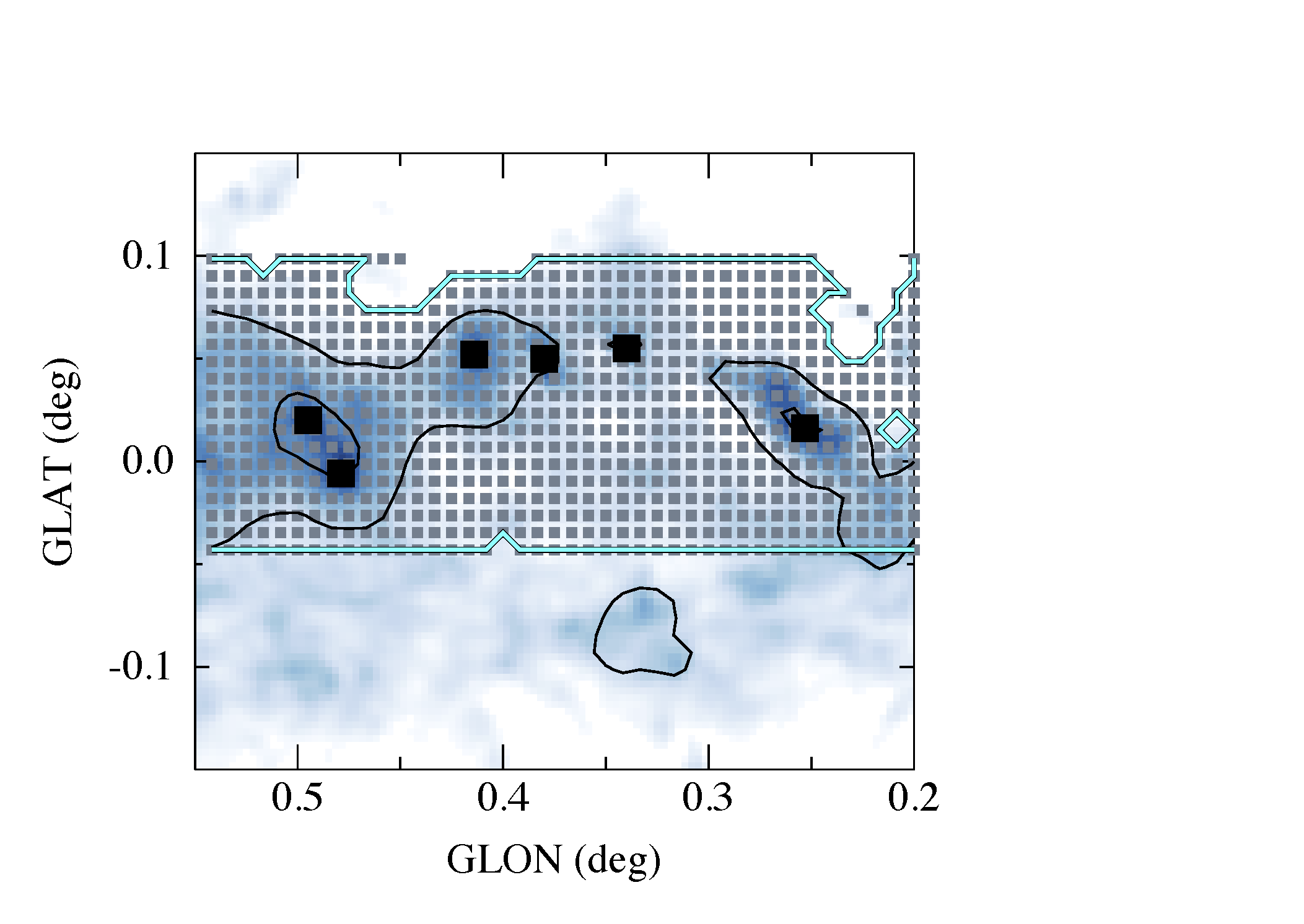}
\end{center}
\caption{A PPV diagram of the dust ridge \citep{lis_1994,lis_2001}. The data points shown are restricted in Galactic latitude between $-0\fdg05<b<0\fdg10$ and in velocity between $-5.0\,{\rm km\,s^{-1}}<v_{\rm LSR}<50.0\,{\rm km\,s^{-1}}$ to focus on emission from the dust ridge only (see \S~\ref{Section:results_dustridge} for more details). The data points have equivalent meaning to those presented in Figure~\ref{Figure:PPV_streams}. Bottom panel: Highlighting the distribution of data points in the top panel. Black squares in the bottom panel indicate the location of (in the direction of increasing Galactic longitude) G0.253+0.016 and clouds B-F. The background image and contours in both panels are equivalent to those in Figure~\ref{Figure:PPV_all}.}
\label{Figure:PPV_dustridge}
\end{figure}

The top panel of Figure~\ref{Figure:PPV_dustridge} is a PPV diagram of the molecular line emission associated with the dust ridge \citep{lis_1994}. Data points are shown over a restricted latitude and velocity range ($-0\fdg05<b<0\fdg10$ and $-5\,{\rm km\,s^{-1}}<v_{\rm LSR}<50\,{\rm km\,s^{-1}}$, respectively) in order to focus solely on emission from the dust ridge molecular clouds. The bottom panel of Figure~\ref{Figure:PPV_dustridge} highlights the corresponding spatial distribution of these data points. The black squares denote the location of some of the highest mass ($M\sim10^{4}-10^{5}$\,\solar) and most compact (radii, $R\sim$\,a few pc) molecular clouds within the Milky Way \citep{lis_2001, immer_2012, longmore_2013b, walker_2015}. 

Spectra exhibiting multiple velocity components are prominent throughout the dust ridge (see Figure~\ref{Figure:ncomp}). Within each dust ridge cloud there is a spread of over $>30$\,\kms \ in centroid velocity (see Table~\ref{Table:positions}). This is reflected in Figure~\ref{Figure:PPV_dustridge}, where we find that although complex, the dust ridge appears to be split into two main features that are almost parallel in velocity (centred on $\sim5$\,\kms \ and 25\,\kms, respectively). 

At the location of G0.253+0.016, and in the immediate surrounding region, we detect four different velocity components ($v_{\rm 0, 1}$\,$\sim$\,7\,\kms, $v_{\rm 0, 2}$\,$\sim$\,21\,\kms, $v_{\rm 0, 3}$\,$\sim$\,38\,\kms, $v_{\rm 0, 4}$\,$\sim$\,70\,\kms; not all of which are observed simultaneously). This is consistent with the observed velocity structure of G0.253+0.016 (e.g. \citealp{lis_1998, kauffmann_2013, bally_2014, johnston_2014, rathborne_2014, rathborne_2015, kruijssen_2015, mills_2015}). 

It has been suggested that cloud-cloud collisions may be responsible for the disturbed velocity structure and prevalence of shocked gas tracers (traced via SiO; e.g. \citealp{kauffmann_2013}) throughout G0.253+0.016 (e.g. \citealp{lis_1998, higuchi_2014, johnston_2014}). Considering each velocity component in the global context of the CMZ leads us to conclude that the prominent component observed at $\sim$\,70\,\kms \ is unlikely to be associated with G0.253+0.016 (hence the choice of velocity range in Figure~\ref{Figure:PPV_dustridge}). Instead, emission from the $\sim$\,70\,\kms \ component is morphologically different to G0.253+0.016, and can instead be linked to the extended high-velocity feature discussed in \S~\ref{Section:results_continuous}. Although the right-hand panels of Figure~\ref{Figure:PPV_streams} show that this feature spatially overlaps with the southern portion of G0.253+0.016, the two features are predicted to be physically separated along the line-of-sight in several different interpretations for the 3-D structure of the CMZ (see \S~\ref{Section:results_models}). This implies that any interaction between the two is unlikely. This does not however, rule out any interaction between the other cloud components associated with G0.253+0.016.

An alternative explanation for the disturbed kinematics may be due to the location of the dust ridge in its orbit around the Galactic centre. In the orbital solution of \cite{kruijssen_2015}, G0.253+0.016 has most recently traversed pericentre, the closest approach to the bottom of the Galactic gravitational potential ($\Delta t_{\rm p, last}$\,=\,0.30$^{+0.30}_{-0.03}$\,Myr, where $\Delta t_{\rm p, last}$ is the time elapsed since last pericentre passage). The tidal interaction between the cloud and the Galactic gravitational potential during pericentre passage may produce complex line-of-sight velocity structures due to the combination of compressive tidal forces, geometric convergence, and shear (Kruijssen et al.~in prep.).

\subsubsection{The Sagittarius B2 molecular cloud complex}\label{Section:results_sgrb2}

The Sgr B2 molecular cloud complex is evident most prominently between $0\fdg50<l<0\fdg85$ and $-0\fdg15<b<0\fdg10$ in position, and between $10.0\,{\rm km\,s^{-1}}<v_{\rm LSR}<70\,{\rm km\,s^{-1}}$ in velocity (see Figure~\ref{Figure:PPV_all}). Figure~\ref{Figure:ncomp} shows that spectral profiles over this spatial area are often best represented by two or three (and in some cases, four) independent velocity components. On average, \scouse \ finds $N_{\rm comp}/N_{\rm fit}\sim2$ within this region, which is greater than the CMZ as a whole ($\sim1.6$; see Table\,\ref{Table:global_stats}). This is an immediate indication of the complexity in the line-of-sight velocity structure. 

\begin{figure*}
\begin{center}
 \includegraphics[trim = 45mm 18mm 30mm 20mm, clip, width = 0.50\textwidth]{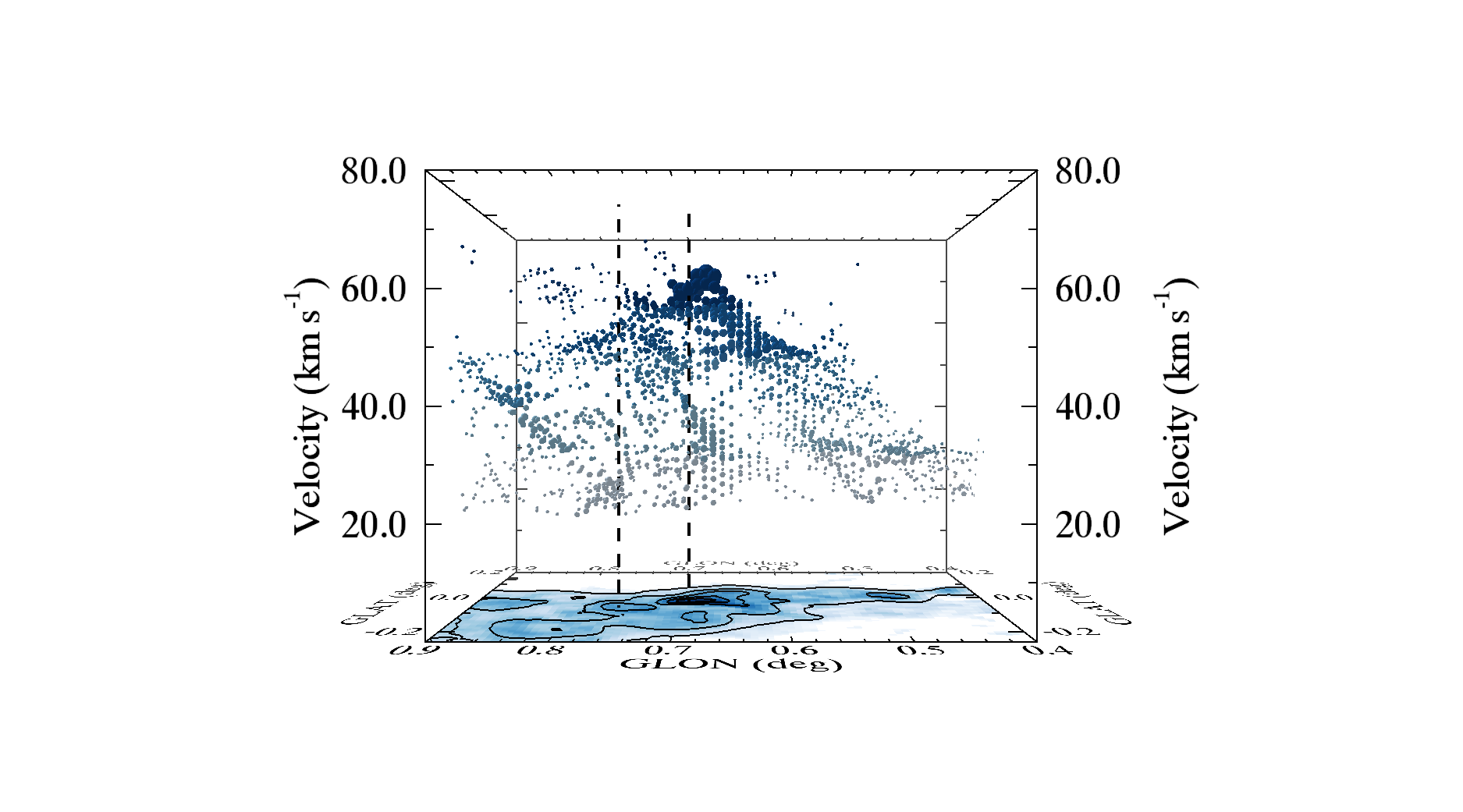}
  \includegraphics[trim = 0mm 5mm 0mm 10mm, clip, width = 0.45\textwidth]{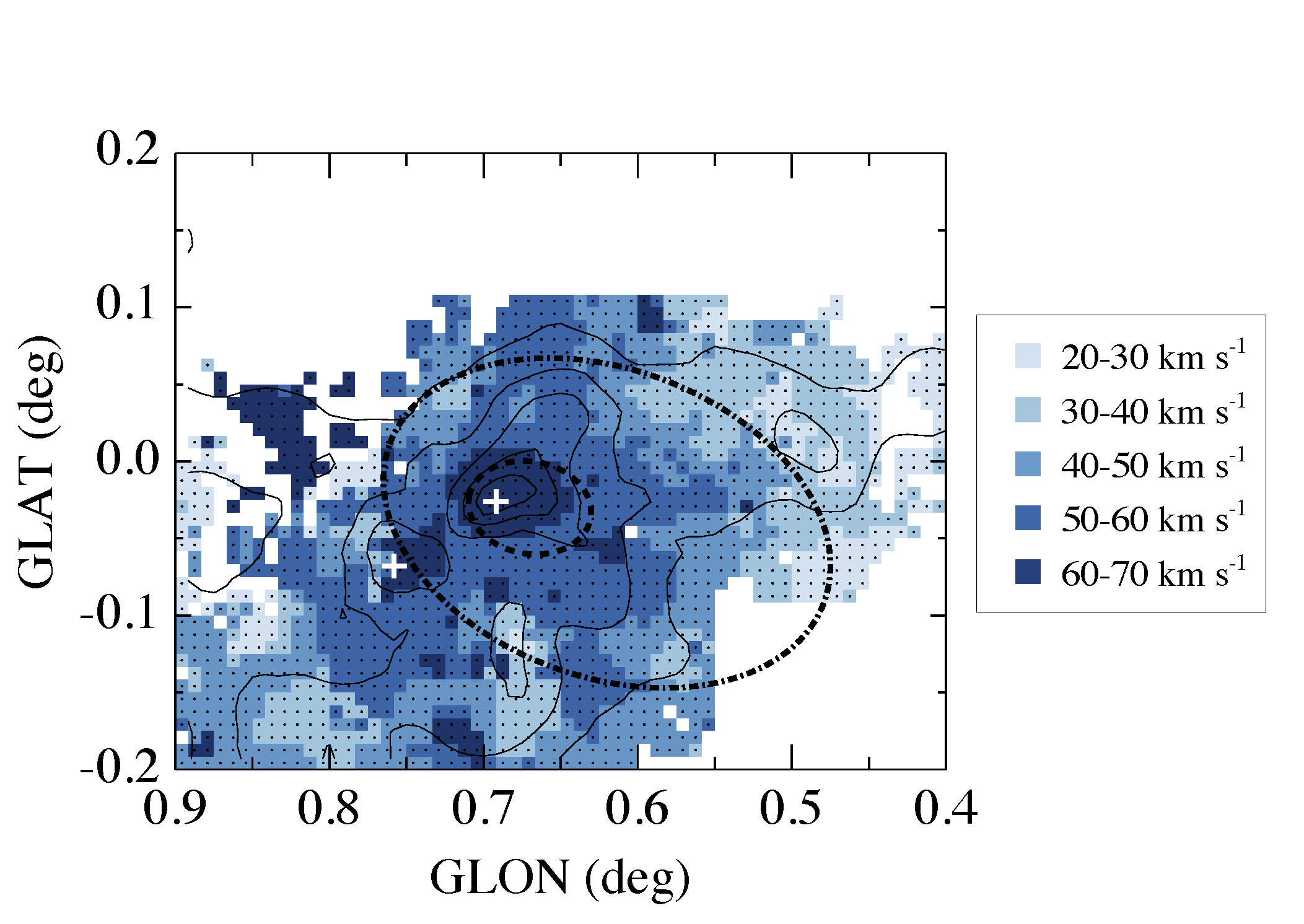}
\end{center}
\caption{Left: A PPV diagram of Sgr B2. Only the HNCO solutions are shown to avoid confusion caused by the self-absorption observed in the \ntwoh \ and HNC spectral line profiles (see \S~\ref{Section:method_cmz}). The data points are colour-coded according to five different velocity ranges (note the key on the right-hand side of the figure): 20-30\,\kms; 30-40\,\kms; 40-50\,\kms; 50-60\,\kms; 60-70\,\kms. As with Figure~\ref{Figure:PPV_all}, the size of each data point is proportional to the peak intensity of the corresponding spectral component. The vertical dashed lines are equivalent to those in Figure~\ref{Figure:PPV_all} and refer to positions P1 and P2 (the corresponding spectra of which are shown in Figure~\ref{Figure:spectra}). Right: A map showing the spatial distribution of the data points shown in the left panel. The black contours are equivalent to those presented in Figure~\ref{Figure:ncomp}. The colours are equivalent to those in the left panel. The black dot-dashed ellipse refers to the location of the ring of emission identified by \citet{bally_1987} between $20-40$\,\kms. The black dashed ellipse highlights the location of the ``hole'' in \tco \ (1-0) emission discussed by \citet{hasegawa_1994}. These images reveal the ``cone-like'' PPV-structure of Sgr B2 discussed in \S\,\ref{Section:results_sgrb2}. This illustrates how a conical PPV-structure can lead to the perception of shells (and corresponding holes) in velocity-integrated emission maps. }
\label{Figure:PPV_sgrb2}
\end{figure*}

The centroid velocity and velocity dispersion of the HNCO data at the location of Sgr B2 main ($\{l,\,b\}=\{0\fdg667,-0\fdg036\}$) are $v_{0}=64.30\pm0.04\,{\rm km\,s^{-1}}$ and $\sigma=10.91\pm0.04\,{\rm km\,s^{-1}}$, respectively. Within the surrounding region, velocity dispersions can be factors of $\sim$a few greater (see Table~\ref{Table:positions}). 

Figure~\ref{Figure:PPV_sgrb2} provides a close up view of the $\{l,\,b,\,v_{\rm LSR}\}$ structure of Sgr B2. In this image we only include the line-centroid velocities of the HNCO spectral features. This is because the \ntwoh \ and HNC line profiles suffer from self-absorption towards the densest region of Sgr B2, and are therefore poor tracers of the line-of-sight velocity (see Figure~\ref{Figure:spectra}). The velocity structure is remarkable. It increases from $\sim20$\,\kms \ at the outer edges of Sgr B2 to $\sim65$\,\kms \ at the column density peak. This corresponds to a gradient of $\sim~2.3$\,\vel, maintained over a projected distance of $\sim~20$\,pc. 

A recurring feature of Sgr B2 observations is a shell-like structure (and corresponding ``hole'') that is most prominent in emission maps integrated over the velocity range $20\,{\rm km\,s^{-1}}~<~v_{\rm LSR}~<~40\,{\rm km\,s^{-1}}$. This feature was first identified by \citet{bally_1988} (see their Figures~5g and 5h). It is centred on $\{l, b\}~=~\{0\fdg62,~\,-0\fdg04\}$ and spans $\sim45\times30$\,pc in diameter. We illustrate the spatial extent of this feature with a black dot-dashed ellipse in the right panel of Figure~\ref{Figure:PPV_sgrb2}. The radius of the inner cavity of the shell decreases with increasing velocity, and a secondary ``hole'' is evident when integrating over the velocity range $50\,{\rm km\,s^{-1}}~<~v_{\rm LSR}~<~60\,{\rm km\,s^{-1}}$ (this is not discussed explicitly by \citealp{bally_1988}, but can be seen in their Figure~5j) . This feature is centred on $\{l, b\}=\{0\fdg67,\,~-0\fdg03\}$, and is smaller by comparison, with a physical extent of $\sim6\times4$\,pc (in diameter). 

The perceived presence of a shell in the emission profile can be explained with closer inspection of Figure~\ref{Figure:PPV_sgrb2}. In the left-hand panel we display all velocity components identified between $20\,{\rm km\,s^{-1}}<v_{\rm LSR}<70\,{\rm km\,s^{-1}}$, revealing that the PPV-structure of Sgr B2 has a ``cone-like'' appearance. We group all data points into five velocity bins (each spanning 10\,\kms), and colour-code each data point according to the relevant velocity range. The right-hand panel displays the spatial distribution of data points within each velocity bin. A conical PPV-structure explains the shells seen in integrated emission maps of Sgr B2. Qualitatively, integrating emission over the velocity range $20\,{\rm km\,s^{-1}}<v_{\rm LSR}<40\,{\rm km\,s^{-1}}$ would display only the base of the cone, resulting in the appearance of a shell (and corresponding hole, indicated by the large dot-dashed ellipse in the right panel of Figure~\ref{Figure:PPV_sgrb2}). Alternatively, integrating over the velocity range $20\,{\rm km\,s^{-1}}<v_{\rm LSR}<60\,{\rm km\,s^{-1}}$ would display all but the tip of the cone, which again would give the impression of a broad shell (incorporating most of the Sgr B2 emission) and a smaller hole (the small dashed ellipse). 

\citet{hasegawa_1994} observe the small hole (dashed ellipse, right panel, Figure~\ref{Figure:PPV_sgrb2}) over a different velocity range to that presented in the right-hand panel of Figure~\ref{Figure:PPV_sgrb2} ($40-50$\,\kms \ versus our $50-60$\,\kms, respectively). Similarly, their observed emission peak lies between $70-80$\,\kms \ (rather than $60-70$\,\kms \ as in Figure~\ref{Figure:PPV_sgrb2}). \citet{sato_2000}, using the same data as \citet{hasegawa_1994}, acknowledge that \tco \ line profile is affected by self-absorption. This will have a significant effect on the inferred kinematics and is likely to be the source of discrepancy in the line-of-sight velocity. 

While the complex velocity structure of Sgr B2 has led to a commensurate number of interpretations of its origin, \citet{hasegawa_1994} cite the morphological similarity between the small hole and the emission feature as evidence for collision between two physically independent molecular clouds (see also \citealp{mehringer_1993, hunt_1999, sato_2000, lang_2010}). In contrast to \citet{hasegawa_1994} however, we do not observe a discontinuity between the velocity of the hole and that of the emission peak. Instead, we find that the PPV-structure of Sgr B2 is \emph{continuous}, as is evidenced by Figure~\ref{Figure:PPV_sgrb2}. Any HNCO emission between $70\,{\rm km\,s^{-1}}~<~v_{\rm LSR}~<~100\,{\rm km\,s^{-1}}$ is linked to the extended high-velocity feature highlighted by the pink contour in Figure~\ref{Figure:pp_pv}. This is an important distinction to make. In our kinematic analysis, the ``hole''  of \citet{hasegawa_1994} can only be reproduced if we remove all velocity components between $60\,{\rm km\,s^{-1}}~<~v_{\rm LSR}~<~70\,{\rm km\,s^{-1}}$).

\subsection{The three-dimensional structure of the CMZ I: Comparing different geometries with the observed gas kinematics}\label{Section:results_models}

Star and cluster formation within the CMZ is largely confined within a Galactocentric radius of 100~pc and may be closely linked to the orbital dynamics of the gas (e.g. \citealp{molinari_2011, longmore_2013b, kruijssen_2015}). Establishing the true three dimensional structure of the CMZ is therefore an important step in understanding star formation within the Galactic centre. Figures~\ref{Figure:pp_models} and \ref{Figure:pv_models} depict three different interpretations for the 3-D structure of the CMZ as they would appear in projection in $\{l,\,b\}$ and $\{l,\,v_{\rm LSR}\}$ space, respectively. The inset image presented in each $\{l,\,v_{\rm LSR}\}$ panel presents a face-on schematic of each interpretation (rotation is in the clockwise direction in each case). In the following sections, we discuss how the prominent gas features discussed in Section~\ref{Section:results_local} contribute to the overall 3-D structure of the CMZ in each interpretation.  

\subsubsection{Two spiral arms}\label{Section:results_spiralarms}

One interpretation is that the CMZ is dominated by two spiral arms \citep{scoville_1974, sofue_1995, sawada_2004}. \citet{sofue_1995} speculate that the two extended PPV-structures discussed in \S~\ref{Section:results_continuous} and displayed in Figure~\ref{Figure:PPV_streams} represent two spiral arms that combine to create a ring-like structure of radius $\sim120$\,pc. Such a configuration is qualitatively supported by observations of external galaxies whereby either loosely wound or more chaotic spirals have been identified within the circumnuclear starburst rings of galaxies with grand design structure (see e.g. Figure~9 of \citealp{peeples_2006}).

%\footnote{Defined by the presence of a symmetric two-arm dust spiral, at least one of which extends all the way to the unresolved nucleus of the galaxy, and at least one of which is dominant \citep{martini_2003, peeples_2006}.}

\begin{figure*}
\begin{center}
\includegraphics[trim = 30mm 10mm 30mm 20mm, clip, width = 0.98\textwidth]{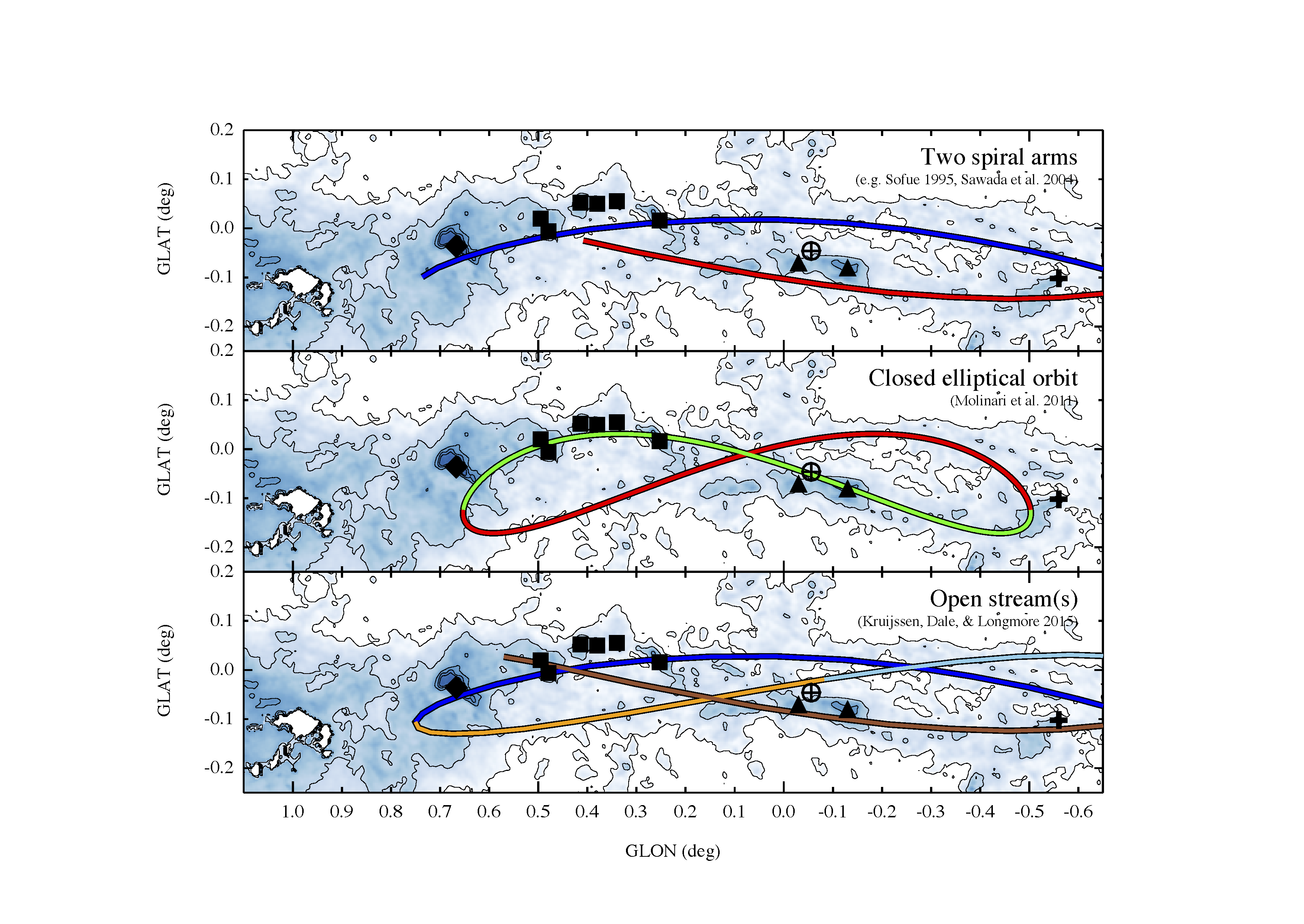}
\end{center}
\caption{The \emph{Herschel}-derived molecular hydrogen column density map of the CMZ (Battersby et al.~in prep.) overlaid with three different interpretations of the 3-D structure of the CMZ (as they would appear in projection on the sky; $\{l,\,b\}$). The top panel depicts the interpretation that the gas and dust distribution of the CMZ is organised into two prominent spiral arms (see \S~\ref{Section:results_spiralarms}). ``Arm\,{\sc i}'' (blue) is situated on the near-side of the Galactic centre, whereas ``Arm\,{\sc ii}'' (red) is situated on the far-side. The two spiral arms are \emph{illustrated} in this figure using Streams~1 (Arm\,{\sc ii}) and 2 (Arm\,{\sc i}) of the \citet{kruijssen_2015} orbital solution (note that the physical interpretation of the \citealp{kruijssen_2015} orbital solution is fundamentally different from the geometry presented by \citealp{sofue_1995}, see \S~\ref{Section:results_openstream}). The latitude of these two streams has been altered such that the arms are approximately point symmetric about Sgr~A*. The central panel depicts the \citet{molinari_2011} orbital geometry. In this interpretation, the gas and dust follow a closed elliptical orbit around Sgr~A* (see \S\,\ref{Section:results_closedellipse}). The green and red portions of the orbit represent the the near and far side (when viewed in projection), respectively. The bottom panel depicts the dynamical orbital solution of  \citet{kruijssen_2015}. In this model, material follows an open eccentric orbit around Sgr~A* (see \S~\ref{Section:results_openstream}). The colours in this final panel refer to those presented by \citet{kruijssen_2015}. Streams~1, 2, 3, and 4 appear in brown, blue, orange, and cyan, respectively. The point-of-interest markers are identical to those presented in the top panel of Figure~\ref{Figure:pp_pv}.}
\label{Figure:pp_models}
\end{figure*}

\citet{sofue_1995} interprets Arm\,{\sc i} (left panels of Figure~\ref{Figure:PPV_streams} and blue in Figures~\ref{Figure:pp_models} and \ref{Figure:pv_models}) as being situated on the near-side of the Galactic centre, further suggesting that it is associated with the dust ridge and Sgr B2. Conversely, Arm\,{\sc ii} (right panels in Figure\,\ref{Figure:PPV_streams} and red in Figures~\ref{Figure:pp_models} and \ref{Figure:pv_models}) is situated on the far-side of the Galactic centre and is associated with Sgr~C. 

This geometry was revisited by \citet{sawada_2004} who estimated the positions of molecular clouds along the line-of-sight by comparing emission and absorption features in CO (1-0) and OH, respectively. \citet{sawada_2004} infer that the CMZ is well represented by an ellipse of physical extent $\sim$\,500\,pc\,$\times$\,200\,pc (see also \citealp{ferriere_2007}). This ellipse is centred on $l\sim0\fdg29$, offset with respect to Sgr~A* by a projected distance of $\sim50$\,pc (with Sgr A* situated at $l\sim-0\fdg056$; \citealp{petrov_2011}), and is inclined by $\sim70\degr$ with respect to the our observed line-of-sight.

Whilst \citet{sofue_1995} make a qualitative comparison between the molecular gas distribution and simulated $\{l,\,v_{\rm LSR}\}$ diagrams produced using a model that assumes spiral arms of gas, \citet{sawada_2004} note that the angular resolution of their observations (10 arcmin; $\sim24$\,pc) is insufficient to resolve the features described in \citet{sofue_1995}. Consequently, the depiction of this configuration in Figures~\ref{Figure:pp_models} and \ref{Figure:pv_models} should serve merely as an \emph{approximation}. The main features of the spiral arm interpretation are instead illustrated by adjusting the locations of streams 1 and 2 of \citet{kruijssen_2015}, which cover a similar region of $\{l,\,b\}$ and $\{l,\,v_{\rm LSR}\}$ space (note however, that the physical nature of the streams differ \emph{fundamentally}, cf. \S~\ref{Section:results_openstream}). The streams are shifted such that they are approximately point symmetric about Sgr~A*. In the face-on schematic presented in the inset image of the top panel of Figure~\ref{Figure:pv_models}, plotted are two spiral arms centred on Sgr~A* (i.e. not at $l\sim0\fdg29$ as in \citealp{sawada_2004} and \citealp{ferriere_2007}), with apocentre radius, $r_{\rm maj}=120$\,pc \citep{sofue_1995}, pericentre radius, $r_{\rm min}=80$\,pc (an eccentric orbit is required to produce non-zero line-of-sight velocities at $l=0\fdg0$), inclined by $70\degr$ with respect to the observed line-of-sight \citep{sawada_2004}, and with a constant pitch angle of $10\degr$. 

\citet{sofue_1995} and \citet{sawada_2004} make several, mutually consistent, predictions for the locations of the more prominent molecular clouds along the line-of-sight. Both studies place Sgr~C on the far-side of the Galactic centre (associated with Arm\,{\sc ii}), and Sgr~B2 on the near-side (associated with Arm\,{\sc i}). Sgr~C and Sgr~B2 are both situated at the lead points of their respective spiral arms (see Figure~10 of \citealp{sofue_1995} and Figure~12 of \citealp{sawada_2004}, respectively). Close inspection of the continuous PPV-structures presented in Figures~\ref{Figure:PPV_streams} and \ref{Figure:pv_models}, leads us to infer that the dust ridge clouds are associated with Arm\,{\sc i} and the 20\,\kms \ and 50\,\kms \ clouds are associated with Arm\,{\sc ii}. This would imply that the 20\,\kms \ and 50\,\kms \ clouds are situated on the far-side of the Galactic centre. This is in disagreement with the observation that both clouds appear as absorption features at 70\,\micron \ (\citealp{molinari_2011}). Although the line-of-sight placement of the 20\,\kms \ and 50\,\kms \ clouds is not discussed explicitly by either \citet{sofue_1995} or \citet{sawada_2004}, this represents an potential inconsistency in the spiral arm interpretation. To be consistent (i.e. with the clouds on the near-side of the Galactic centre and the arm on the far-side), there would have to be no physical connection between the 20\,\kms \ and 50\,\kms \ clouds and the high-velocity gas stream that represents Arm\,{\sc ii}. This seems unlikely given the large-scale coherency and continuity of the gas stream presented in the right panels of Figure\,\ref{Figure:PPV_streams}.

\citet{sawada_2004} suggest that Arm\,{\sc ii} (red, Figure~\ref{Figure:pv_models}) may be an extension of the $1\fdg3$-cloud complex (not included in this work). One may infer from Figure~\ref{Figure:pv_models} that Arm\,{\sc ii} continues between $0\fdg7<l<1\fdg1$ and $50.0\,{\rm km\,s^{-1}}<v_{\rm LSR}<100.0\,{\rm km\,s^{-1}}$ (we have highlighted this material with the pink contour in Figure~\ref{Figure:pp_pv}). This possibility is illustrated with a thin red line in the face-on schematic diagram presented in Figure~\ref{Figure:pv_models}. 

The extension of Arm\,{\sc ii} is in contention with \citet{tsuboi_1999}, who instead infer a connection between Arm\,{\sc i} (blue, Figure~\ref{Figure:pv_models}) and the emission at longitudes $>0\fdg7$. \citet{tsuboi_1999} refer to this as the Galactic centre bow, envisaging a single-arm structure situated on the nearside of the Galactic centre. Differences such as these can arise when identifying structure in either velocity-integrated channel maps or position-integrated PV diagrams. This is particularly important in the CMZ, where broad velocity dispersions can lead to independent features blending in velocity. By comparison, Figures~\ref{Figure:PPV_all} and \ref{Figure:pv_models} plot only the centroid velocity of spectral components identified by \scouse, and are therefore less susceptible to confusion. Because of this, the phase-space separation between the high-velocity gas at $l>0\fdg7$ and that associated with Arm\,{\sc i} is evident, which disfavours the Galactic centre bow interpretation of \citet{tsuboi_1999}.

\subsubsection{A closed elliptical orbit}\label{Section:results_closedellipse}

An alternative explanation is that the CMZ represents a elliptical ring of gas and dust. \citet{binney_1991} propose that the observed gas distribution and non-circular motions within $-10\degr<l<10\degr$ and $-0\fdg5<b<0\fdg5$ may be generated by the presence of a non-axisymmetric potential, or bar. The presence of a galactic bar leads to the formation of different orbital families; those elongated along the bar, `$x_{1}$', and those perpendicular to the bar (embedded within the former), `$x_{2}$' (e.g. \citealp{contopoulos_1977, binney_1991, athanassoula_1992}). \citet{binney_1991} compare large-scale emission from H\,{\sc i} gas in $\{l,\,v_{\rm LSR}\}$ space with a progressive series of $x_{1}$ orbits modelled for a bar inclination of $16\degr$, highlighting the similarity between the two, and infer that clouds within the CMZ move on elliptical $x_{2}$ orbits. 

\begin{figure*}
\begin{center}
\includegraphics[trim = 10mm 5mm -15mm 10mm, clip, width = 0.85\textwidth]{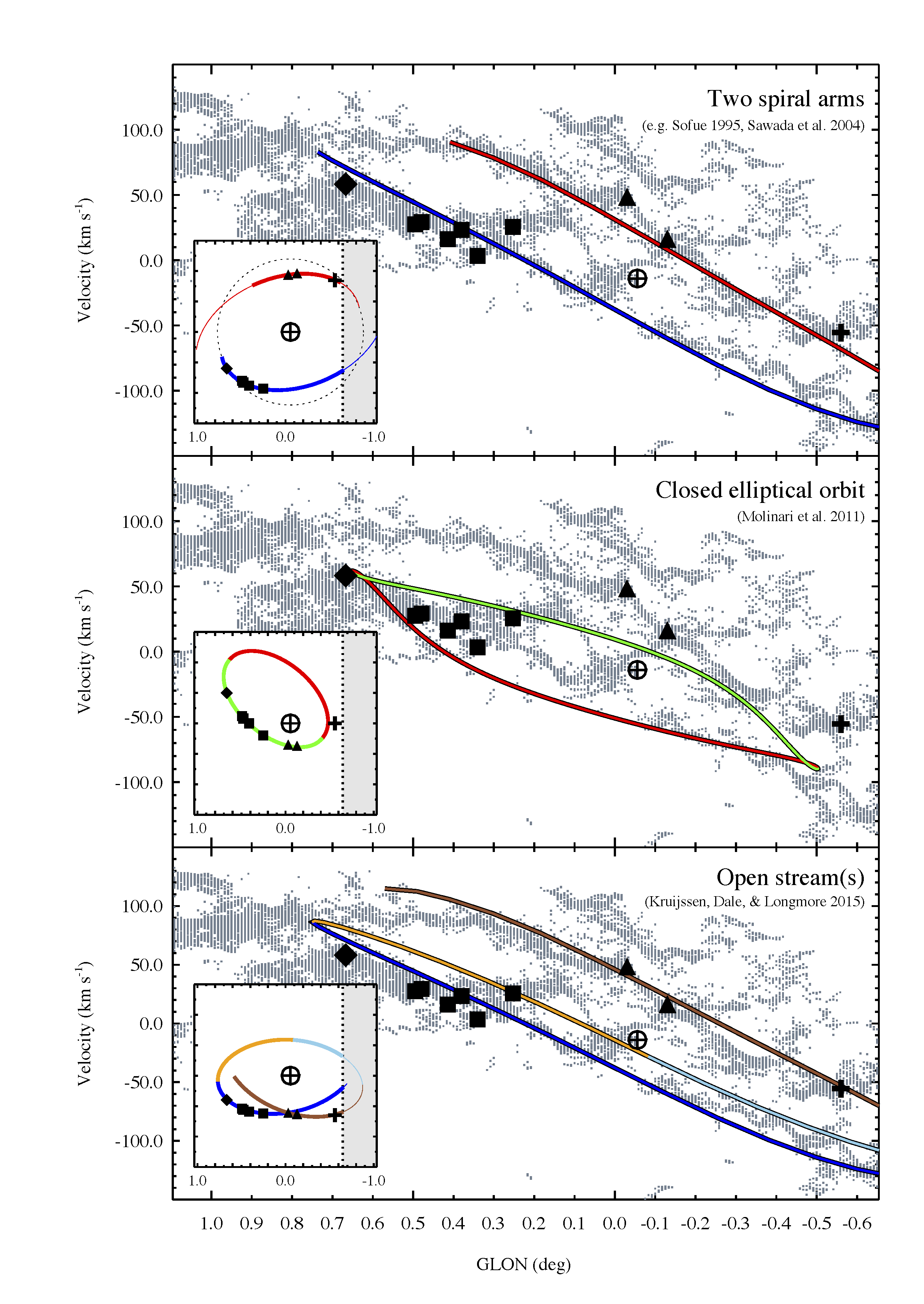}
\end{center}
\caption{Three different interpretations for the 3-D structure of the CMZ as they would appear in $\{l,\,v_{\rm LSR}\}$ space. Within each panel the inset figure represents a schematic of the top-down view of the respective interpretation. The top panel depicts the spiral arm interpretation (see Figure~\ref{Figure:pp_models} and \S~\ref{Section:results_spiralarms}). The line-of-sight velocities of the two arms have been altered such that they are approximately point symmetric about Sgr~A*. Focussing on the schematic diagram, the thick lines refer approximately to those in the $\{l,\,v_{\rm LSR}\}$ diagram. The thin portion of each line represents possible extensions of these features not depicted in either the  $\{l,\,v_{\rm LSR}\}$ diagram, or the $\{l,\,b\}$ diagram in Figure~\ref{Figure:pp_models} (e.g. \citealp{sawada_2004, johnston_2014}). The dotted circle is used to emphasise the deviation from circular geometry. The central panel depicts the \citet{molinari_2011} parameterization (see \S~\ref{Section:results_closedellipse}). The bottom panel displays the \citet{kruijssen_2015} orbital solution (see \S\,\ref{Section:results_openstream}). Streams~1, 2, 3, and 4 are coloured brown, blue, orange, and cyan, respectively. The grey pixels and symbols in each panel are identical to those in the bottom panel of Figure~\ref{Figure:pp_pv}. The $x$ and $y$-axis of each inset figure refers to Galactic longitude and offset position along the line-of-sight (in parsec ranging from $-150$\,pc to $150$\,pc), respectively. The shaded region within each inset figure reflects the longitude limit of the data presented in this paper.}
\label{Figure:pv_models}
\end{figure*}

The central panels of figures~\ref{Figure:pp_models} and \ref{Figure:pv_models} depict the parametric orbital configuration of \citet{molinari_2011}. This geometry was created using dust continuum observations from the \emph{Herschel} Hi-GAL survey and line-of-sight velocities at 20 locations throughout the CMZ extracted from the CS (1-0) data presented by \citet{tsuboi_1999}. In this interpretation, the gas dynamics of the CMZ are attributed to a closed elliptical $x_{2}$ orbit, the vertical oscillation of which reproduces the prominent $\infty$-symbol shaped distribution of dust identified in \emph{Herschel} observations at 250\,\micron \ \citep{molinari_2011}. 

In Figures~\ref{Figure:pp_models} and \ref{Figure:pv_models} the front-side of the orbit is illustrated in green and the rear-side in red. The front side of the orbit contains the dust ridge clouds (G0.253+0.016 and Clouds B-F) and Sgr B2. This is inferred from their enhanced 70\,\micron \ absorption seen in the \emph{Herschel} images. Additionally, the front side of the orbit contains the 20\,\kms \ and 50\,\kms \ clouds. This is in contrast to the spiral arm interpretation discussed in \S~\ref{Section:results_spiralarms}. Sgr B2 and Sgr C are located close to apocentre on the opposite sides of the Galactic centre, where clouds linger longest as they move along the line-of-sight \citep{binney_1991}. 

In the \citet{molinari_2011} parameterization, the gas and dust distribution from Sgr C to Sgr B2 (including the 20\,\kms, 50\,\kms, and dust ridge clouds) is continuous, resulting in the $\infty$-symbol morphology (see central panel of Figure~\ref{Figure:pp_models}). However, while their geometry suitably describes projected images of the CMZ, \citet{kruijssen_2015} identified a significant discontinuity in the $\{l,\,v_{\rm LSR}\}$ distribution of molecular gas thought to be associated with this portion of the \citet{molinari_2011} orbit. This leads \citet{kruijssen_2015} to conclude that the gas and dust between the 20\,\kms \ and 50\,\kms \ clouds and the dust ridge cannot be physically connected. 

While Figure~\ref{Figure:pv_models} supports the conclusion that the dust ridge is not connected to the gas stream containing the 20\,\kms \ and 50\,\kms \ clouds, our new analysis method allows us to follow the coherence of the latter gas stream to higher longitudes. As such, our results go further than the notion of a ``discontinuity'' as described by \citet{kruijssen_2015} (see their Figure~2, bottom panel). Instead, we find that the observed molecular gas distribution is better described by two parallel (in $\{l,\,v_{\rm LSR}\}$ space) features that do not physically connect Sgr C and Sgr B2 in the \emph{specific} way described by the \citet{molinari_2011} orbital configuration. Moreover, Figure~\ref{Figure:pv_models} shows that the spectral components associated with the dust ridge clouds and those associated with the 20\,\kms \ and 50\,\kms \ clouds are largely offset in velocity from the \citet{molinari_2011} orbit (even when incorporating the range of velocities presented in Table\,\ref{Table:positions}). In fact, the intensity-weighted velocities of Clouds B--F are more consistent with the $\{l,\,v_{\rm LSR}\}$ coordinates of the far-side of the orbit (although they do not match in latitude). These discrepancies lead us to conclude that, the parameterization presented by \citet{molinari_2011} is unable to successfully describe the observed molecular gas distribution of the CMZ. 

There is another potential caveat which applies more generally to the concept of a closed elliptical orbit. \citet{kruijssen_2015} suggest that at longitudes close to $l\sim0\fdg25$, there exists three (extended and continuous) PPV-structures or streams. The left and right panels of Figure~\ref{Figure:PPV_streams} show that we clearly identify two of these streams. The third is more difficult to distinguish unambiguously because, although it is separated in latitude from the extended low-velocity PPV structure (see left panels of Figure~\ref{Figure:PPV_streams}), it is almost indistinguishable in velocity (see the blue and orange streams in the bottom panels of Figures~\ref{Figure:pp_models} and \ref{Figure:pv_models}). For any given longitude, a closed elliptical orbit describing the distribution of CMZ material should only comprise two principal components, i.e. the near- and far-side. Therefore, the simultaneous presence of three independent gas streams over a range of longitudes, if confirmed (either via high angular resolution observations or proper motion measurements), cannot be explained by an elliptical orbit alone. 

\subsubsection{Open stream(s)}\label{Section:results_openstream}

\citet{kruijssen_2015} have used \amm \ (1,1) emission-line observations from the H$_{2}$O southern Galactic Plane Survey (HOPS; \citealp{walsh_2011,purcell_2012}) in conjunction with a modified gravitational potential (implied by the stellar mass distribution from \citealp{launhardt_2002}) to develop a dynamical model describing the gas dynamics of the CMZ. The bottom-panels of Figures~\ref{Figure:pp_models} and \ref{Figure:pv_models} depict their best-fitting orbit in $\{l,\,b\}$ and $\{l,\,v_{\rm LSR}\}$ space, respectively. 

There are several key differences between the dynamical model of \citet{kruijssen_2015} and the parametric geometry of \citet{molinari_2011}. Firstly, the orbit is open, as would be the case for an extended mass distribution with an axisymmetric gravitational potential. Although the presence of the Galactic bar results in a non-axisymmetric potential on $\sim{\rm kpc}$ scales, there is no evidence for such asymmetries on scales equivalent to the CMZ gas stream. As a consequence, \citet{kruijssen_2015} state that the orbit is likely to be open, with vertical oscillations ensuring that gas structures survive for several orbital periods without disruption due to self-interaction. Secondly, the orbital velocity is variable, which is consistent with the eccentricity of the orbit. Finally, Sgr A* is not displaced towards the near-side of the stream. 

The dynamical solution of \citet{kruijssen_2015} is described using a continuous orbit split into four segments or ``streams''. The inset image in the bottom panel of Figure~\ref{Figure:pv_models} depicts the configuration of the streams as seen from above the Galactic plane. In contrast to the spiral arm geometry (\S~\ref{Section:results_spiralarms}), the 20\,\kms \ and 50\,\kms \ are situated on the near-side of the Galactic centre (as they are in the \citealp{molinari_2011} interpretation), and associated with Stream~1 (brown; Figure~\ref{Figure:pp_models}). The dust ridge clouds are associated with Stream~2 (blue) and are also on the near-side. Due to the difference in the line-of-sight velocity of these two streams, which disfavours any immediate continuity between the two (see \S~\ref{Section:results_closedellipse}), the resultant orbit follows in the order 2-3-4-1 (blue-orange-cyan-brown). Stream~2 begins upstream, Streams~3 and 4 are on the far-side of the Galactic centre, and Stream~1 represents a downstream `tail'. 

\begin{figure*}
\begin{center}
\includegraphics[trim = 30mm 5mm 30mm 5mm, clip, width = 0.48\textwidth]{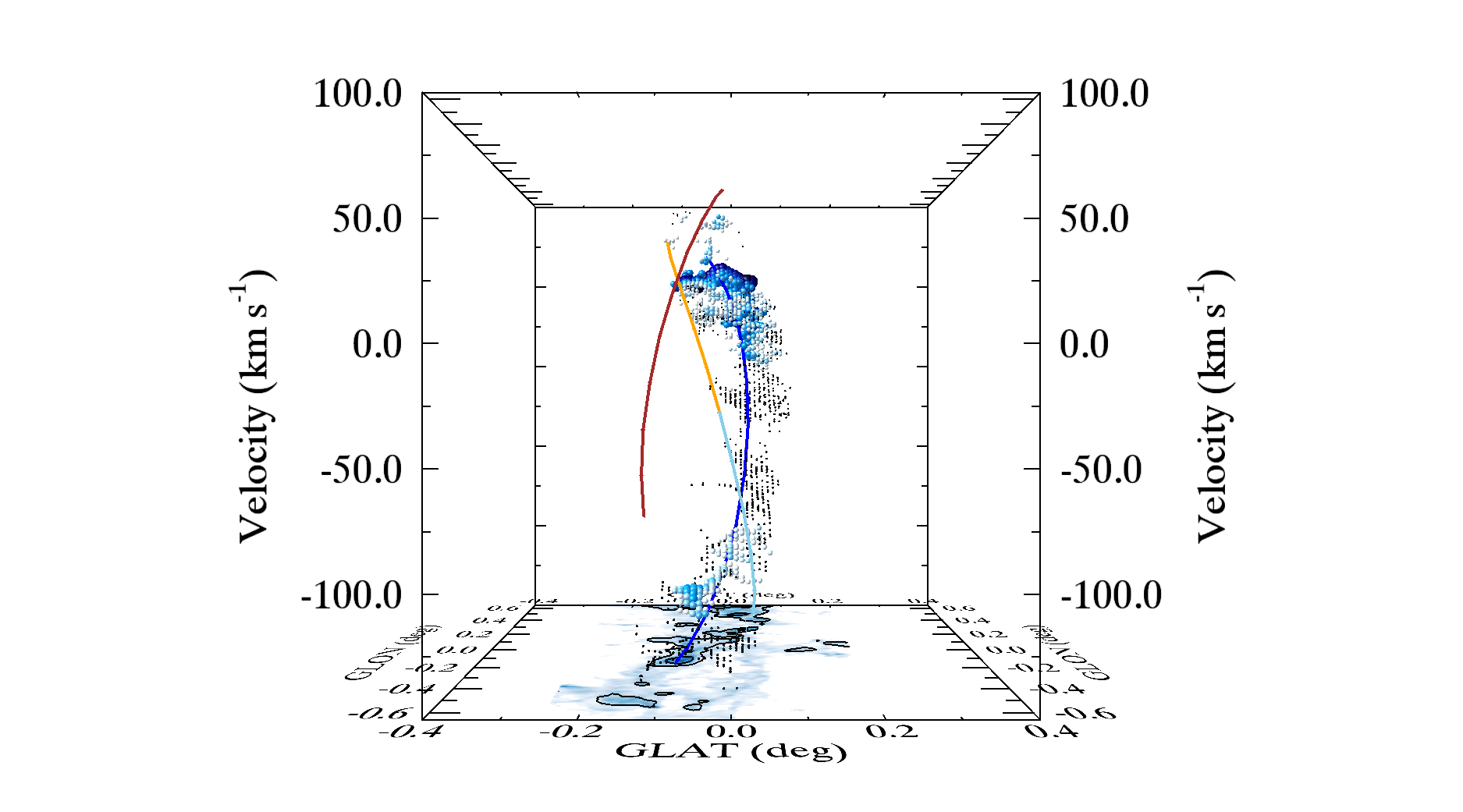}
\includegraphics[trim = 30mm 5mm 30mm 5mm, clip, width = 0.48\textwidth]{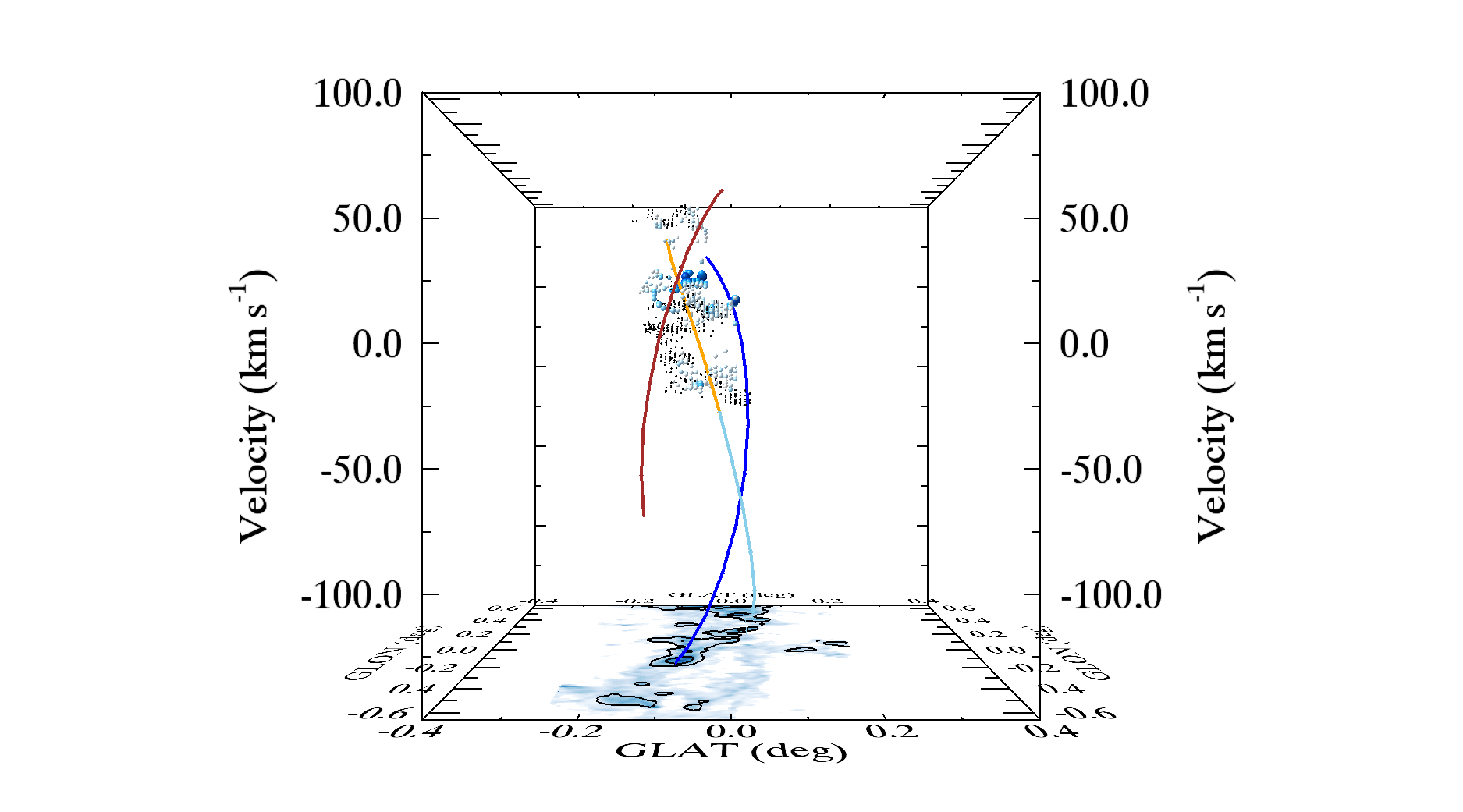}
\includegraphics[trim = 30mm 5mm 30mm 5mm, clip, width = 0.48\textwidth]{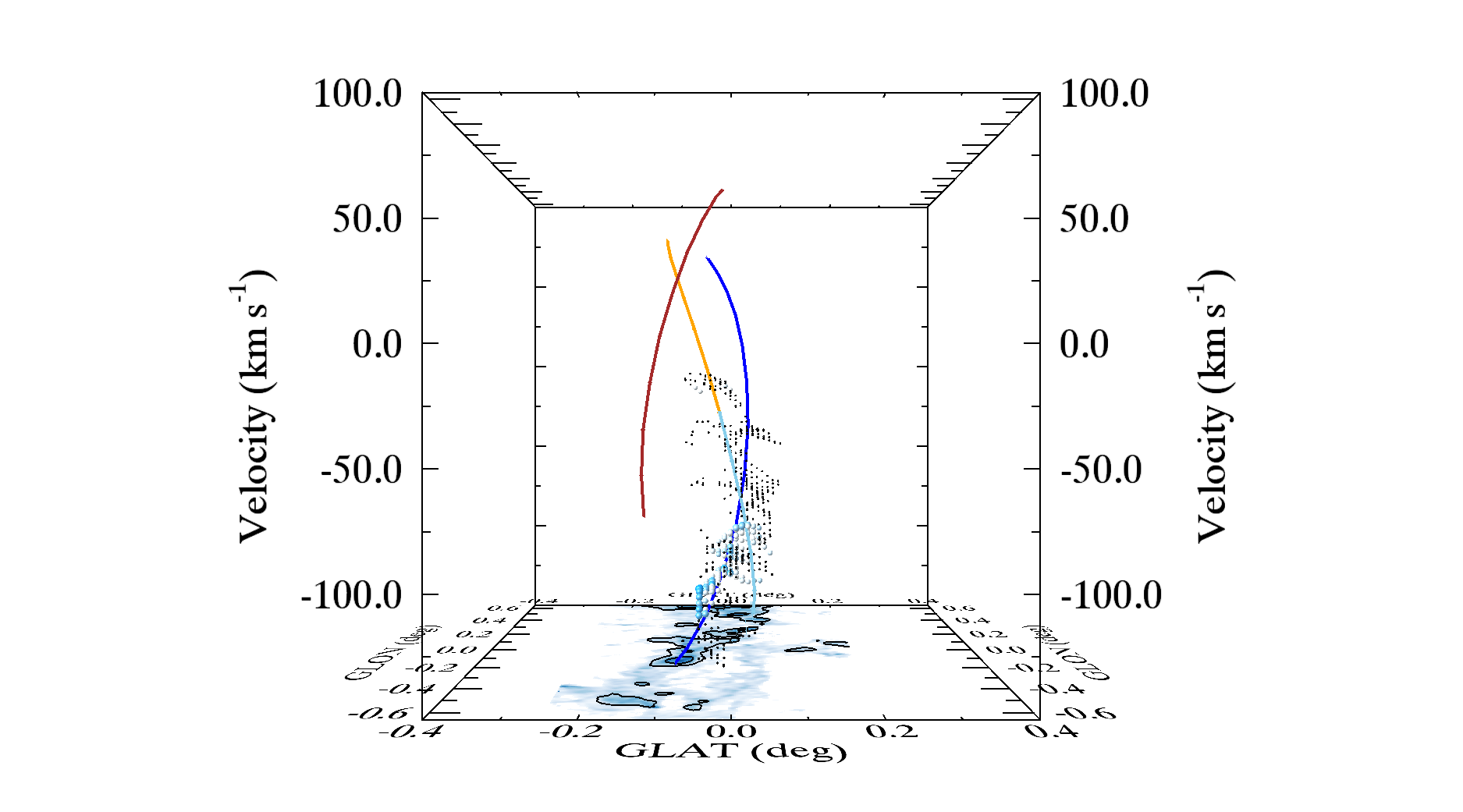}
\includegraphics[trim = 30mm 5mm 30mm 5mm, clip, width = 0.48\textwidth]{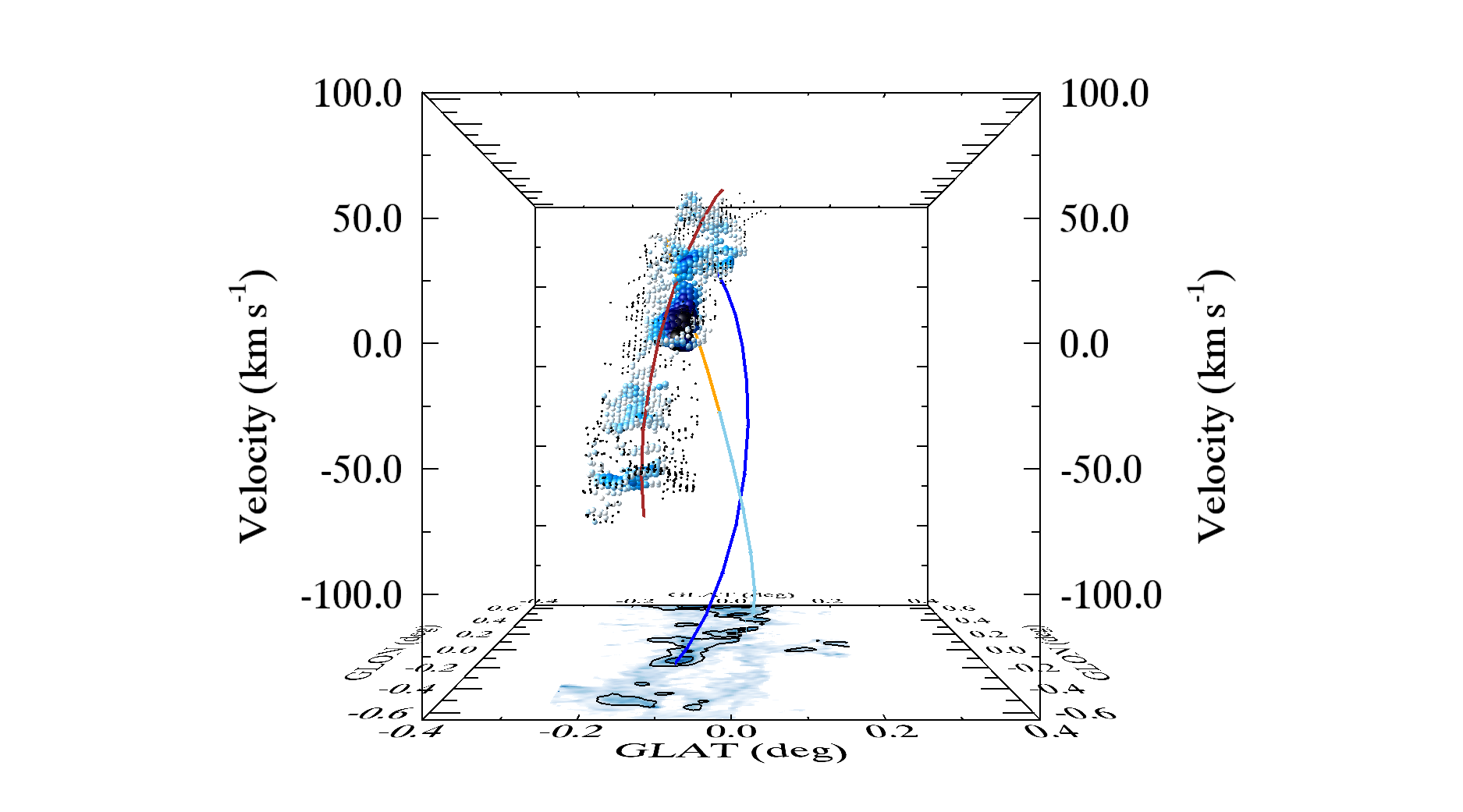}
\end{center}
\caption{PPV diagrams shown in the $\{b,\,v_{\rm LSR}\}$ projection. Each sub-figure highlights the velocity components identified within 10\,pc and 20\,\kms \ of the four streams with describe the \citet{kruijssen_2015} orbital solution. Clockwise from top-left, following the order of the orbital solution (cf. the inset image of Figure\,\ref{Figure:pv_models}), the streams displayed are: Stream~2 (blue); Stream~3 (orange); Stream~4 (cyan); Stream~1 (brown). This figure highlights that the association of velocity components with the streams can in some cases be degenerate (as discussed in \S~\ref{Section:results_openstream} and by \citealp{kruijssen_2015}). The pixel sizes and colours are equivalent to those in Figure~\ref{Figure:PPV_streams}. The image at the base of each plot and the contours are equivalent to those presented in Figure~\ref{Figure:PPV_all}.}
\label{Figure:PPV_bv}
\end{figure*}

Although the placement of the dust ridge clouds on the near-side of the Galactic centre is consistent with the spiral arm geometry presented in \S~\ref{Section:results_spiralarms}, a fundamental difference between the two involves the nature of the low-velocity feature presented in the left panels of Figure~\ref{Figure:PPV_streams}. Whereas \citet{sofue_1995} and \citet{sawada_2004} (also \citealp{bally_1988, tsuboi_1999}) interpret this as a continuous structure, in the \citet{kruijssen_2015} model it comprises both Streams~2 (blue) and 4 (cyan). This is an important distinction to make since Stream~2 resides on the near-side of the Galactic centre whereas Stream~4 is situated on the far-side, and the two are \emph{not} contiguous. In this sense, the $90$\,pc gap in HNCO emission (see \S~\ref{Section:results_continuous}) may be due to the varying physical conditions at different locations in the orbit.

In Figure~\ref{Figure:PPV_bv} we investigate this further using the kinematic information extracted using \scouse. Each panel of Figure~\ref{Figure:PPV_bv} is a PPV-diagram shown in the $\{b,\,v_{\rm LSR}\}$ projection, focusing on a different stream (clockwise from top-left the order displayed is 2-3-4-1 or blue-orange-cyan-brown). The velocity components shown are those which lie within 10\,pc (in both $l$ and $b$) and 20\,\kms \ from the respective stream. 

The top- and bottom-left hand panels focus on Streams~2 (blue) and 4 (cyan), respectively. At low longitudes (those in the foreground), the data points cannot be linked to either stream unambiguously. Similarly, there is a close relationship between Stream~2 (blue) and Stream~3 (orange; top-right panel) at positive longitudes (those in the background). This arises due to the small separation in velocity between the respective streams (see the bottom panel of Figure~\ref{Figure:pv_models}). The latter ambiguity however, is partially alleviated by the difference in latitude between the two streams (with Stream~2 situated at greater latitudes; see bottom panels of Figure~\ref{Figure:pp_models}).

\citet{kruijssen_2015} discuss the degeneracy between Stream~2 (blue) and Stream~4 (cyan) explicitly, and note that the best-fitting orbital solution is unaffected, as the gas fits both streams. However, it is worth stating that both of the ambiguities discussed above are related to the far-side of the gas stream. This may indicate that gas associated with the streams on the far-side may be more tenuous than the front, possibly signifying a near-far side asymmetry in the gas distribution in addition to the well-documented longitudinal asymmetry of the CMZ (see \S~\ref{Section:results_global}). While the \citet{kruijssen_2015} model allows for this interpretation, establishing proper motions of the negative-longitude, high-latitude gas clouds would remove any residual ambiguity (see \S~\ref{Section:results_obstests}).

\subsection{The three-dimensional structure of the CMZ II: Future work, confronting theory with observations}\label{Section:results_obstests}

Establishing the true 3-D structure of the CMZ is complicated by our edge-on view through the Galactic plane. To fully understand the nature of the Galactic centre gas stream, we require more complete information (ideally in 6-D space; 3-D position and 3-D motion). Fundamental differences between the orbital configurations presented in \S~\ref{Section:results_models} make it possible to distinguish between them observationally. This can be achieved by measuring proper motions of masers and combining them with their projected locations and line-of-sight velocities. 

One such difference regards the direction of motion of the 20\,\kms \ and 50\,\kms \ clouds in the plane of the sky. The spiral arm interpretation predicts proper motions in the direction of decreasing Galactic longitudes  (see Arm\,{\sc ii}; red in Figure~\ref{Figure:pp_models}). This is in contrast to both the closed elliptical orbit of \citet{molinari_2011} and the open stream of \citet{kruijssen_2015}, which predict proper motions in the direction of increasing Galactic longitude.

Another key determinant is the proper motion of Sgr C. The \citet{molinari_2011} interpretation predicts that Sgr C is very close to apocentre. This suggests that the motion would largely be along the line-of-sight. In contrast however, both the spiral arm interpretation and the open stream predict motion in the plane of the sky, but in \emph{opposite directions}. In the case of the spiral arm interpretation, the motion in the plane of the sky would be directed towards decreasing Galactic longitudes whereas the open stream predicts the opposite.

As discussed in \S~\ref{Section:results_openstream}, establishing proper motions of the negative longitude, high-latitude gas clouds would also remove the current ambiguity regarding the nature of the gas stream presented in the left-panels of Figure~\ref{Figure:PPV_streams}. The orbital solution of \citet{kruijssen_2015} predicts that the proper motions of these clouds are directed towards negative longitudes (associated with Stream~4), opposite to those of the dust ridge (associated with Stream\,2). If confirmed, this would demonstrate that the emission presented in the left panels of Figure~\ref{Figure:PPV_streams}, in fact, comprises to two physically independent gas streams. This is in contrast to the spiral arm interpretation, which predicts proper motions in the direction of increasing longitudes throughout the entirety of this singular and continuous feature.

So far, the only published maser proper motion values in the Galactic centre have been determined for the water masers in Sgr B2 (${\mu_{l}, \mu_{b}} = {2.3 \pm 1.0, -1.4 \pm 1.0}$\,mas\,yr$^{-1}$; \citealp{reid_2009}). These values are consistent with the predictions of \citet{kruijssen_2015} while the \citet{molinari_2011} geometry predicts a too small velocity in the plane of the sky for Sgr B2 \citep{kruijssen_2015}. CH$_{3}$OH and H$_{2}$O masers, best suited for proper motion measurements of star-forming regions in molecular clouds, have been detected in Sgr B2, Sgr C, the 20\,\kms \ cloud, south of G0.253+0.016, and in the dust ridge clouds C and E (e.g. \citealp{caswell_2010, walsh_2014}; Lu et al., in prep.). Proper motion measurements of these masers with the VLBI Exploration of Radio Astrometry (VERA) interferometer and the Very Long Baseline Array (VLBA) are currently underway.

A final avenue for future work concerns the gas at $l>~0\fdg7$ and $v_{\rm LSR}>50$\,\kms, depicted by the pink contour in Figure~\ref{Figure:pp_pv}. This gas appears to connect smoothly to the $1\fdg3$ cloud complex in $\{l,\,b,\,v_{\rm LSR}\}$ space (Longmore et al.~in prep.). The $1\fdg3$ cloud complex contributes significantly to the longitudinal asymmetry in the gas distribution of the CMZ. \citet{rodriguez_2008} suggest that the $1\fdg3$ cloud complex is the main accretion site of material onto the CMZ, and that asymmetric accretion may explain observed gas distribution. An alternative hypothesis is that the longitudinal asymmetry of the CMZ (including the $1\fdg3$ cloud complex) is the result of gravitational instabilities whose length-scale is $\lambda\sim200$\,pc \citep{krumholz_2015}.

A current matter of contention is how exactly the gas at $l>~0\fdg7$ relates to the extended PPV-structures presented in Figure~\ref{Figure:PPV_streams}). In the spiral arm interpretation, Arm\,{\sc ii} represents a direct continuation of the gas at $l>~0\fdg7$ (with plane of the sky motion in the direction of decreasing Galactic longitude). An alternative explanation is that the $1\fdg3$ complex and Bania's Clump 2 are moving along the innermost self-intersecting $x_{1}$ orbit \citep{binney_1991, morris_1996, rodriguez_2006, bally_2010}. Finally, \citet{kruijssen_2015} do not draw any (immediate) physical connection between the gas at $l>0\fdg7$ and their Stream~1 (whose motion in the plane of the sky is directed towards increasing Galactic longitude). This is because for Stream~1 to smoothly connect to the $1\fdg3$ degree cloud complex, it would have to originate from significantly beyond Sgr C (due to energy conservation). 

When viewed in $\{l,\,v_{\rm LSR}\}$ projection, these two gas features appear to be continuous (see e.g. \citealp{johnston_2014}). However, by definition, $\{l,\,v_{\rm LSR}\}$-diagrams lack latitude information. In Figure~\ref{Figure:pp_pv}, we see that the $l>0\fdg7$ gas (pink contour) covers a much larger latitude range than the gas associated with the extended high-velocity PPV-structure (red contour in Figure~\ref{Figure:pp_pv}). The latitude range of the $l>0\fdg7$ gas is covers the full extent of the observations presented here ($\sim60$\,pc). By comparison, the extended PPV-structures cover a latitude range equivalent to $\sim7-15$\,pc (this is also noted in \citealp{rodriguez_2008}). While this distinct change in the scale height of the gas may be due to projection (potentially supporting the spiral arm or closed elliptical orbit interpretations), it may also imply that the two bodies of gas are not contiguous. Instead, the material at $l>0\fdg7$ may be a gas reservoir which resides several orbital revolutions upstream. \citet{krumholz_2015} predict that this gas will undergo continued angular momentum loss driven by acoustic instabilities over a further 5~Myr before entering the 100\,pc gas stream. Establishing the proper motion of gas associated with the high-velocity PPV-structure will remove this ambiguity, and may have wider implications for understanding the asymmetric gas distribution of the CMZ and the role of the $1\fdg3$ cloud complex.

\section{Summary and Conclusions}\label{Section:conclusions}

We have presented \scouse, a line-fitting routine whose purpose is to efficiently analyse large quantities of spectral line data in a systematic way. Using a combination of molecular species taken from the Mopra central molecular zone (CMZ) survey of \citet{jones_2012}, we present a new and detailed description of the molecular gas kinematics within the inner 250\,pc of the Galaxy. Our main findings are summarised below:

\begin{enumerate}

\item We compare the line-of-sight kinematics derived using \scouse \ with those obtained using moment analysis. Where only one velocity component is identified, the results agree well ($\sim98$ per cent of centroid velocity measurements differ by $<6$\,\kms). As the number of identified spectral components increases, significant deviations are observed (only $\sim36$, $17$, and $7$ per cent of centroid velocity measurements agree for 2, 3, and 4 component fits, respectively). Since multiple component fits are required to represent the line profiles of $\sim45$ per cent of the (HNCO) data, we conclude that moment analysis inadequately describes the molecular gas kinematics of the CMZ. \\

\item Consistent with previous works we note that the distribution of gas in the CMZ is asymmetric about Sgr~A* (with a greater proportion of material lying at positive longitudes). We confirm the presence of a large-scale velocity gradient, with a magnitude of $\mathscr{G}\sim0.6$\,\vel \ in a direction $\Theta_{\mathscr{G}}\sim56.9^{\circ}$ east of Galactic north. \\

\item We confirm that gas motions within the CMZ are inherently supersonic. Internal velocity dispersions range from $\sim2.6-53.1$\,\kms, with a median value of 9.8\,\kms. This median velocity dispersion corresponds to a Mach number of $\mathcal{M}_{\rm 3D}\sim28-37$ (assuming a fiducial temperature range for the dense gas of $60-100$\,K). \\

\item The kinematic properties of several sub-regions and molecular clouds are investigated independently. The following describes our main findings:

	\begin{itemize}
		
		\item {\textbf{Continuous PPV-structures within the CMZ:}} The molecular gas distribution is dominated by two extended features which are approximately parallel in $\{l,\,v_{\rm LSR}\}$ space. HNCO emission is more prominent throughout the high-velocity PPV-structure than the low-velocity PPV-structure. It remains an open question as to whether the low-velocity PPV-structure represents a continuous stream (where the 90\,pc gap in HNCO emission reflects varying chemical and physical conditions along the structure), or whether this feature comprises two physically independent gas streams.\\
			
		\item {\textbf{The Sagittarius C molecular cloud complex (and surrounding region):}} We investigate the possibility that supernova explosions are responsible for the production of shell-like features within the Sgr C molecular cloud complex. Careful inspection of the data in $\{l,\,b,\,v_{\rm LSR}\}$ space leads us to conclude that such features may be caused by the projected alignment of several (independent and extended) molecular gas streams situated along the line-of-sight. Consequently we provide a cautionary note against structure identification in velocity-integrated emission maps.  \\
		
		\item {\textbf{The 20\,km\,s$^{-1}$ and 50\,km\,s$^{-1}$ clouds:}} Consistent with other works, we find that the clouds are coherently connected in $\{l,\,b,\,v_{\rm LSR}\}$ space. The separation in projected distance and velocity implies a velocity gradient of $\sim2.3$\,\vel. Our kinematic analysis finds that the bulk of the molecular line emission associated with the 20\,\kms \ and 50\,\kms \ clouds is just a small segment of the extended high-velocity PPV-structure discussed above. This is in agreement with the orbital solution of \citet{kruijssen_2015}, which places the clouds at a Galactocentric radius of $\sim60 {\rm pc}$. \\
		
		\item {\textbf{The dust ridge molecular clouds:}} Each dust ridge cloud spans a range of $>30$\,\kms \ in centroid velocity. Emission associated with the dust ridge is divided into two main features that are almost parallel in velocity (centred on $\sim5$\,\kms \ and $\sim25$\,\kms, respectively). Considering all spatially coincident velocity components in the global context of the CMZ leads us to conclude that a prominent component observed at $\sim70$\,\kms \ is unlikely to be associated with the dust ridge. Instead, this emission can be linked to the extended high-velocity PPV-structure discussed above. \\
		
		\item {\textbf{The Sagittarius B2 molecular cloud complex:}} The kinematics of the Sgr B2 region are complex. The number of velocity components identified per position is $\sim2$, compared to the CMZ average of 1.6. Sgr B2 contains a significant fraction ($\sim45$ per cent) of the broad velocity dispersion gas ($13.3\,{\rm km\,s^{-1}}<\sigma<53.1\,{\rm km\,s^{-1}}$). The kinematic structure is remarkable, increasing from $\sim20$\,\kms \ at the edges to $\sim65$\,\kms \ at the centre, giving the complex a conical appearance in $\{l,\,b,\,v_{\rm LSR}\}$ space. Such a structure would provide an explanation for the presence of shell-like features (and corresponding ``holes'') observed within velocity-integrated emission maps of Sgr B2, which have previously been cited as evidence for cloud-cloud collisions (e.g. \citealp{hasegawa_1994, sato_2000}). \\
		
	\end{itemize}
 
\item The molecular gas kinematics are compared against three interpretations for the 3-D structure of the CMZ. We conclude that:

	\begin{itemize} 

		\item \textbf{Two spiral arms:} The spiral arm interpretation (e.g. \citealp{sofue_1995, sawada_2004}) provides a qualitatively good reproduction of the observed molecular gas distribution. However, this interpretation currently lacks a physical model that relates specifically to the CMZ. Additionally, the appearance of the 20\,\kms \ and 50\,\kms \ clouds as absorption features at 70\,\micron \ \citep{molinari_2011} is difficult to reconcile with the geometry and is a particular area of uncertainty.\\

		\item \textbf{A closed elliptical orbit:} The parameterization of the elliptical orbit presented by \citet{molinari_2011} is unable to reproduce the observed gas distribution in $\{l,\,b,\,v_{\rm LSR}\}$ space. Additionally, there is some evidence to suggest that three independent gas streams are observed simultaneously over a range of Galactic longitudes. If verified, this empirical conclusion cannot be explained by an elliptical orbit alone. \\

		\item \textbf{An open stream:} The open stream describing the dynamical orbital solution of \citet{kruijssen_2015} reproduces the observed $\{l,\,b,\,v_{\rm LSR}\}$ gas distribution well. Additional observations are required to remove ambiguities in the association of gas emission to either the near- or far-side of the gas stream. However, out of the three investigated, an open stream is the only interpretation that does not show discrepancies with the observational data. \\

	\end{itemize}

\item Proper motion measurements of masers will help to observationally distinguish between different orbital configurations. We describe the main differences in the direction of motion predicted by the different interpretations at several locations throughout the orbit. We propose several regions to focus future efforts. These include: the 20\,\kms \ and 50\,\kms \ clouds, Sgr C, the negative longitude/high-latitude clouds, and the $1\fdg3$ cloud complex. 
 
\end{enumerate}

Looking to the future, it is important to understand how the gas in this region couples to the wider Galactic environment. Similar analysis of the observed gas kinematics at larger galactocentric radii, and comparison to the predictions of large-scale numerical simulations (e.g. \citealp{renaud_2013, renaud_2015, emsellem_2015, krumholz_2015, sormani_2015a, sormani_2015b, sormani_2015c, suzuki_2015}) offer a promising way forward in this direction.

\section*{Acknowledgements}
% Entry for the table of contents, for this guide only
\addcontentsline{toc}{section}{Acknowledgements}
We thank the anonymous referee for their comments which have helped to improve the clarity of this paper. JDH would like to thank Paola Caselli and Cormac Purcell for discussions which have helped to improve the implementation of \scouse, and Jens Kauffmann for providing comments which have helped to improve the presentation of the paper. The data were obtained using the Mopra radio telescope, a part of the Australia Telescope National Facility which is funded by the Commonwealth of Australia for operation as a National Facility managed by CSIRO. The University of New South Wales (UNSW) digital filter bank (the UNSW-MOPS) used for the observations with Mopra was provided with support from the Australian Research Council (ARC), UNSW, Sydney and Monash Universities, as well as the CSIRO. JMDK is funded by a Gliese Fellowship.

%%%%%%%%%%%%%%%%%%%%%%%%%%%%%%%%%%%%%%%%%%%%%%%%%%



\bibliographystyle{mnras}
\bibliography{References/references} % if your bibtex file is called example.bib

%%%%%%%%%%%%%%%%%%%%%%%%%%%%%%%%%%%%%%%%%%%%%%%%%%

%%%%%%%%%%%%%%%%% APPENDICES %%%%%%%%%%%%%%%%%%%%%

\appendix
\section{Towards a more complete picture of CMZ gas kinematics}\label{App:merge}

Gas within the CMZ spans a wide range of physical properties. Any single molecular line tracer will therefore only be observable over a finite range of excitation conditions. Consequently, in using only a single molecular line tracer, our view of the line-of-sight velocity structure of the CMZ is limited. In Section~\ref{Section:results_global} we show that although HNCO traces a significant fraction of CMZ gas, there are also significant gaps in the information (see Figure~\ref{Figure:PPV_all}). 

In Section~\ref{Section:results_local}, we expand on this and make use of multiple molecular line transitions in order to develop a more complete picture of the line-of-sight velocity structure of the CMZ. Our selected (additional) molecular lines are \ntwoh \ and HNC. Both of these lines exhibit hyperfine structure. Since the hyperfine structure is blended due to the large line-widths, we can use \scouse \ to extract the kinematics. This allows us to determine the centroid velocity in regions where the gas is optically thin. However, we acknowledge that this influences the measurement of the velocity dispersion. While therefore it is possible to use such measurements to investigate the \emph{relative} change in velocity dispersion throughout the CMZ, using absolute values of the dispersion should be avoided. 

Taking the above into consideration, we use this kinematic information in a qualitative sense, to `fill in the gaps' in the HNCO emission. In spite of this, Figures~\ref{Figure:PPV_streams}--Figure~\ref{Figure:PPV_sgrb2} reveal a remarkable continuity between the line-of-sight velocity structure of the different molecular species. While small scale deviations in the velocity between molecular lines may be present on a pixel-to-pixel level (of the order $\sim$\, a few \kms), owing their origin to differences in a combination of opacity, excitation, and chemistry, these are clearly less important on the scales of interest to this study ($\sim250$\,pc in projected distance, and $\sim300$\,\kms \ in velocity).

In this appendix we describe the method used in order to merge the data files output by \scouse \ in order to create the combined view of the line-of-sight velocity presented in Section\,\ref{Section:results_local}. Initially, \scouse \ is run independently on the different molecular line data cubes. The merging process begins with a control catalogue and a merger catalogue. In the first iteration of this process, the best-fitting solutions to the HNCO emission comprise the control catalogue and the \ntwoh \ solutions make up the merger catalogue. With each subsequent iteration, the control catalogue will contain a composite mixture of the kinematic information merged during the previous iteration. In this way, one can continue to add to the final solution file the kinematic information from several different molecular lines.

\begin{figure}
\begin{center}
\includegraphics[trim = 60mm 50mm 40mm 50mm, clip, width = 0.48\textwidth]{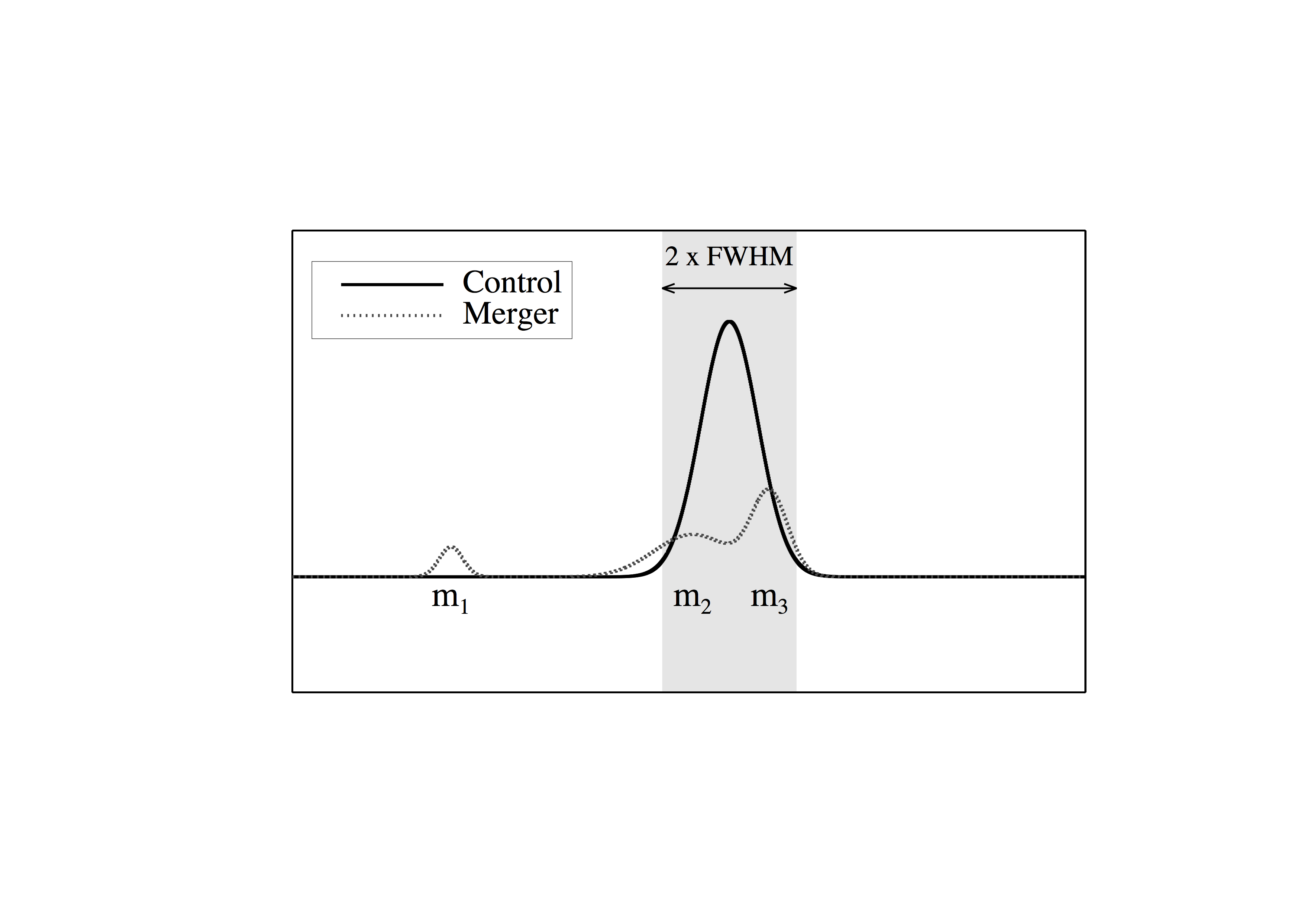}
\end{center}
\caption{A cartoon illustrating scenario~\ref{merge:step4} outlined in Appendix~\ref{App:merge}. The dark solid line represents the best-fitting solution contained in the control catalogue, which comprises a single velocity component. The lighter dotted line represents the best-fitting solution contained in the merger catalogue, which comprises three velocity components, ${\rm m}_{1}$, ${\rm m}_{2}$, and ${\rm m}_{3}$. In this specific example component ${\rm m}_{1}$ is retained since it is well-separated in velocity from the control component (and therefore does not represent a pseudo-component introduced due to the high-optical depth of the line). All components within the shaded region (${\rm m}_{2}$ and ${\rm m}_{3}$) are removed from the analysis. }
\label{Figure:merge_schem}
\end{figure}

Following the creation of the two catalogues we cycle through each position in the mapped region. We first of all remove all velocity components from the merger catalogue whereby $I^{\rm m}_{\rm peak}~<~I_{\rm tol}\sigma^{\rm m}_{\rm rms}$. We set $I_{\rm tol}=4$ and $I_{\rm tol}=8$ for the \ntwoh \ and HNC data files, respectively. This is to ensure that only high signal-to-noise components are included in the process. Following this, at each location there are four possible outcomes:

\begin{enumerate}

\item There are no available best-fitting solutions available (either in the control or merger catalogues). In this instance, we do nothing.\\

\item The location has an associated best-fitting solution in the control catalogue, but not in the merger catalogue. In this instance, the control catalogue solution is retained.\\

\item The location has an associated best-fitting solution in the merger catalogue, but not in the control catalogue. In this instance, the merger catalogue solution is retained.\\

\item There are best-fitting solutions available in both the control and merger catalogues. In this instance, all components where $|v^{\rm c}_{0, i}~-~v^{\rm m}_{0, j}|~<~[{8{\rm ln}(2)}]^{1/2}\sigma^{\rm c}_{\rm i}$, are removed from the merger catalogue (i.e those separated by less than the FWHM line-width from the control component). The components from the control catalogue and all remaining components from the merger catalogue are retained. A cartoon of this process is shown in Figure~\ref{Figure:merge_schem}. In this specific example, the control component and component ${\rm m}_{1}$ are retained, and components ${\rm m}_{2}$ and ${\rm m}_{3}$ are removed from the merger catalogue. \label{merge:step4}

\end{enumerate}

Once these steps are completed for every position in the map, the two files are merged together. As a final step we repeat this process this time using the merged HNCO and \ntwoh \ data file as the control catalogue, and the HNC data file as the merger. 

\begin{figure*}
\begin{center}
\includegraphics[trim = 20mm 40mm 0mm 80mm, clip, width = 0.98\textwidth]{Tex/Figures/Figs/Figure_number_of_components_hnco.png}
\includegraphics[trim = 20mm 40mm 0mm 80mm, clip, width = 0.98\textwidth]{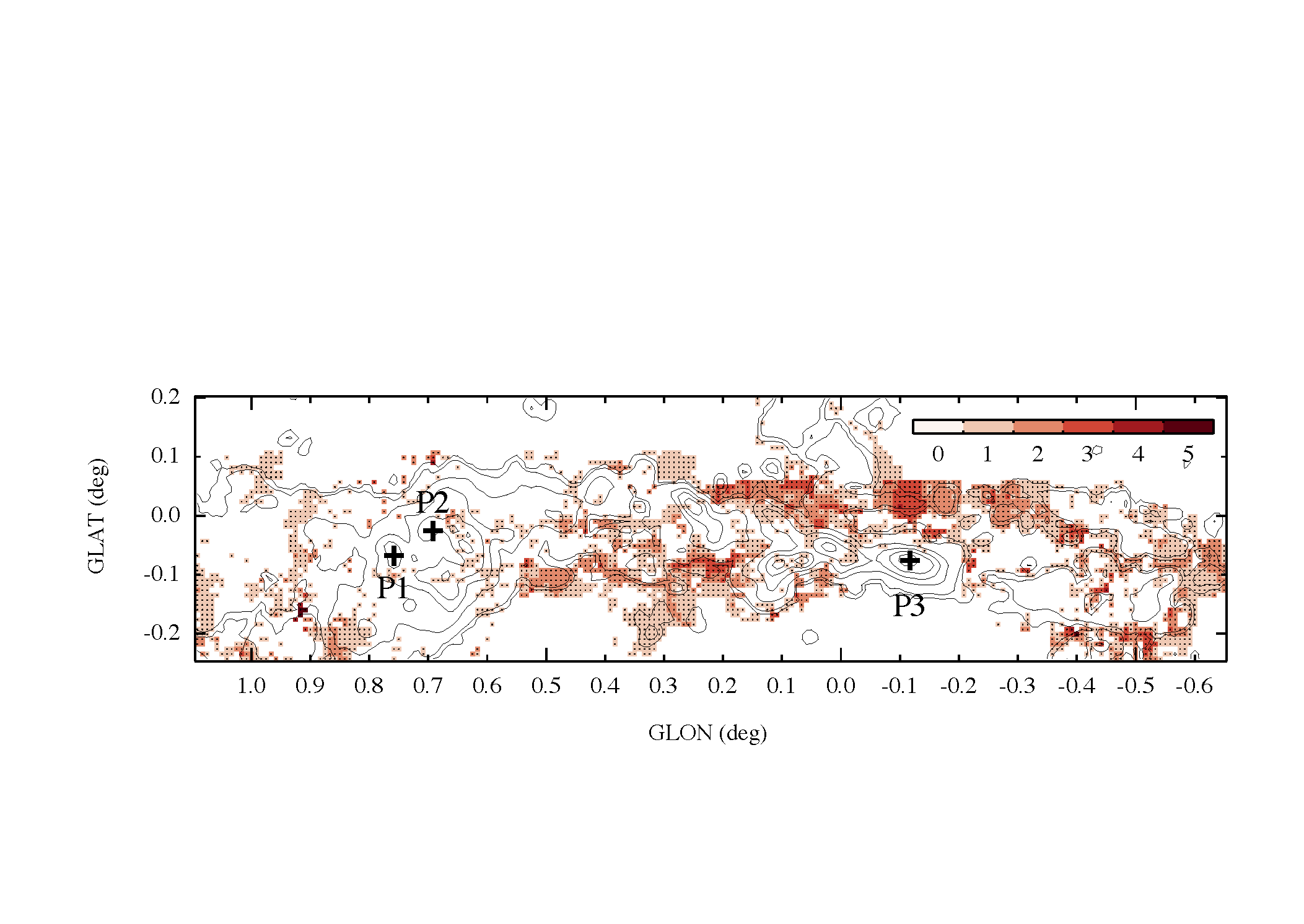}
\includegraphics[trim = 20mm 40mm 0mm 80mm, clip, width = 0.98\textwidth]{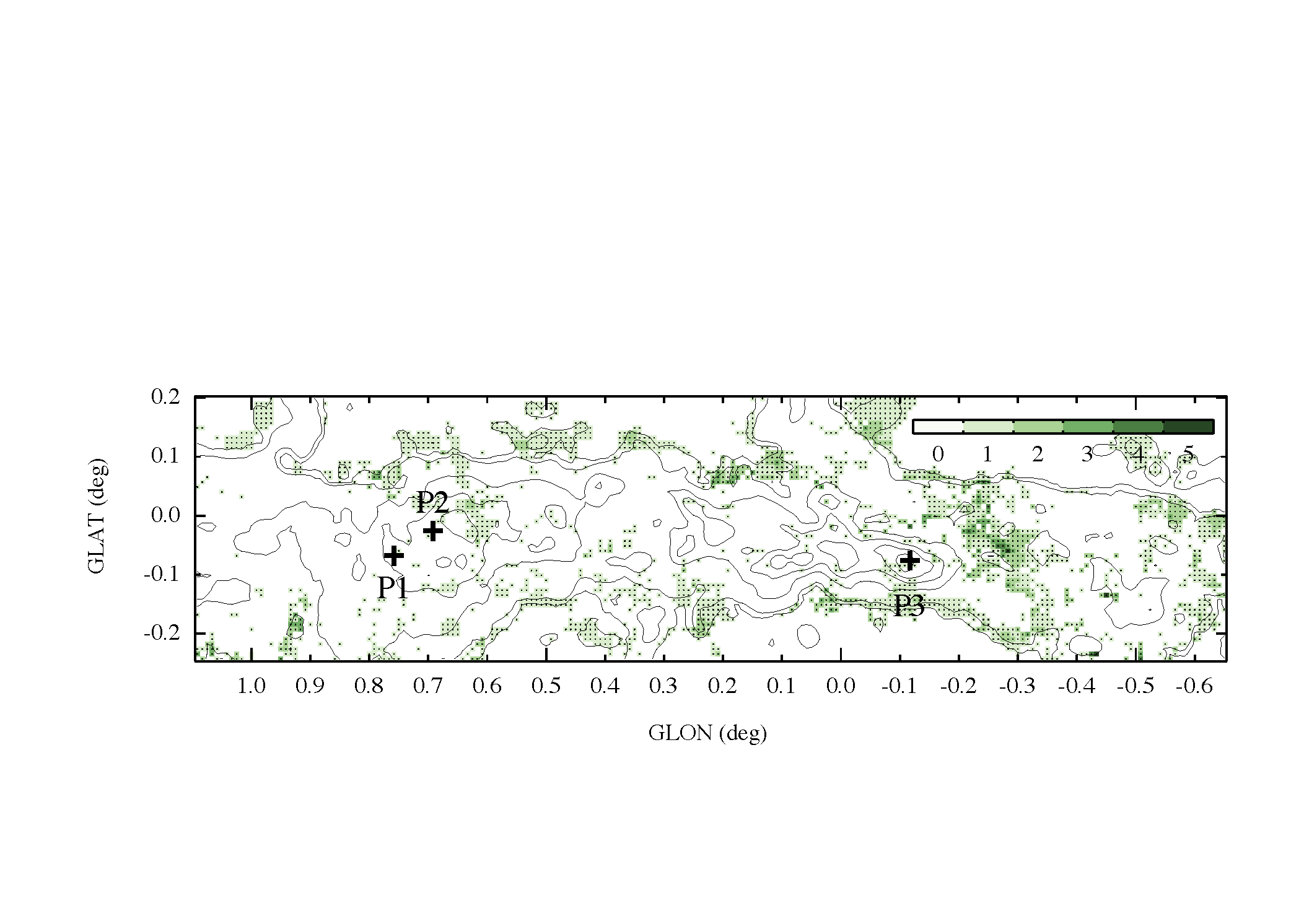}
\end{center}
\caption{Illustrating the result of the merger process. The top panel is equivalent to Figure~\ref{Figure:ncomp}. Each coloured pixel refers to a location with a corresponding best-fit solution. Each pixel is colour-coded according to the number of velocity components identified at that location. The central and bottom panels are the same but for the \ntwoh \ and HNC data, respectively (note that only components selected during the merging process outlined in Appendix~\ref{App:merge} are shown). Contours in all cases show the integrated intensity of the respective molecular line. Contour levels are, 1, 5, 25, 45, 65, and 85 per cent of peak (integrated) emission.}
\label{Figure:merge}
\end{figure*}

Figure~\ref{Figure:merge} visualizes the result of the merger process. The top panel is equivalent to Figure~\ref{Figure:ncomp}, each coloured pixel refers to the location of a best-fitting solution in the HNCO data. The colour of each pixel refers to the number of velocity components at that location. The central and bottom panels are the same but for the \ntwoh \ and HNC data, respectively. The images show that the \ntwoh \ and HNC fill in the gaps in the HNCO data. Where equivalent velocity components are identified in all three data files, only the HNCO components are retained. 

\section{Estimating the supernova rate of a star cluster}\label{App:SNrate}

Estimating the supernova rate of a star cluster is sensitive to several factors. Firstly, it depends on the number of massive stars, which in turn depends on the initial mass function (IMF) and the cluster mass, $M_{\rm cl}$. Secondly, it depends on the age of the cluster, since the lifetimes of stars are a strong function of their initial mass. 

To calculate the supernova rate within a cluster of mass $M_{\rm cl}$, we use a Chabrier IMF \citep{chabrier_2003} to generate a distribution of stellar masses with a fixed total mass of $M_{\rm cl}$. The Geneva evolutionary tracks of \citet{ekstrom_2012} are then used to determine the lifetimes of each star. Since these tracks are computed at discrete masses which are sparsely sampled at higher masses, a fifth-order polynomial is fit to the mass-lifetime relation to avoid interpolation artefacts. The supernova rate is then determined by simply counting the number of stars which ``disappear'' per time-step of 10$^{5}$ years. To estimate the impact of stochasticity, we run these simulations 100 times and measure the 1-$\sigma$ standard deviation in the supernova rate.

Figure~\ref{Figure:SNrate} shows the results of this analysis (which are further discussed in \S~\ref{Section:results_sgrc}). The overall shape of the supernova rate with cluster age is flat to within a factor of $\sim2$. This is due to the slopes of the IMF and the mass-lifetime relation being comparable. As one would expect, the stochastic effects are perfectly consistent with Poisson noise (shaded region). The rates shown are for the non-rotating Geneva models, though the impact of using different suites of evolutionary tracks (e.g. rotating) is minor (i.e. much smaller than the level of stochasticity).

%%%%%%%%%%%%%%%%%%%%%%%%%%%%%%%%%%%%%%%%%%%%%%%%%%

% Don't change these lines
\bsp	% typesetting comment
\label{lastpage}
\end{document}